\documentclass[
final
, nomarks
]{dmtcs-episciences}

\usepackage[square,sort,comma,numbers]{natbib}

\usepackage[svgnames,dvipsnames]{xcolor}
\usepackage{amssymb}
\usepackage[T1]{fontenc}
\usepackage{mathtools}
\usepackage{breqn}
\usepackage{xfrac}
\usepackage{braket}
\usepackage{physics}
\usepackage{tikz}
\usepackage{xparse}
\usepackage{float}
\usepackage{varwidth}
\usepackage[inline]{enumitem}
\usepackage{hyperref}
\usepackage[utf8]{inputenc}
\usepackage{subfigure}

\usepackage{xurl}

\usepackage{amsthm}

\usepackage[noabbrev,capitalize,nameinlink]{cleveref}
\usepackage{changepage}
\usepackage[normalem]{ulem}
\usepackage{booktabs}
\usepackage{array}
\usepackage{graphicx}
\usepackage{soul}
\usepackage[shortcuts]{extdash}
\usepackage{appendix}

\hypersetup{%
  bookmarksnumbered,
  pdfencoding=unicode,
}

\floatstyle{boxed}
\newfloat{algorithm}{tbp}{loa} % Creates the "algorithm" float
\floatname{algorithm}{Algorithm} % Sets its displayed name

\usetikzlibrary{automata, positioning, decorations.pathmorphing, calc, quotes, decorations.pathreplacing, backgrounds} %, arrows.meta}

\tikzset{
    every state/.append style={
        execute at begin node=$,
        execute at end node=$
    },
    initial text = 
}

\usepackage{thmtools}
\usepackage{thm-restate}
\declaretheorem{theorem}
\declaretheorem[sibling=theorem]{lemma,corollary}

\theoremstyle{definition}
\newtheorem{fact}[theorem]{Fact}

\crefname{fact}{fact}{facts}
\Crefname{fact}{Fact}{Facts}

\NewDocumentEnvironment{delineate}{m}{\textcolor{cyan!70!black!}{> > > > Begin: #1 > > > >}}{\textcolor{red!70!black!}{< < < < End: #1 < < < <}}

\mathchardef\mhyphen="2D

\renewcommand{\P}[1]{\mathcal{P}\paren*{#1}}

\DeclarePairedDelimiter{\ceil}{\lceil}{\rceil}
\DeclarePairedDelimiter{\floor}{\lfloor}{\rfloor}
\DeclarePairedDelimiter\paren\lparen\rparen
\DeclarePairedDelimiter\aparen\langle\rangle

\DeclarePairedDelimiter\sqparen{[}{]}

\DeclareMathOperator{\EX}{\mathbb{E}}
\newcommand{\exv}[1]{\ensuremath{\EX\sqparen*{#1}}}
\newcommand{\given}{\ensuremath{\:\middle|\:}}

\newcommand{\justOH}{\ensuremath{\mathit{O}}}
\newcommand{\OH}[1]{\ensuremath{\justOH\paren*{#1}}}

\newcommand{\OHMEGA}[1]{\ensuremath{\Omega\paren*{#1}}}
\newcommand{\OHMEGAp}[2][]{\ensuremath{\Omega\paren[#1]{#2}}}

\newcommand\langclassformat[1]{\begingroup\ensuremath{\rm #1}\endgroup}
\newcommand\restrictionformat[1]{\begingroup\ensuremath{\rm #1}\endgroup}

\newcommand\consX{\restrictionformat{con}}
\newcommand\logX{\restrictionformat{log}}

\newcommand\polyX{\restrictionformat{poly}}

\newcommand\rtX{\restrictionformat{rt}}
\newcommand\owayX{\restrictionformat{1way}}

\newcommand\Xspace[1]{\restrictionformat{#1\mhyphen\allowbreak{}space}}
\newcommand\Xrandom[1]{\restrictionformat{#1\mhyphen\allowbreak{}random\mhyphen\allowbreak{}bits}}
\newcommand\Xtime[1]{\restrictionformat{#1\mhyphen\allowbreak{}time}}
\newcommand\Xinput[1]{\restrictionformat{#1\mhyphen\allowbreak{}input}}

\newcommand{\vgeneric}[1]{\langclassformat{VER(\scriptstyle #1 \textstyle)}}
\newcommand{\vsri}[3]{\vgeneric{\Xspace{#1},\allowbreak\Xrandom{#2},\allowbreak\Xinput{#3}}}

\newcommand{\vtsr}[3]{\vgeneric{\Xtime{#1},\allowbreak\Xspace{#2},\allowbreak\Xrandom{#3}}}
\newcommand{\vsr}[2]{\vgeneric{\Xspace{#1},\allowbreak\Xrandom{#2}}}
\newcommand{\langclass}{\vsri{\consX}{\consX}{\rtX}}

\newcommand\ze{\restrictionformat{0}}
\newcommand\co{\restrictionformat{con}}
\newcommand\lo{\restrictionformat{log}}
\newcommand\li{\restrictionformat{linear}}
\newcommand\po{\restrictionformat{poly}}

\newcommand\rexlabel{\ensuremath{_\%}}
\newcommand\rex{\restrictionformat{\rexlabel}}
\newcommand\pexlabel{\ensuremath{^\%}}
\newcommand\pex{\restrictionformat{\pexlabel}}
\newcommand\pub{\restrictionformat{\mhyphen\allowbreak{}public}}
\newcommand\pri{\restrictionformat{\mhyphen\allowbreak{}private}}
\newcommand\spa{\restrictionformat{\mhyphen\allowbreak{}space}}
\newcommand\ran{\restrictionformat{\mhyphen\allowbreak{}coins}} % random\mhyphen\allowbreak{}bits}}
\newcommand\tim{\restrictionformat{\mhyphen\allowbreak{}time}}

\newcommand{\IPhighlabel}{\langclassformat{IP}}
\newcommand{\IPlabel}{\langclassformat{IP_*}}

\newcommand{\IPhigh}[1]{\ensuremath{\langclassformat{\IPhighlabel}(#1)}}
\newcommand{\IP}[1]{\ensuremath{\langclassformat{\IPlabel}(#1)}}

\newcommand{\CDEBhighlabel}{\langclassformat{CDEB}}

\newcommand{\CDEBhigh}[2][1]{\langclassformat{\CDEBhighlabel(\scalebox{#1}{$#2$})}}

\newcommand{\ODFAK}[1]{\langclassformat{1DFA\paren*{#1}}}
\newcommand{\ONFAK}[1]{\langclassformat{1NFA\paren*{#1}}}

\newcommand{\eL}[1]{\mathcal{L}\paren*{#1}}

\newcommand{\TNFAK}[1]{\ensuremath{\eL{\tnfak{#1}}}}
\newcommand{\TNFAS}{\TNFAK{*}}
\newcommand{\TNFAKT}[2]{\ensuremath{\eL{\tnfak{#1}, #2}}}
\newcommand{\TNFAST}[1]{\TNFAKT{*}{#1}}

\newcommand{\TNFASL}{\TNFAST{\li\tim}}

\newcommand{\PP}{\langclassformat{P}}
\newcommand{\NP}{\langclassformat{NP}}

\newcommand{\NL}{\langclassformat{NL}}
\newcommand{\REG}{\langclassformat{REG}}
\newcommand{\LOGCFL}{\langclassformat{LOGCFL}}
\newcommand{\NC}{\langclassformat{NC}}
\newcommand{\NCx}[1]{\ensuremath{\langclassformat{NC}^{#1}}}

\NewDocumentCommand{\DTIME}{ o m }{\langclassformat{DTIME}\IfValueTF{#1}{\paren[#1]{#2}}{\paren*{#2}}}

\newcommand{\Complete}{\-/\langclassformat{complete}}

\newcommand{\dfa}{\ensuremath{\mathrm{dfa}}}
\newcommand{\nfa}{\ensuremath{\mathrm{nfa}}}
\newcommand{\pfa}{\ensuremath{\mathrm{pfa}}}
\newcommand{\afa}{\ensuremath{\mathrm{afa}}}

\newcommand{\odfa}{\ensuremath{\mathrm{1\dfa}}}
\newcommand{\odfak}[1]{\ensuremath{\odfa\paren*{#1}}}

\newcommand{\tafa}{\ensuremath{\mathrm{2\afa}}}

\newcommand{\onfa}{\ensuremath{\mathrm{1\nfa}}}
\newcommand{\onfak}[1]{\ensuremath{\onfa\paren*{#1}}}
\newcommand{\tnfa}{\ensuremath{\mathrm{2\nfa}}}
\newcommand{\tnfak}[1]{\ensuremath{\tnfa\paren*{#1}}}

\newcommand{\onfaklnrr}[2]{\ensuremath{\ensuremath{\mathrm{1\mathrm{ns}\nfa}}\paren*{#1, #2}}}
\newcommand{\ONFAKLNRR}[2]{\ensuremath{\langclassformat{\mathrm{1\mathrm{NS}\mathrm{NFA}}}\paren*{#1, #2}}}

\newcommand{\machine}{\ensuremath{\mathrm{\pfa\mhyphen v}(\Xspace{\consX},\allowbreak\Xinput{\rtX})}}

\newcommand{\ith}[2][th]{\ensuremath{#2}\text{#1}}

\newcommand{\inpAlp}{\Sigma}
\newcommand{\certAlp}{\Gamma}

\newcommand{\certAlps}{\certAlp^{*}}

\newcommand{\langformat}[1]{\ensuremath{#1}}
\newcommand{\twin}{\langformat{L_{\mathrm{\langformat{twin}}}}}
\newcommand{\twinlong}{\ensuremath{\twin = \Set{w\#w | w \in \Set{0, 1}^*}}}
\newcommand{\pal}{\langformat{L_{\mathrm{\langformat{pal}}}}}
\newcommand{\nonpal}{\langformat{L_{\mathrm{\langformat{nonpal}}}}}
\newcommand{\lmatch}{\langformat{L_{\mathrm{\langformat{match}}}}}
\newcommand{\lfrac}{\ensuremath{L_{\frac23}}}

\newcommand\rev[1]{\ensuremath{#1^{R}}}

\newcommand{\lend}{\ensuremath{\rhd}}
\newcommand{\rend}{\ensuremath{\lhd}}

\newcommand{\Accept}{\emph{Accept}}

\newcommand{\accept}{\emph{accept}}
\newcommand{\reject}{\emph{reject}}

\newcommand{\accepts}{\emph{accepts}}
\newcommand{\rejects}{\emph{rejects}}

\newcommand{\err}{\ensuremath{\varepsilon}}

\setlist{itemsep=0pt}

\newlist{differences}{enumerate}{1}
\setlist[differences]{
    label=\arabic{*}.,
    ref=\arabic{*}
}
\crefname{differencesi}{difference}{differences}
\Crefname{differencesi}{Difference}{Differences}

\newsavebox{\stepsbox}
\newlength{\annotationwidth}
\newcommand\annotatesteps[2]{%
\begin{lrbox}{\stepsbox}%
\begin{minipage}{\linewidth}%
\raggedright%
#1%
\end{minipage}%
\end{lrbox}%
\par\usebox{\stepsbox}%
\rlap{\tikz[baseline]%, framed, inner frame sep=0pt]
{\draw[decorate, decoration={brace, mirror, amplitude=1ex, raise=1ex}] (0, -\dp\stepsbox) -- node[right=2ex]{\begin{minipage}{\annotationwidth-3ex}\raggedright\small #2\end{minipage}} (0, \ht\stepsbox)}}\par
}

\newcommand\mybar{\smash{\kern1pt\rule[1.5pt-\dp\strutbox]{0.1pt}{\baselineskip}\kern1pt}}

\newcommand\titem[1]{\item \bgroup
\setbox0=\vbox{\hsize=\linewidth%
\hangindent=0.4em% 1.5em% Controls the indent of the text
  #1}%
\bgroup
\dimen0=\ht0
\noindent\rlap{\begin{minipage}[b]{0.3em}% {1em}% Controls the indent of the line-wrap symbol
  \strut
  \loop\ifdim\dimen0 > \baselineskip
    \advance\dimen0 by -\baselineskip
    \null\hfill%\llap{\mybar}%\llap{$\vcenter{\hbox{\scalebox{0.55}[0.75]{$\hookrightarrow$}}}$}
    \ifdim\dimen0 > \baselineskip
        \newline
    \fi
  \repeat
\end{minipage}}\egroup\box0
\egroup}

\usepackage{tracefnt}

\newlist{turinglist}{itemize}{4}
\setlist*[turinglist]{
    itemsep=0em,
    parsep=0em,
    partopsep=0em,
    topsep=0em,
    labelsep=0em,
    label=,
    labelwidth=0em,
    align=parleft,
    itemindent=0em,
}
\setlist*[turinglist,1]{leftmargin=0em, first=\raggedright} %leftmargin=0.5em}
\setlist*[turinglist,2,3,4]{leftmargin=1em}

\newlist{turingmultienum}{enumerate}{4}
\setlist*[turingmultienum]{
    itemsep=0em,
    parsep=0em,
    partopsep=0em,
    topsep=0em,
    itemindent=0em,
    labelsep=0em,
    align=parleft,
    label*=\bfseries\arabic*.
}
\setlist*[turingmultienum,1]{leftmargin=2em, labelwidth=\leftmargin-0.5em, ref=\arabic*, first=\raggedright}

\setlist*[turingmultienum,2,3,4]{topsep=0em, leftmargin=1em}
\setlist*[turingmultienum,2]{labelwidth=1.5em+\leftmargin, ref=\arabic{turingmultienumi}.\arabic*}
\setlist*[turingmultienum,3]{labelwidth=2.5em+\leftmargin, ref=\arabic{turingmultienumi}.\arabic{turingmultienumii}.\arabic*}
\setlist*[turingmultienum,4]{labelwidth=3.5em+\leftmargin, ref=\arabic{turingmultienumi}.\arabic{turingmultienumii}.\arabic{turingmultienumiii}.\arabic*}

\crefname{turingmultienumi}{Stg.}{Stgs.}
\Crefname{turingmultienumi}{Stage}{Stages}
\crefname{turingmultienumii}{Stg.}{Stgs.}
\Crefname{turingmultienumii}{Stage}{Stages}
\crefname{turingmultienumiii}{Stg.}{Stgs.}
\Crefname{turingmultienumiii}{Stage}{Stages}
\crefname{turingmultienumiiii}{Stg.}{Stgs.}
\Crefname{turingmultienumiiii}{Stage}{Stages}

\newlist{turingenum}{enumerate}{1}
\setlist[turingenum]{
    itemsep=-0.5em,
    parsep=0.5em,
    partopsep=-0.5em,
    labelsep=.5em,
    leftmargin=!,
    labelwidth=1.5em,
    label=\bfseries\arabic{*}.,
    ref=\arabic{*}
}

\crefname{turingenumi}{Stg.}{Stgs.}
\Crefname{turingenumi}{Stage}{Stages}

\newenvironment{turing}[2]
 {\begin{enumerate}[labelsep=0pt,align=left,parsep=0pt,leftmargin=\widthof{$#1={}$}+\parindent,listparindent=0pt,labelwidth=\widthof{$#1={}$}]
  \item[$#1={}$]``\ignorespaces#2
  \begin{turingenum}[
    partopsep=0em,
    topsep=0em,
    leftmargin=\labelwidth %\widthof{$#1={}$}-0.6em %+1.5em
  ]}
 {\unskip''\end{turingenum}\end{enumerate}}

\newcommand{\bitem}[1]{\item\begin{adjustwidth}{1em}{0pt}\ignorespaces#1\end{adjustwidth}}
\newcommand{\bbitem}[1]{\item\begin{adjustwidth}{2em}{0pt}\ignorespaces#1\end{adjustwidth}}

\NewDocumentCommand{\branch}{ m o }{\makebox{\textsc{[#1]}\IfValueT{#2}{ Branch #2:}}}
\NewDocumentCommand{\branchpr}{ m o }{\branch{\,Probability: \ensuremath{#1}\,}[#2]}
\NewDocumentCommand{\branchperc}{ m o }{\branch{#1\% prob.}[#2]}

\makeatletter
\def\squiggly{\bgroup \markoverwith{\lower3.9\p@\hbox{\sixly \scalebox{1.2}[0.65]{\char58}}}\ULon}
\makeatother

\def\mysout{\leavevmode\bgroup\def\ULthickness{1pt}\ULdepth=-.4ex\ULset}

\newcommand{\stkout}[1]{\begingroup\ifmmode\text{\mysout{\ensuremath{#1}}}\else\mysout{#1}\fi\endgroup}
\newcommand{\utkanadd}[2][50]{\begingroup\color[Hsb]{210,1,1}%\tikz[overlay]{\draw[color=blue!#1!cyan!50!black,line width=7,opacity=0.25] (0,0.1) circle (0.5);}
#2\endgroup}

\newcommand{\utkanrem}[1]{\begingroup\color{red!80!black!70}\tikz[overlay]{\draw[color=red!50!black!70,line width=7,opacity=0.25] (0,0.1) circle (0.5);}\stkout{#1}\endgroup}

\newcommand{\utkanworry}[1]{\textcolor{red!45!black!90}{\ifmmode\smash[b]{\squiggly{#1}}\else\squiggly{#1}\fi}}

\renewcommand{\textvisiblespace}[1][.7em]{%
  \makebox[#1]{%
    \kern.07em
    \vrule height.5ex
    \hrulefill
    \vrule height.5ex
    \kern.07em
  }% <-- don't forget this one!
}

\DeclareMathAlphabet{\mathsl}{\encodingdefault}{\familydefault}{m}{sl}
\SetMathAlphabet{\mathsl}{bold}{\encodingdefault}{\familydefault}{bx}{sl}

\author{M. Utkan Gezer\thanks{Utkan Gezer's participation in this work is supported by the Turkish Directorate of Strategy and Budget under the TAM Project number 2007K12-873.}
  \and A. C. Cem Say}

\title{$\mathrm{P}$ has polynomial-time finite-state verifiers\thanks{This research was partially supported by Bo\u{g}azi\c{c}i University Research Fund Grant Number 19441.  An earlier version of this paper~\cite{GS23} was presented in the 24th Italian Conference on Theoretical Computer Science, Palermo, Italy, September 13--15, 2023.  This is a substantially extended version.}}

\affiliation{
  Department of Computer Engineering,
  Bo\u{g}azi\c{c}i University,
  \.{I}stanbul, % onlarin orneklerine gore cikardim ama bunu yine de koyasim var
  T\"{u}rkiye
  }


\newcommand{\eg}{\textit{e.g.}}

\newcommand{\ie}{\textit{i.e.}}

\newcommand{\False}{\texttt{false}}
\newcommand{\True}{\texttt{true}}

\DeclareMathOperator\bin{bin}
\newcommand\binleq[1]{\ensuremath{\mathit{BIN}_{\leq #1}}}
\newcommand\maj[1]{\ensuremath{\mathit{MAJ}_{#1}}}
\newcommand\pprefix[1]{\ensuremath{\mathit{PROPER\mhyphen{}PREFIX}\paren*{#1}}}
\newcommand\prefix[1]{\ensuremath{\mathit{PREFIX}\paren*{#1}}}

\newcommand\TICK{\textsf{TICK}}
\newcommand\HALVE{\textsf{HALVE}}
\newcommand\DOUBLE{\textsf{DOUBLE}}

\makeatletter
\AtBeginDocument{% <-- add this
    \catcode`_=12
    \begingroup\lccode`~=`_
    \lowercase{\endgroup\let~}\sb
    \mathcode`_="8000
    \immediate\write\@auxout{\catcode`_=12 }% <-- add this
    \immediate\write\@auxout{\catcode`^=12 }% <-- add this
}
\makeatother

\makeatletter
\def\@testdef #1#2#3{%
  \def\reserved@a{#3}\expandafter \ifx \csname #1@#2\endcsname
 \reserved@a  \else
\typeout{^^Jlabel #2 changed:^^J%
\meaning\reserved@a^^J%
\expandafter\meaning\csname #1@#2\endcsname^^J}%
\@tempswatrue \fi}
\makeatother

\newcommand\paragraphEnd{\vspace{3.25ex plus 1ex minus .2ex}}

\begin{document}

\publicationdata{vol. 27:3}{2025}{15}{10.46298/dmtcs.13854}{2024-07-01; 2024-07-01; 2025-02-25}{2025-10-02}

\maketitle

\begin{abstract}
Interactive proof systems whose verifiers are constant-space machines have interesting features that do not have counterparts in the better studied case where the verifiers operate under reasonably large space bounds. The language verification power of finite-state verifiers is known to be sensitive to the difference between private and public randomization.
These machines also lack the capability of imposing worst-case superlinear bounds on their own runtime, and long interactions with untrustable provers can involve the risk of being fooled to loop forever.
% No analogue of the theorem that guarantees  perfect completeness (the property that the verifier will accept members of the language under consideration with probability 1) is known for finite\-/state machines. We address these issues by  
We  analyze such verifiers under different bounds on the numbers of private and public random bits that they are allowed to use. This separate accounting for the private and public coin budgets as resource functions of the input length provides interesting characterizations of the collections of the associated languages. When the randomness bound is constant, the verifiable class is $\mathrm{NL}$ for private-coin machines, but equals just the regular languages when one uses public coins. Increasing the public coin budget while keeping the number of private coins constant augments the power: We show that the set of languages that are verifiable by such machines in expected polynomial time (with an arbitrarily small positive probability of looping) equals the complexity class $\mathrm{P}$. This hints that allowing a minuscule probability of looping may add significant power to polynomial-time finite-state automata, since it is still not known whether those machines can verify all of $\mathrm{P}$ when required to halt with probability 1, even with no  bound on their private coin usage. We also show that logarithmic-space machines which hide a constant number of their coins are limited to verifying the languages in $\mathrm{P}$.
\end{abstract}

\section{Introduction}

In addition to providing a new perspective on the age\-/old concept of proof, and offering  possibilities for weak clients to check the correctness of difficult computations that they delegate to powerful servers, interactive proof systems also play an important role in the characterization of computational complexity classes~\cite{GMR89,GKR15,Sha92}.
% !!!!İlki89 diyor ama 85 versiyonuna gidiyor, 89 journal versiyonu yapalım!!sonuncusu: https://dl.acm.org/doi/pdf/10.1145/146585.146609!!! Utkan: Yaptim
These systems involve a computationally weak verifier (a probabilistic Turing machine with small resource bounds) engaging in a dialogue with a  strong but possibly malicious prover, whose aim is to convince the verifier that a common input string is a member of the language under consideration. If the input is a non\-/member, the prover may well lie during this exchange to mislead the verifier to acceptance, or to trick it into running forever instead of rejecting. Interestingly, this setup allows the weak machines to be able to verify (that is, to determine the membership status of any given string with low probability of being fooled) a larger class of languages than they can manage to handle in a stand\-/alone fashion, \ie, without engaging with a prover. 

Several specializations of the basic model described above have been studied until now. One parameter is whether the prover can directly see the outcomes of the random choices made by the verifier or not. A \emph{private\-/coin} system hides the results of the verifier's coin flips from the prover, and the verifier only sends information that it deems necessary through the communication channel. \emph{Public\-/coin} systems, on the other hand, hide nothing from the prover, who can be assumed to observe the coin flips and deduce the  resulting changes to the configuration of the verifier as they unfold. It is known~\cite{DS92} that private\-/coin systems are more powerful (\ie, can verify more languages) than public\-/coin ones when the verifiers are restricted to be constant\-/space machines, but this distinction vanishes when the space restriction is lifted~\cite{GS86}.

In this paper, we study the capabilities of constant\-/space verifiers (essentially, two\-/way probabilistic finite\-/state automata) which are allowed to hide some, but not necessarily all, of their coin flips from the prover. The separate accounting for the private and public coin budgets as resource functions of the input length provides new characterizations of the collections of associated languages in terms of well known complexity classes.

The rest of the paper is structured as follows: \Cref{sec:prel} provides the preliminaries. A definition of interactive proof systems which allows separate accounting of private and public coin usage is given in \Cref{subs:vers}, together with notation that generalizes the standard IP classes in an  accordingly parameterized manner. Since some of our constructions involve verifiers which do not halt with probability $1$, we introduce notation that enables us to talk about resource bounds that  apply in the (highly probable) cases where the machines do halt. %\Cref{subs:deb} reviews debate systems,  a generalization of alternation %\utkanadd{(akin to how interactive proof systems are to nondeterminism)} 
%that will be used in some of our characterizations.
Our main result uses a technique that depends on the relationship between logarithmic\-/space computation and multihead finite automata. \Cref{subs:nfaks} contains a quick review of that fact.  \Cref{sec:clock}  describes two methods for implementing probabilistic ``alarm clocks'' for later use in our proofs. \Cref{subs:afas} presents the definition of the two\-/way alternating finite automaton model, which will be employed in the proof of  \Cref{thm:conspubliconly}.

We start \Cref{sec:prelresults} by noting that  finite\-/state verifiers which are only allowed to flip $\OH{1}$ private coins are already known to outperform deterministic verifiers, and consider the analogous question about public coins. It turns out (\Cref{thm:conspubliconly}) that finite\-/state verifiers tossing only $\OH{1}$ public coins can verify only regular languages. This section also includes a technical result (\Cref{thm:strictboundislinear}) establishing that one has to consider bounds on expected, rather than worst\-/case,  %\utkanadd{[Utkan: ``expected bound'' KELİMELERİN ANLAMINI DÜŞÜNÜNCE MANTIKLI BİR TAMLAMA DEĞİL GİBİ. ``expected usage'' OLUR, ``statistical bound'' OLUR, AMA BOUND EXPECT EDILEMEZ]} on the 
runtimes of finite\-/state verifiers to be able to conduct a meaningful study of superlinear resource bounds.

Our main result (\Cref{thm:P}, \Cref{sec:results}) is a demonstration of the equality of the set of languages verifiable by polynomial\-/time finite\-/state machines that flip $\OH{1}$ private coins and polynomially many public coins to the complexity class $\PP$. (The verifiers in question are allowed to be fooled by malicious provers to run forever without halting, but the probability of this error can be bounded below any desired positive value. The bounds on the runtime and number of public coin flips apply to halting computations.)
%\utkanadd{and only in the ``expected'' sense}.) 
The proof builds on previous work~\cite{con89,GKR15,SY14,GS22} on logarithmic\-/space verification, prover\-/aided simulation of multihead automata by single\-/head probabilistic finite automata, and specialized loop avoidance techniques for such verifiers. Notably, \Cref{thm:P} hints that allowing a minuscule probability of looping may add significant power to  polynomial\-/time finite\-/state machines, since it is still not known whether those machines  can verify all of $\PP$ when required to halt with probability $1$, even with no  bound on their private coin usage. We also show that logarithmic\-/space machines which hide a constant number of their coins are limited to verifying the languages in $\PP$, even with arbitrarily large time and public\-/coin budgets.

\Cref{sec:conc} lists some open problems.  Detailed constructions involved in some of the proofs are presented in the Appendix to avoid cluttering the main text.
%We show that allowing these machines to use even only a constant number\footnote{That is, the number of such private coin flips is fixed, regardless of the length of the input.} of private coins enlarges the class of verified languages considerably. We adapt several previous results from the literature to this framework. Our main result (\Cref{thm:P}) is  a new characterization of the complexity class $\PP$ as the set of languages that can be verified by such machines in polynomial time with arbitrarily small error.   

\section{Background}\label{sec:prel}

\subsection{Interactive proof systems}\label{subs:vers}
%HEM CONSTANT-SPACE, HEM DE LOGSPACE (KARŞILIKLI KONUŞMALI(BUNUN TANIMI YS14'TEN BAKILABİLİR, COMMUNİCATİON CELL FALAN)) VE BİRAZ PRIVATE BİRAZ PUBLİC COİNLU VERIFIERLARIN TANIMI, ACCEPTANCE, ERROR FALAN TANIMI, DİL SINIFI NOTASYONU

We start by providing definitions of interactive proof systems and related language classes that are general enough to cover finite\-/state verifiers with both private and public coins, as well as the more widely studied versions with greater memory bounds~\cite{DS92,GMR89,B85}. %!!!BURAYA muggles paperının reflerinde bulabileceğin  Goldwasser et al. [1989] ve Babai [1985] reflerini EKLEYELİM!!! Utkan: Ekledim

% https://tex.stackexchange.com/a/120231/69346
\NewDocumentCommand{\vartextvisiblespace}{ O{.7em} O{.7ex} }{%
  \makebox[#1]{%
    \kern.07em
    \vrule height#2
    \hrulefill
    \vrule height#2
    \kern.07em
  }% <-- don't forget this one!
}

\newcommand{\estring}{\ensuremath{\lambda}}
\newcommand{\blanksymb}{\ensuremath{\text{\vartextvisiblespace}}}

\NewDocumentCommand{\qacc}{ e{_} o }{\ensuremath{q_{\IfValueT{#2}{#2,}\mathrm{acc}\IfValueT{#1}{,#1}}}}
\NewDocumentCommand{\qrej}{ e{_} o }{\ensuremath{q_{\IfValueT{#2}{#2,}\mathrm{rej}\IfValueT{#1}{,#1}}}}

\NewDocumentCommand{\Qpub}{ e{_} o }{\ensuremath{Q_{\IfValueT{#2}{#2,}\mathrm{pub}\IfValueT{#1}{,#1}}}}
\NewDocumentCommand{\Qpri}{ e{_} o }{\ensuremath{Q_{\IfValueT{#2}{#2,}\mathrm{pri}\IfValueT{#1}{,#1}}}}
\NewDocumentCommand{\Qcom}{ e{_} o }{\ensuremath{Q_{\IfValueT{#2}{#2,}\mathrm{com}\IfValueT{#1}{,#1}}}}
\NewDocumentCommand{\bpub}{ e{_} o }{\ensuremath{b_{\IfValueT{#2}{#2,}\mathrm{pub}\IfValueT{#1}{,#1}}}}
\NewDocumentCommand{\bpri}{ e{_} o }{\ensuremath{b_{\IfValueT{#2}{#2,}\mathrm{pri}\IfValueT{#1}{,#1}}}}

\NewDocumentCommand{\Qex}{ e{_} o }{\ensuremath{Q_{\IfValueT{#2}{#2,}\exists\IfValueT{#1}{,#1}}}}
\NewDocumentCommand{\Qun}{ e{_} o }{\ensuremath{Q_{\IfValueT{#2}{#2,}\forall\IfValueT{#1}{,#1}}}}

% \utkanadd{
An \emph{interactive proof system (IPS)} for some language $L$ is a protocol between  a \emph{verifier} and a \emph{prover}. The verifier is a probabilistic Turing machine tasked with determining whether the prover's argument is sufficiently convincing to conclude that the input string is a member of $L$. %!!!!!ŞU CÜMLEYİ SİLİNCE BİR ŞEY KAYBETMEYİZ DİYE DÜŞÜNEREK COMMENTLEDİM:!!! \utkanadd{Utkan: Bence dogru olmak zorunda da degildi zaten. Sadece oyle kabul etsek de sorun olmuyordu.} %The prover always claims that the input string is in $L$, regardless of whether this is actually the case or not. 
The messages of the prover at any step during its communication with the verifier are determined by a  (not necessarily computable) function of  the input string and the transcript of the communication up to that point. This function maximizes the probability that the verifier accepts if the input is in $L$, and minimizes the probability that it rejects otherwise.

\newcommand{\randmark}{\ensuremath{\tikz[x=1.2ex,y=0.7ex,ultra thin]{
\draw[->] (0,0) .. controls (0.5,0) and (0.1,1) .. (1,1);
\draw[ultra thick,white] (0,1) .. controls (0.5,1) and (0.1,0) .. (1,0);
\draw[->] (0,1) .. controls (0.5,1) and (0.1,0) .. (1,0);
}}}

The verifier has 
%private access to  !!!!BÖYLE DERSEN PROVER İNPUTU GÖRMÜYOR 
a read\-/only input tape, a read\-/write work tape, and %shared access (shared with the prover) to 
a read\-/write communication cell for interacting with the prover.  It is modeled as a 6\=/tuple $\paren*{Q,\Sigma,\Phi,\Gamma,\delta,q_0}$, with the following components:
% }
% \utkanrem{A \emph{verifier} in an \emph{interactive proof system (IPS)} is a 6-tuple $(Q,\Sigma,\Phi,\Gamma,\delta,q_0)$, where}
\begin{enumerate}
    \item $Q$ is the finite set of states. The  following subsets of $Q$ are not necessarily disjoint, unless specified otherwise: % \utkanrem{such that $Q = \Qpri \cup \Qpub \cup \Set{\qacc, \qrej}$ where}
    \begin{itemize}
        \item $\Qpri$ is the set of states that flip private coins. %, as well as public coins,
        \item $\Qpub$ is the set of states that flip public coins. % \utkanrem{$\Qpri \subseteq \Qpub$, and} % are not associated with private coins, such that $\Qpri \cap \Qpub = \emptyset$, and
        \item $\Qcom$ is the set of communication states, \ie, those that
        write to the communication cell.  %A public coin flip entails communication of the outcome, therefore 
        $\Qpub \subseteq \Qcom$.
        \item $\Set{\qacc, \qrej}$ % \utkanrem{$\qacc, \qrej \notin \Qpri \cup \Qpub$}
        are the accept and reject states, respectively. $\qacc, \qrej \notin \Qpri %\cup \Qpub
        \cup \Qcom$.%!!!!BU İFADEDEN Qpub ÇIKSA DA BİRAZ KISALSA????!!!!!!!!!! \utkanadd{Utkan: bilemedim, ayni sey tabii ki, yine de sanki acik acik yazilmasi daha sade. eger ikinci satira tasmamasini saglamak icinse, dokuman formatini submit edecegimiz yerin formatina degistirince muhtemelen bu da degisecek (ya cikarmis olmak yeterli olmayacak ya da cikarmamis olsak bile tek satirda olacak)}
    \end{itemize}
    \item $\inpAlp$ is the input alphabet.
    \item $\Phi$ is the work tape alphabet, containing the special blank symbol \blanksymb.
    % \utkanrem{which is guaranteed to include the special blank symbol, \blanksymb{}.}
    \item $\certAlp$ is the communication alphabet. $\blanksymb \in \certAlp$.
    % $\certAlp_{\randmark} = \certAlp \sqcup \Set{0_{\randmark}, 1_{\randmark}}$
    \item $\delta$ %: Q \times \inpAlp^\prime \times \Phi \times \certAlp^\prime \times C \to Q \times \Delta$ 
    is the transition function, described below.
    \item  $q_0$ is the initial state. $q_0 \in Q$.
\end{enumerate}

% \utkanadd{
The computation of a verifier on an input string $w \in \Sigma^*$ is initialized as follows:
\begin{itemize}
    \item The input tape contains $\lend w \rend$, where $\lend, \rend \notin \Sigma$ are the left and right end\-/markers, respectively. The input tape head  starts on the left end\-/marker.
    \item The work tape is filled with blank symbols, and the work tape head is positioned at the beginning of the tape.
    \item The communication cell is also blank.
\end{itemize}
% }

% \utkanrem{The computation of a verifier is initialized with $\lend w \rend$ written on its read\-/only input tape, where $w \in \Sigma^*$ is the input string, and $\lend, \rend \notin \Sigma$ are the left and right end\-/markers, respectively. The input head of the verifier is initially on the left end\-/marker.
% %first symbol of $w$.
% The read\-/write work tape is initially filled with blank symbols, with the work tape head positioned at the beginning of the tape.  Apart from these two tapes, the verifier also has access to a communication cell, which is a single\-/cell tape that is initially blank.}

\newcommand\query{\ensuremath{\mathit{query}}}
\newcommand\qneg[1]{\ensuremath{\overline{#1}}}

Let $\Sigma_{\bowtie} = \Sigma \cup \Set{\lend, \rend}$.
% Let $\certAlp_{\estring} = \certAlp \cup \Set{\estring}$, where $\estring$ is the zero\-/length ``empty string'' that is the identity element of string concatenation (\ie, $x\lambda = \lambda x = x$ for any string $x$).
Let $\Delta = \Set{-1, 0, +1}$ be the set of possible head movements, where $-1$ means ``move left'', $0$ means ``stay put'', and $+1$ means ``move right''.  
Let $\qneg{A}$ denote the complement of a set $A$, \eg, $\qneg{\Qpri} = Q \setminus \Qpri$. % !!!!BU İŞ İÇİN EĞRİ DEĞİL DÜZ ÇİZGİ STANDART, DEĞİL Mİ???!?!?!!!!!!!!! Utkan: Evet, bence de öyle. Eskiden 'complement EKSİ {qacc, qrej}' şeklinde tanımlamıştım, herhalde ondan ötürü böyle eğri kalmış. Dediğiniz gibi yaptım.
%\utkanrem{$\qneg{\Qpri} = Q \setminus \Qpri \setminus \Set{\qacc, \qrej}$, and let $\qneg{\Qpub}$ and $\qneg{\Qcom}$ be defined similarly.}
The computation of the verifier is governed by its transition function $\delta$, which is defined in parts in the manner depicted in \Cref{tab:transitionfunction} and explained below.

\begin{table}[!ht]
    \centering
    \caption{Parts of the verifier's transition function.}\label{tab:transitionfunction}
    \renewcommand*{\arraystretch}{1.4}
    $$
    \begin{array}{l l}
        \text{\textbf{Case}} & \text{\textbf{Mapping}}\\\midrule
        q \in \Qpri \cap \Qpub &
        \delta\paren{q,\sigma,\phi,\gamma,\bpri,\bpub} =
        % \paren{q',\phi',\paren{\gamma',\bpub},d_i,d_w}\\%\midrule
        \paren{q',\phi',\gamma',d_i,d_w}\\%\midrule
        q \in \Qpri \cap \qneg{\Qpub} \cap \Qcom &
        \delta\paren{q,\sigma,\phi,\gamma,\bpri} =
        \paren{q',\phi',\gamma',d_i,d_w}\\%\midrule
        q \in \Qpri \cap \qneg{\Qpub} \cap \qneg{\Qcom} &
        \delta\paren{q,\sigma,\phi,\gamma,\bpri} =
        \paren{q',\phi',d_i,d_w}\\%\midrule
        % % % %
        q \in \qneg{\Qpri} \cap \Qpub &
        \delta\paren{q,\sigma,\phi,\gamma,\bpub} =
        % \paren{q',\phi',\paren{\gamma',\bpub},d_i,d_w}\\%\midrule
        \paren{q',\phi',\gamma',d_i,d_w}\\%\midrule
        q \in \qneg{\Qpri} \cap \qneg{\Qpub} \cap \Qcom &
        \delta\paren{q,\sigma,\phi,\gamma} =
        \paren{q',\phi',\gamma',d_i,d_w}\\%\midrule
        q \in \qneg{\Qpri} \cap \qneg{\Qpub} \cap \qneg{\Qcom}
        \setminus \Set{\qacc, \qrej} &
        \delta\paren{q,\sigma,\phi,\gamma} =
        \paren{q',\phi',d_i,d_w}%\\\bottomrule
    \end{array}
    $$
\end{table}

%These parts of the transition function dictate the following orderly procedure, carried out by the verifier 
At each step of its computation, the verifier does the following:
\begin{itemize}
    \item It reads the symbols $\paren{\sigma, \phi, \gamma} \in \Sigma \times \Phi \times \Gamma$ from its input tape, work tape, and communication cell  respectively.  If its current state $q$ is in $\Qpri$, it tosses a private coin, obtaining the outcome $\bpri \in \Set{0, 1}$.  If $q$ is in $\Qpub$, it tosses a public coin, obtaining the outcome $\bpub \in \Set{0, 1}$.
    \item The value of $\delta$ corresponding to this tuple (see the   template of the corresponding part of $\delta$ in \Cref{tab:transitionfunction}) dictates that the machine will switch to state $q' \in Q$,  write $\phi' \in \Phi$ to its work tape, and move its input and work tape heads in the directions  $d_i, d_w \in \Delta$,  respectively.  The verifier 
    %determines the symbol to write to
    overwrites the communication cell with $\gamma' \in \Gamma$ if
    % $q \in \Qcom \cap \qneg{\Qpub}$, and with $\paren{\gamma',\bpub} \in \Gamma$ !!!!!BUNA IN GAMMA DEMESEK DAHA MI HAYIRLI PEDAGOJİK OLARAK? BÖYLECE GAMMA'YA İKİNCİ ELEMANI 0 OLAN BİR TUPLE KOYSALAR BİLE AMBIGUITY OLMAZ!!!!!!!! if $q \in \Qpub$.
    $q \in \Qcom$.  If $q \in \Qpub$, the outcome of the public coin flip (\bpub) is also  communicated to the prover automatically through a separate channel.% \utkanadd{in this stage}.
\end{itemize}

%The prover comes into play for a brief moment, immediately after 
Each time the verifier writes a symbol to the shared communication cell, the prover overwrites that symbol with a (possibly different) member of $\Gamma$, determined as a function of $w$, the history of public coin outcomes, and the communication symbols written to the communication cell up to that point. %\footnote{\utkanadd{Note that the verifier might, at times, want to simply query the prover for a new communication symbol without having anything to say.  To initiate this interaction, the verifier may write a special communication symbol to the communication cell, signifying just that.}} 
% !!!!BU PARAGRAFIN GERİSİNDE YAZDIKLARINI COMMENTLEYİP REWRITE ETTİM; NASIL??!!!!!!!! Utkan: ok gibi; prover'in neyi bilip neyi bilmedigini daha iyi mi aciklasak diye dusundum
Since  the prover is supposed to embody the optimal function to convince (or, when  the input string is not in the language $L$, to deceive) the verifier, one can assume that the prover knows the algorithm of the verifier. Note, however, that the prover does not see the private coin outcomes, tape head positions, work tape content, and internal state of the verifier.

    We define the \emph{configuration} of a verifier at any given time as the tuple composed of its state, the contents of its work tape, the symbol in the communication cell, and the positions of its input and work tape heads.
A verifier halts with acceptance (rejection) when it executes a transition entering \qacc{} (\qrej).  Any transition that moves the input head beyond an end\-/marker delimiting the string written on the read\-/only input tape leads to a rejection, unless that last move enters \qacc. 
% !!!!ŞU CÜMLE REF2 İÇİN YETER Mİ???!!! 
Any transition that attempts to move the work head off the left end of its tape also leads to rejection. 
% !!!OK??!!!!!!!!!!!!!! \utkanadd{UTKAN: Bence yeterli.}
% The head on the certificate tape is defined to be one\-/way, since it is known~\cite{AB09} that allowing two\-/way access to that tape can lead to ``unfair'' accounting of the space usage.
Note that the verifier may possibly never halt, in which case it is said to be looping.

% To describe the language recognized by a verifier and also the error involved due to the randomness employed, we consider acceptances and rejections of the verifier when paired with the ``best'' prover (or \emph{one of the best}, when there are many satisfying that condition).  The best prover is the function that, for each input string, maximizes the acceptance probability of the verifier as long as there exists a function that makes that probability greater than $\sfrac23$ given its interaction with the verifier so far, and minimizes the rejection probability otherwise.\footnote{The choice of $\sfrac23$ (and accordingly the $\sfrac13$ on the next paragraph) is arbitrary. It could be any constant strictly between $\sfrac12$ and $1$.  We disallow $\sfrac12$ to exclude verifiers recognizing languages with errors infinitesimally close to $\sfrac12$ from our definition.}
% The best prover can trivially be assumed, without loss of generality, to be one of the infinitely many provers that have an undoubted assumption about the internal structure of the verifier they will be paired against, and that structure then matches the structure of the actual verifier it is paired against.
% Thus, it is assumed that the best prover ``sees'' the configuration of its verifier throughout the interaction, unless and until the verifier starts using private coins.

\newcommand{\verr}{\ensuremath{\varepsilon}}
\newcommand{\verracc}{\ensuremath{\verr^+}}
\newcommand{\verrrej}{\ensuremath{\verr^-}}
\newcommand{\verrloop}{\ensuremath{\verr^{\text{loop}}}}

% The largest set of strings that a verifier $V$ can accept paired with a prover $P$ in an IPS is the set of strings that $V$ accepts with a probability strictly greater than $0.5$ as a result of its interaction with $P$.
% The language $L$ of a verifier $V$ in an IPS is the maximum set of strings that are accepted with a probability higher than $0.5$ for every prover $P$ that $V$ might be paired against. Note that such an $L$ indeed includes every other set of strings $V$ would be accepting with high probability when paired against any other prover.

We say a verifier $V$ in an IPS \emph{verifies a language $L$ with %\utkanadd{(bounded)} 
error $\verr = \max\paren*{\verracc, \verrrej}$} if there exist numbers $\verracc, \verrrej < \sfrac12$ satisfying the following: %\utkanrem{where}
%\footnote{The choice of $\sfrac13$ is arbitrary. It could be any constant strictly between $0$ and $\sfrac12$.  We disallow $\sfrac12$ to exclude verifiers recognizing languages with errors infinitesimally close to $\sfrac12$ from our definition.}
\begin{itemize}
    \item  There exists a prover $P$ such that, for all input strings $w \in L$,  $V$ halts by accepting with probability at least $1 - \verracc$ when started on $w$ and interacting with $P$. % \utkanrem{, and,}
    \item For all provers $P^*$ and for all input strings $w \notin L$, $V$ halts by rejecting with probability at least $1-\verrrej$ when started on $w$ and interacting with $P^*$.
\end{itemize}
The terms $\verracc$ and $\verrrej$ bound the two possible types of error corresponding to failing to accept and reject, respectively.

We will be using the notation
% !!!!!!!YILDIZSIZ!!!!!!!!!! Utkan: Yaptim
\IPhigh{\restrictionformat{resource}_1, \restrictionformat{resource}_2, \dotsc, \restrictionformat{resource}_k} to denote the class of languages that can be verified with error $\verr$ (for some $\verr < \sfrac{1}{2}$) %\utkanrem{with arbitrarily small (but possibly positive) errors}
by machines that operate within the resource bounds indicated in the parentheses. These may represent budgets for runtime, work tape (space) usage, and number of public and private random bits, given as a function of the length of the input string, in asymptotic terms.  We reserve the symbol $n$ to denote the length of the input string. The terms %\ze, 
\co, \lo, \li, and \po{} will be used to represent the well\-/known types of functions to be considered as resource bounds, with ``\co'' standing for constant functions of the input length, the others being self evident, to form arguments like ``\po\pri\ran''
% !!!!BUNUN YERİNE poly-private-coins YAZALIM!!!!! Utkan: yazdim
or ``\co\spa''.\footnote{Constant\-/space verifiers which do not use their work tape at all are also called \emph{finite\-/state verifiers}. (A Turing machine that is restricted to scan \OH{1}
% !!!!GÜZEL NOTASYON!! Utkan: Yaptim
work tape cells can be simulated by  a finite\-/state machine.) Stand\-/alone finite\-/state verifiers (that do not ``listen'' to what the prover says) are called \emph{probabilistic finite automata}.}  The symbol $\infty$ will be used to indicate that no upper bound limits the usage of a resource under consideration, as in ``$\infty\tim$''. The absence of a specification for a given type of resource (\eg, private coins) %\utkanrem{for the runtime and working memory usage}
shall indicate that that type of resource is simply unavailable to the verifiers of that class.%\footnote{\utkanrem{Verifiers that use only some private coins and no  public coins can be described in the framework given above by simply specifying their transition functions to be insensitive to the value of the public random bit argument, \ie, $\delta(q,\sigma,\phi,\gamma,\bpri,0) = \delta(q,\sigma,\phi,\gamma,\bpri,1)$ and $\delta(q,\sigma,\phi,\gamma,0) = \delta(q,\sigma,\phi,\gamma,1)$ for all values of $q,\sigma,\phi,\gamma$, and $\bpri$.}} 
%\utkanrem{is unbounded in terms of those resources, whilst the same for the private and public random bits should indicate that there are no such random bits at the verifier's disposal. The symbol $\infty$ will be used to indicate that a verifier is unbounded in its private or public random bit usage, \eg, $\infty\pub\ran$.}

By default, a given resource budget should be understood as a worst\-/case bound, indicating that it is impossible for the verifier to exceed those bounds.  
To indicate bounds in terms of statistical expectations, we will add the ``\rexlabel'' denotation as a subscript, such as ``\po\rex\tim'' to indicate polynomial expected runtime.  
%Note that such a statistical bound implies that the probability of the resource being used in higher amounts approaches to zero, since the expected usage analyis would be diverging otherwise.
% Importantly, some of the  bounds we consider  will be ``probabilistic'' in the sense that the verifiers in question will be guaranteed to operate within that bound with some (high) probability that may be less than 1. BU LAFA DİKKAT; BUNU ORADA DA IP()NOTASYONUNU KULLANDIĞIMIZDA HATIRLATMALIYIZ!!!!!!!!!!!!!!!
Some of the interactive protocols to be discussed 
%\utkanadd{of our interest} 
have the property that the verifier has some small probability of being fooled to run forever by a malicious prover trying to prevent it from rejecting the input. The designer of such protocols can set a parameter $\verrloop$, representing an upper bound to this probability, to any desired small positive value. The denotation ``\pexlabel'' will be used, this time as a superscript, to mark that the indicated amount corresponds to such a machine's expected consumption of a specific resource with the remaining (high) probability that is at least $1-\verrloop$ for any input string.  For instance, ``\po\pex\tim'' will indicate that the verifier's expected runtime is  polynomially bounded with probability almost, but possibly not exactly, $1$.

%To confer an unbounded amount of a resource to the verifier while we still expect it to use only a limited amount of it most of the time (statistically, with some constant probability less than 1 that can be set arbitrarily high), we will qualify the bound with the ``\pexlabel'' sign.  

Some verifier algorithms we describe in this article will have  ``dials'' that allow one  to tune them to set their error bounds to any desired (positive) constant.  Such verifiers will be qualified as verifying their language with \emph{(arbitrarily) low error}. %, or simply, \emph{low error}.  %Contrarily, some other verifier algorithms to be considered do not support such control  and have  error bounds that cannot be improved beyond a limit.  %Those will be named  \emph{high error} verifiers.
To denote  classes of languages that are verifiable with arbitrarily low error, 
%instead of low, 
we will be using the $\IP{\ldots}$ denotation instead of $\IPhigh{\ldots}$.

%\footnote{Note that the ``expected'' runtimes appearing in our  IP complexity class notation  defined in \Cref{sec:prel} are supposed to be interpreted as follows:  The probability of the associated verifiers committing the ``error'' of being tricked by a malicious prover to run forever rather than rejecting some input strings can be reduced below any desired  positive threshold $\varepsilon$. The verifier's runtime can thus be infinite with probability at most $\varepsilon$. With the remaining (large) probability $1-\varepsilon$, its expected runtime is bounded as indicated.  If one lifts the restriction that $\varepsilon$ can be pushed to be arbitrarily low, and only requires it to be below $\frac{1}{2}$, these machines can verify all languages in \NL{}~\cite{SY14}. }

% BUNA GEREK YOK!!!!!!!!!!!!!!!!!
% The ``one\-/way'' mode, where the input  head is not allowed to move left, will be indicated by the parameter ``\ow\inp'', whereas the further restriction to real\-/time movement, where the head is not allowed to pause at any step during its left\-/to\-/right scan, will be indicated by ``\rt\inp''. 

We will also be considering verifiers with one\-/sided error: A verifier $V$ for language  $L$ is said to have \emph{perfect completeness} if there exists a prover whose interaction with $V$ leads it to accept any string in $L$ with probability $1$, and no prover can convince $L$ to accept a string that is not in $L$ with probability $1$.

In one of our proofs in \Cref{sec:results},
we will be considering finite\-/state verifiers  with multiple input tape heads that the machine can move independently of one another. This type of verifier can be modeled easily by modifying the tuples in the transition function definitions above to accommodate more scanned input symbols  and input head directions. %\utkanadd{, and removing the parts involving the work tape}. !!!!HAYIR, BUNLARIN WORK TAPE'İ VAR AMA KULLANMIYOR DEMİŞİZ ALG2'DE!!!!!
% \Cref{subs:nfaks,sec:hartmanis}
\Cref{subs:nfaks}
provides more information on finite automata with multiple input heads and their relationships with the  standard Turing machine model.

\subsection{Multihead finite automata and log-space machines}\label{subs:nfaks}
%\begin{definition}\label{def:tnfak}

A \emph{$k$\=/head nondeterministic finite automaton} (\tnfak{k}) is a nondeterministic finite\-/state machine with $k$ read\-/only heads that move on an input string flanked by two end\-/marker symbols. Each head can be made to stay put or move to an adjacent tape cell in each computational step.  Formally, a \tnfak{k}  is a 4\=/tuple $(Q,\Sigma,\delta,q_0)$, where
\begin{enumerate}
    \item $Q$ is the finite set of internal states, which includes the two halting states $\qacc$ and $\qrej$,
    \item $\Sigma$ is the finite  input alphabet,
    \item $\delta \colon Q \times \Sigma_{\bowtie}^k \to \P{Q \times \Delta^k}$ is the transition function describing the sets of alternative moves the machine may perform at each execution step, where each move is associated with a  state to enter and a  movement direction for each head, given the machine's current state and the list of symbols that are currently being scanned by the $k$ input heads, and $\Sigma_{\bowtie}$ and $\Delta$ are as defined previously in \Cref{subs:vers}, and%:
    % \begin{itemize}
    %     \item $\Delta = \Set{-1, 0, +1}$ is the set of possible head movements, where $-1$ means ``move left'' $0$ means ``stay put'' and $+1$ means ``move right'',
    %     \item $\Sigma_{\bowtie} = \Sigma \cup \Set{\lend, \rend}$, where $\lend, \rend \notin \Sigma$ are respectively the left and  right end\-/markers, placed automatically to mark the boundaries of the input,
    % \end{itemize}
    \item $q_0 \in Q$ is the initial state.
    % \item $\qacc \in Q$ is the final state at which the machine halts and accepts, and
    % \item $\qrej \in Q$ is the final state at which the machine halts and rejects.
\end{enumerate}

Given an input string $w \in \Sigma^*$, a \tnfak{k} $M = \paren*{Q, \Sigma, \delta, q_0}$  begins execution from the state $q_0$, with $\lend w \rend$ written on its tape, and all $k$ of its heads on the left end\-/marker.
%BUNLARI HEP FIRST SEMBOL OF w'DEN LEFT ENDMARKERA ÇEVİRDİM!!!!!!!!!!!!!!!!!!!!!!!!!!!!!!!!!!!!!!!!
At each step, $M$ nondeterministically updates its state and head positions according to the choices dictated by its transition function. Computation halts if one of the states \qacc{} or \qrej{} has been reached, or a head has moved beyond either end\-/marker.

%Each different sequence of choices $M$ may take corresponds to a different \emph{computation history}, \ie, a  sequence of tuples describing all the state and head positions that $M$ goes through in that particular eventuality.
%and moves its \ith{i} head to $d_i$, where $\paren*{s, d_1, \dotsc, d_k}$ is one of the possibly many options in $\delta\paren*{q, x_1, \dotsc, x_k}$, where $q$ is the current state of $M$ and each $x_i$ is the symbol currently under the \ith{i} head.

$M$ is said to \emph{accept} $w$ if there exists a sequence of nondeterministic choices where it reaches  the state \qacc{}, given $w$ as the input.  $M$ is said to \emph{reject} $w$ if every sequence of choices either reaches  \qrej{}, ends with a  transition whose associated set of choices is $\emptyset$, or by a head moving beyond an end\-/marker without a halting state being entered.  $M$ might also loop on the input $w$, neither accepting nor rejecting it.

The \emph{language recognized by $M$} is the set of strings that it accepts.
%

% BİRKAÇ SATIR TASARRUF İÇİN KAFA SAYISINI UMURSAMAYAN   LINEAR-TIMELI CLASS DOĞRUDAN DA TANIMLANABİLİR!!!!!!!!!!!!!!!!!!!!!!!!!!!!!!!!! \utkanadd{BIR KISALTMA YAPTIM}

Let \TNFAS{} denote the set of languages that have a \tnfak{k} recognizer (for some $k > 0$), and \TNFASL{} denote the set of languages that have a \tnfak{k} recognizer running in linear time, regardless of the nondeterministic choices it makes.
Our characterization of \PP{} (\Cref{sec:results}) makes use of the equivalence of multiple input heads and logarithmic amounts of memory, discovered by Hartmanis~\cite{H72}. 
%an interesting relationship~\cite{H72} between multihead finite automata and logarithmic\-/space Turing machines, which will be detailed in % \Cref{sec:hartmanis}. In this subsection,
%\Cref{thm:hartmanisextended}. We first provide the necessary definitions and establish the link between these machines and probabilistic finite\-/state verifiers. 
The following theorem, whose proof is given in \Cref{sec:hartmanisextendedproof}, reiterates one direction of that result in detail, and will be useful for our purposes.%and also contains an analysis for the overhead in time incurred during the simulation.

\begin{restatable}{theorem}{hartmanisextended}\label{thm:hartmanisextended}
    Any language recognized by a Turing machine that
    % \utkanrem{, for inputs of length $n \ge 2$,}
    uses at most
    % \utkanrem{$\ceil*{\log n}$} \utkanrem{$\floor*{\log n}$}
    $\floor*{\log(n+2)}$ space with a work tape alphabet of size at most $2^c$ (for some integer constant $c > 0$) and within $t(n)$ time can also be recognized by a $\paren*{c+5}$\=/head finite automaton within
    % \utkanrem{$t(n) \cdot \paren*{(2c+3)n + c  + 1}$}
    $t(n) \cdot \paren*{\paren*{2c + \sfrac{3}{2}}n + 2c + 2}$
    time.
\end{restatable}

%\utkanadd{The proof is given in \Cref{sec:hartmanisextendedproof}. BUNU KALDIRMISSINIZ SANIRIM AMA YANLISLIKLADIR DIYE TAHMIN ETTIM}

\newcommand{\head}{\ensuremath{h}}

\newcommand{\errprematuretimeout}{\ensuremath{\err_{\mathrm{premature}}}}

\subsection{Probabilistic clocks}\label{sec:clock}

A logarithmic\-/space Turing machine can time its own execution to satisfy any desired polynomial bound by counting up to that bound in the logarithmic space available. The constant\-/space machines we construct in \Cref{sec:results} will employ a different technique  using randomness, which is illustrated in the following lemma, to obtain the same bound on expected runtime. 

\begin{restatable}{lemma}{polyclock}\label{lem:polyclock}
    For any integer %constant
    $t > 0$, integer\-/valued function $f(n) \in \OH{n^t}$, and desired ``error'' bound $\errprematuretimeout > 0$, there exists a probabilistic finite automaton with expected runtime in \OH{n^{t+1}}, such that the probability that this machine halts in fewer than $f(n)$ time\-/steps is \errprematuretimeout.
\end{restatable}

% \paragraph{Proof idea.}
% \begin{proof}[Proof idea.]\let\qed\relax
% Assume, for the sake of simplicity, that $t$ is an integer.  We program a probabilistic finite\-/state automaton to make $t$ random walks with its input head, each starting from the first symbol on the input and ending at either one of the end\-/markers.  If all the walks have ended on the right end\-/marker, the machine halts.  Otherwise, the process is repeated.  The analysis shows that such a machine has all the necessary characteristics in its runtime.
% \end{proof}

%% W/O/APPENDIX VERSION
% The detailed proof can be found in~\cite[Appendix~A.2]{GS23a}.
%% W/APPENDIX VERSION
The proof of \Cref{lem:polyclock}, based on the idea~\cite{DS92,AW02} of the machine performing a sequence of random walks with its head on the input tape, is given in \Cref{sec:polyclockproof}.

In the proof of \Cref{lem:dir2NEW} in \Cref{sec:results}, we describe a probabilistic log\-/space machine that uses a timer which is expected to run out after at least $2^{2^{p(n)}}$ ``ticks'', where $p$ is some polynomial, and $n$ is the length of the machine's input string. We describe the main idea, which was inspired by a similar construction in~\cite{C93b}, below; a detailed analysis can be found in \Cref{sec:dir2appendices}.

It will be helpful to imagine that we have fair coins of two colors, say, red and blue. The timer mechanism consists of a subroutine that embodies a single tick, and a global Boolean variable that tracks whether all the red coins flipped since the last reset of this variable came out heads. %\utkanadd{UTKAN: ya da alternatif olarak ``all \utkanrem{the} red coins flipped since the last reset of this variable \utkanadd[65]{have come} out heads''; ilk alternatifte ``since'' nitelemesi red coins'e baglaniyor, ikinci alternatifte fiile baglaniyor}.  
Each call of this subroutine, which is named \TICK{}, %\utkanadd{UTKAN: Bu hep düz-yazı TICK mi kalacak yoksa şimdi veya ileride bold/sans-serif/vs.\ bir şey yapmayı düşünür müyüz? Bir de bu subroutine aslında $T$ adında bir read-write variable'a ihtiyaç duyuyor, onu indicate edecek bir şey yapmamız mümkün mü?  TICK(T) yazsak çok iyi olmayacak, çünkü normalde (yaygın programalma dillerinde) böyle çağrılan fonksiyonlar parametre olarak aldığı T'yi değiştiremiyor, değiştirilebilir olarak iletmek için özel şeyler yapmak gerekiyor.}!!!!FONT İŞİ BENCE GEREKSİZ AMA KOLAY BİR MODİFİKASYON YAPMAK İSTERSEN OK. \utkanadd{Makroya aldım, eskisi ok'di, yine de şimdi bi sans-serif yapmayı denedim, istediğiniz gibi değiştirebiliriz.} % T'DEN DE BURADA HİÇ BAHSETMEYELİM; GENELDE BU GİBİ DİREKT HAKEM RAPORUNDA OLMAYAN ŞEYLERİ DEĞİŞTİRMEME PRENSİBİNDEN ŞAŞMAYALIM!!!!!!!! \utkanadd{Hakemlere giden halinde "TICK" ismi hiç yoktu, şu an yeni var ettiğimiz TICK konusunda baştan sona hakem raporunun dışında bir iş yapmaktayız. Yani demek istediğim, (T)'li hali de (T)'siz hali de "hakem raporunda olmayan şeyleri değiştirmeme prensibinden" şaşıyor, biri diğerinden daha az şaşmıyor.}
flips one red coin and $p(n)$ blue coins. If one or more blue coins come out tails, the subroutine call ends with the timer still running. If all the blue coin outcomes are heads, but the Boolean variable indicates that one or more red coins flipped since the last reset came out tails, the variable is reset, and the subroutine ends with the timer still running. If all the blue coin outcomes are heads, \emph{and} if all the red coins flipped since the last reset also came out heads, the timer runs out. 

The expected number of ticks is at least double exponential in $p(n)$, %\utkanadd{UTKAN: bunu soyleyebilecegimizi sanmiyorum. benim tespit edebildigim kadariyla ``at least'' demeliyiz.}
since the timer runs out when an all\-/heads sequence of red coins is observed, and the expected length of this critical sequence is exponential in $p(n)$.

% \utkanadd{

\subsection{Alternating finite automata}\label{subs:afas}

A \emph{two\-/way alternating finite automaton} (\tafa{}) is a 4\=/tuple $(Q,\Sigma,\delta,q_0)$, %\hl{$q_0$'i buraya koyup da $\Qpub$, $\Qex$, $\qacc$ gibi $Q$'nun diger parcalarini koymamamiz garip olmamis mi?} 
where
\begin{enumerate}
    \item $Q$, the finite set of states, is the union of the following disjoint subsets:
    \begin{itemize}
        \item $\Qex$ is the set  existential states,
        \item $\Qun$ is the set of universal states,
        \item $\Set{\qacc, \qrej}$ are the accept and reject states, respectively,
    \end{itemize}
    \item $\Sigma$ is the finite  input alphabet,
    \item $\delta \colon Q \times \Sigma_{\bowtie} \to \P{Q \times \Delta}$, which is identical in format to the corresponding item in the definition of a \tnfak{1} (\Cref{subs:nfaks}), is the transition function describing the sets of alternative moves that the machine may perform at each execution step, 
    %where each move is associated with a state to enter and \hl{how} % UTKAN: changed from "whether", is it formally appropriate?
    %or not to move the head, given the machine's current state and the symbol that is currently being scanned by the input head, and $\Sigma_{\bowtie}$ and $\Delta$ are as defined previously in \Cref{subs:vers,subs:nfaks}, 
    and%:
    % \begin{itemize}
    %     \item $\Delta = \Set{-1, 0, +1}$ is the set of possible head movements, where $-1$ means ``move left'' $0$ means ``stay put'' and $+1$ means ``move right'',
    %     \item $\Sigma_{\bowtie} = \Sigma \cup \Set{\lend, \rend}$, where $\lend, \rend \notin \Sigma$ are respectively the left and  right end\-/markers, placed automatically to mark the boundaries of the input,
    % \end{itemize}
    \item $q_0 \in Q$ is the initial state.
    % \item $\qacc \in Q$ is the final state at which the machine halts and accepts, and
    % \item $\qrej \in Q$ is the final state at which the machine halts and rejects.
\end{enumerate}

%is a nondeterministic finite\-/state machine with a single read\-/only head that moves on an input string flanked by two end\-/marker symbols. Each head can be made to stay put or move to an adjacent tape cell in each computational step.  Formally, a \tnfak{k} 

As a computational model, the \tafa{} is a generalization the \tnfak{1}. A \tnfak{1} is simply a \tafa{} whose set of universal states is empty, as will be evident from the definition (adapted from~\cite{DSY15}) of string acceptance by \tafa{}'s to be presented below.

\newcommand\suc[1]{\ensuremath{\vdash_{#1}}}
The \emph{configuration} of a \tafa{} $M = \paren*{Q, \Sigma, \delta, q_0}$ at any given time is the pair composed of $M$'s state and its input tape head position.  In its initial configuration, %\utkanadd[80]{UTKAN: Burada herhalde bir typo var.  Bir de böyle yazınca $\alpha_0$'yı $q_0$ gibi özel bir notasyon olarak tanıtıyormuşuz gibi olmuş.  Bence burada $\alpha_0$ hiç demeyelim ve aşağıda $\alpha_0$ kullandığımız tek yerde de yerine $(q_0, \lend)$ yazalım.} 
$M$ is in state $q_0$, with the head positioned on the left end\-/marker.  A configuration $(q,i)$ %\utkanadd[80]{UTKAN: $p$ yerine $i$ kullanmayı düşünebiliriz.}
is said to be \emph{accepting} (resp.\ \emph{rejecting}) if $q$ is \qacc{} (resp.\ \qrej{}).  A configuration is said to be \emph{universal} (resp.\ \emph{existential}) if its state is universal (resp.\ existential). % !!!!ŞİMDİ OK MİYİZ??!!! UTKAN: Ok'iz hocam, teşekkürler %Otherwise, it is said to be . %\utkanadd[80]{UTKAN: Yukarıdaki tanıma göre bu doğru değil.  $\qacc$ ve $\qrej$ universal da existential da değil.}!!!İŞTE O YÜZDEN DE O KONFİGÜRASYONLAR EXISTENTIAL!!!!!
Given an input string $w$, a configuration $\beta$ is said to be a \emph{successor} of a configuration $\alpha$ of $M$, denoted $\alpha \suc{M, w} \beta$,
% (or simply $\alpha \suc[] \beta$)
if $\beta$ can follow $\alpha$ immediately according to $\delta$.  Note that a configuration may have multiple successors, or none at all.%!!!!!!SON CÜMLEYİ NEDEN SİLMİYORUZ????!!!  \utkanadd[80]{UTKAN: "Successor" kelimesi bence tekillik çağrıştırıyor.  Yani mesela bugüne kadar yazılmış tüm metinlerdeki "successor" kelimelerine baksak, bana öyle geliyor ki kayda değer bir çoğunluğu tekil bir şeyden bahsediyordur.  Şimdi ChatGPT'ye de sordum, o da öyle düşünüyor: \url{https://chatgpt.com/share/6799662e-5d8c-8007-ac48-f40ee0a08049}  Bu doğal çağrışımdan arındırmak adına o açıklamayı yapmalıyız diye düşündüm.  "Successor" yerine "child" da diyebiliriz, o zaman böyle bir çağrışım en baştan hiç oluşmamış olur.}

\newcommand\conflabelf[1]{\ensuremath{l_{#1}}}
$M$ is defined to \emph{accept} %(resp.\ \emph{reject}) 
$w$ if and only if % \utkanadd[80]{UTKAN: Burada "and only if" diyerek sadece "aksi takdirde accept etmemiştir" demiş oluyoruz, "aksi takdirde reject etmiştir" gibi bir şey demiyoruz, değil mi? 2nfa(k) tanımımızda benzer yerlere "and only if" yazmamışız hiç.} !!!! EVET, TEK DERDİMİZ ACCEPT'İ TANIMLAMAK. BURADA SORUN YOKSA MAVİLİKTEN KURTAR PLZ!!!!!!!!!!! Tamamdır
$\conflabelf{M,w}\paren*{\paren*{q_0, 0}}$ equals \True{}, 
%(resp.\ \False{}), 
where the function $\conflabelf{M,w}$
% (or simply $\conflabelf[]$)
is defined recursively as follows:
% !!!!!!!!!BU or simply'lere NE GEREK VAR??!??!?!!  \utkanadd[80]{UTKAN: Simple halleri sayesinde $\conflabelf[M,w]$ tanımını aşağıdaki gibi yazabiliyoruz, daha sade oluyor. "Simple hallerine ne gerek var" değil de "simple \textbf{olmayan} hallerine ne gerek var" mı demek istediniz? Simple halleri matematiksel olarak ill-defined.  Doğruluk değeri context'e (tam olarak $M$'ye ve $w$'ya) bağlı. Yani mesela aynı $\alpha$ ve $\beta$ için input string $w$ iken $\alpha \suc[] \beta$ doğru, input string $w'$ iken aynı ifade yanlış olabilir.  Çünkü mesela $(q_{123}, 2) \suc[] (q_{234}, 3)$ ifadesi "\texttt{>u\uline{t}kan<}" input stringi için doğru, "\texttt{>g\uline{e}zer<}" stringi için yanlış olabilir (mesela makine $q_{123}$ state'indeyken \texttt{t} okuyunca $q_{234}$ state'ine girip kafasını sağa oynatıyordur ama \texttt{e} okuyunca başka bir state'e girip sola gidiyordur).  Benzer şekilde söz konusu makine değişince de sağdaki tuple soldaki tuple'ı succeed ederken/etmezken etmez/eder hale gelebilir.}!!!!!!!SIMPLE'LARI BOŞVERİP KARIŞIK HALİYLE YAZALIM; İKİ DEFACIK ZATEN!!!!!!
%\utkanremlong[]{
%\begin{equation*}
 %   \conflabelf{M,w}(\alpha) = \begin{cases}
  %      \True, & \text{if $\alpha$ is accepting},\\
   %     \False, & \text{if $\alpha$ is rejecting},\\
    %    \bigvee\limits_{\alpha \suc{M,w} \beta} \conflabelf{M,w}(\beta), & \text{if $\alpha$ is existential},\\
     %   \bigwedge\limits_{\alpha \suc{M,w} \beta} \conflabelf{M,w}(\beta), & \text{if $\alpha$ is universal}.
   % \end{cases}
%\end{equation*}
%}
%\utkanadd{
\begin{equation*}
    \conflabelf{M,w}(\alpha) = \begin{cases}
        \True, & \text{if $\alpha$ is accepting},\\
        \False, & \text{if $\alpha$ is rejecting},\\
        \displaystyle\smashoperator[r]{\bigvee_{\alpha \suc{M,w} \beta}} \conflabelf{M,w}(\beta), & \text{if $\alpha$ is existential},\\
        \displaystyle\smashoperator[r]{\bigwedge_{\alpha \suc{M,w} \beta}} \conflabelf{M,w}(\beta), & \text{if $\alpha$ is universal}.
    \end{cases}
\end{equation*}
%}
%\utkanadd{
%\begin{equation*}
 %   \conflabelf{M,w}(\alpha) = \begin{cases}
  %      \vphantom{\displaystyle\smashoperator[r]{\bigvee_{\alpha \suc{M,w} \beta}}} \True, & \text{if $\alpha$ is accepting},\\
   %     \vphantom{\displaystyle\smashoperator[r]{\bigvee_{\alpha \suc{M,w} \beta}}} \False, & \text{if $\alpha$ is rejecting},\\
    %    \displaystyle\smashoperator[r]{\bigvee_{\alpha \suc{M,w} \beta}} \conflabelf{M,w}(\beta), & \text{if $\alpha$ is existential},\\
     %   \displaystyle\smashoperator[r]{\bigwedge_{\alpha \suc{M,w} \beta}} \conflabelf{M,w}(\beta), & \text{if $\alpha$ is universal}.
 %   \end{cases}
%\end{equation*}
%}
%$M$ is said to loop on the input $w$, if it neither accepts nor rejects it.

%END OF ALTERNATIVE 3

The \emph{language recognized by $M$} is the set of input strings that it accepts.
%

%We recall the following facts about two\-/way alternating finite automata~\cite{LLS78}: Each non\-/halting state in such an automaton is either \emph{existential} or \emph{universal}. 

A good way to understand the relationship of a \tafa{} with the strings it accepts is to visualize a ``prover'' (whose aim is to make the machine accept the input) as dictating which move to make when the transition function allows multiple outgoing choices from a configuration with an existential state, and a ``refuter'' (with the opposite aim)  as similarly dictating moves from configurations with universal states. The input string $w$ is accepted by the machine if and only if the prover has a winning strategy,  whereby it can lead the machine from the initial configuration associated with $w$ to an accepting configuration, no matter what moves are made by the refuter, in this  perfect information game. %We say that the machine \emph{recognizes} the set of strings that it accepts.

% BİRKAÇ SATIR TASARRUF İÇİN KAFA SAYISINI UMURSAMAYAN   LINEAR-TIMELI CLASS DOĞRUDAN DA TANIMLANABİLİR!!!!!!!!!!!!!!!!!!!!!!!!!!!!!!!!! \utkanadd{BIR KISALTMA YAPTIM}
It is known~\cite{LLS78} that the class of all languages recognized by \tafa{}'s equals \REG{}, \ie, the class of regular languages.
% }

\section{Private vs.\ public coins and worst-case time bounds}\label{sec:prelresults}

In their seminal paper, Dwork and Stockmeyer~\cite{DS92} showed that finite\-/state verifiers employing only private coins are strictly more powerful than those using only public coins, even when the former machines are bounded to operate in polynomial expected time: 

\begin{fact}\label{fact:DSpal}
    \begin{equation*}
        \IPhigh{\co\spa, \infty\pub\ran, \infty\tim} \subsetneq \IP{\co\spa, \po\rex\pri\ran, \po\rex\tim}.
    \end{equation*}
\end{fact}
% !!!!!SAĞ TARAFI YILDIZLI YAP LÜTFEN!!!! Utkan: Yaptim
%FACT BİLMEMNE: IP($\infty$-time, con-space, public-coins) 
%			PROPERSUBSET IP(poly-time, con-space, private-coins) (BUNU BİZİM NOTASYONA PLZ; THM NUMARASINA GEREK YOK)

A witness for the inequality in \Cref{fact:DSpal} is the 
language of palindromes,
\begin{equation*}
    \pal = \Set{w | w \in \Set{0, 1}^*, w = \rev{w}},
\end{equation*}
where $\rev{x}$ denotes the reverse of  string $x$. The constant\-/space verifier provided for $\pal$
by Dwork and Stockmeyer  uses only a constant number of private coins, irrespective of the length of the input. This inspired %(DID IT?) yes
the study~\cite{SY14,GS22,GDES23} of machines that flip a fixed number of coins, leading to the following results:

\begin{fact}\label{fact:GS22}
    $\TNFASL \subseteq \IP{\co\spa, \co\pri\ran, \li\pex\tim}$~\cite{GS22}.\footnote{Recall from the definition of our  IP complexity class notation   in \Cref{sec:prel} that the verifier's runtime can be infinite with at most a small probability  $\verrloop$, and its expected runtime is bounded as indicated with the remaining large probability.}%  If one lifts the requirement that $\varepsilon$ can be pushed to be arbitrarily low, and only requires it to be below $\frac{1}{2}$, these machines can verify all languages in \NL{}~\cite{SY14}.}
\end{fact}

\begin{fact}\label{fact:SY14}
    $\IPhigh{\co\spa, \co\pri\ran, \infty\tim} = \NL$~\cite{SY14}.
\end{fact}

In contrast to \Cref{fact:SY14}, we now show that the ability to use constant number of \emph{public} coin flips provides no additional language verification power over determinism to constant\-/space machines.

\begin{theorem}\label{thm:conspubliconly}
$\IPhigh{\co\spa, \co\pub\ran, \infty\tim} = \REG$. %\utkanadd{\REG{} ve \NL{} notasyonlarını hiç tanıtmamışız!}
\end{theorem}
\begin{proof}
    Let $V$ be a finite\-/state machine that  verifies a language $L$ with some error bound $\verr < \sfrac{1}{2}$, flipping at most    $r>0$   public coins for any input. Let the prover that interacts with $V$ be named $P$. % Since all coins  are  public, we assume that $P$ will be able to infer $V$'s configuration at every step of their interaction  and that $V$ sends no further information through the communication cell, without loss of generality.

We will construct a two\-/way alternating finite automaton $M$ that recognizes $L$. This will be enough to conclude that $L$ is regular, by the fact~\cite{LLS78} that two\-/way alternating finite automata recognize all and only the regular languages.

The construction is based on the following idea: 
Consider the set of all possible bit sequences of length $r$. Each execution of $V$ would use (a prefix of) one of these sequences as its public coin outcomes during its interaction with the prover.
Any input string $w$ is in $L$ if and only if a majority of the members of this set of sequences leads $V$ to acceptance during this interaction. $M$ will be designed so that it accepts $w$ if and only if % UTKAN: This was missing.
it verifies that this condition is satisfied. Note that simply simulating $V$'s behavior sequentially, on one coin sequence after another, would not work, since this would give the simulated prover the opportunity to cheat by violating the condition that its responses to $V$ given two coin sequences with the same prefix (like $p0$ and $p1$, for some prefix $p\in \{0,1\}^*$) should be identical up to the point when the coin sequences finally differ. $M$ will use the ``parallel computation'' nature of alternation to enforce this consistency on the simulated prover.

$M$ starts with an existential state, where it guesses a prefix\-/free set $S$ of public coin sequences (of length at most $r$) whose probabilities add up to a value greater than $\sfrac{1}{2}$. (Since there are only finitely many sets of bit sequences %of length at most $r$
with this property, this choice can be made in a single step.)
%\utkanadd{UTKAN: bunu boyle yazacaksak eger, secilen sequence kumesinde hicbirinin bir digerinin prefixi olmamasini da garanti ettirmeliyiz, veya prefixler var olabilecekse de ``probability toplami'' hesabina ``kumede bir prefixi bulunan sequencelarin'' olasiliklarinin dahil edilmeyecegini yazmaliyiz.  (buna ek olarak bir sonraki maviyle yazdigim kisimdaki eklemem de gerekli oluyor.)  ALTERNATIF OLARAK: bence ``of length at most $r$'' yerine ``of length $r$'' (exactly $r$) yazsak her sey kolayca cozuluyor.  ispatin kalanindaki mevcut anlati zaten simulasyon bitince refuterin $M$'yi takip ettirdigi sequence(lar)in artiklarinin kalmasini dert etmiyor.  boyle yapinca ``probabilities add up to a value greater than \sfrac12'' yerine ``exactly $2^{r-1}+1$ tane'' de diyebiliyoruz; boyle demek constructionin bu asamasindaki gereksiz existential branchleri ($2^{r-1}+1$ elemandan fazla iceren kumeleri) eliyor ama demeyebiliriz de.}
%{a set of $2^{r-1}+1$ public coin sequences of length $r$. (Since there are finitely many such sets, $\binom{2^r}{2^{r-1}+1}$ to be exact, this choice can be made in a single step.)}
Recall that such choices of moves from existential states can be viewed as being made by a prover, which we will name $P_M$. By this first choice, $P_M$ would be claiming that all of these coin sequences would lead $V$ to acceptance in an interaction with $P$.

$M$ attempts to verify this claim: $M$'s transition function is based on that of $V$; essentially, $M$ uses alternation to create 
a parallel simulation of $V$ (and its interaction with $P$) on all the coin sequences in the set $S$ obtained at the start.
At every step, $M$ keeps a record of the (simulated) coin sequence used up to that point in the present branch of its computation.
Whenever its simulation of $V$ reaches a new coin\-/flipping state, $M$  checks this record and \rejects{} if it determines that it is in a branch that would use a coin sequence that is not  a member of $S$.
If only one of the outcomes of the coin to be flipped at the present state is consistent with $S$, $M$ plugs that bit into $V$ to advance the simulation.
If, on the other hand, \emph{both} of the 0 and 1 outcomes are consistent with $S$, $M$ makes a universal choice (controlled by the refuter $R_M$) to determine which bit to feed $V$.
%Each step of $V$ corresponding to a coin flip both of whose outcomes will be considered in this verification\footnote{Note that some coin flips  of $V$ may \emph{not} be associated with  universal branching in $M$. Such branchings occur for a coin only if \emph{both} of the 0 and 1 outcomes are included in the set of sequences guessed at the beginning.} corresponds to a universal choice (controlled by the ``refuter'' $R_M$).
%{a ``tree'' of ``parallel'' simulations of $V$ (and its interaction with $P$) corresponding to all prefixes of the coin sequences in the set obtained at the start.  At each step of $V$ that corresponds to a coin flip, if the prefixes corresponding to both outcomes (\ie, $p0$ and $p1$, given $p$ as the sequence of outcomes so far) are in the obtained set, then both of them will be considered with a universal fan\-/out.  Otherwise, the branch will continue with the outcome corresponding to the only prefix that exists in the obtained set, without fanning out.}
%\utkanadd{$M$ \emph{rejects} immediately if $V$'s simulation reaches a coin flipping state despite neither a $0$ or a $1$ outcome is included in the set of sequences that $P_M$ promised would lead $V$ to acceptance. BU SEQUENCE'IN KISA KACTIGI DURUMLAR ICIN. YA DA SEQUENCE KISA ISE ``bundan sonra refuter ne dese yas'' VEYA ``bundan sonra her ne random bit gelirse gelsin seni ikna edecegim'' ANLAMINA GELECEK SEKILDE BIR ONCEKI FOOTNOTE'U DEGISTIREBILIRIZ.}
Each communication symbol sent by $P$ to $V$  is obtained by an existential choice (as if $P_M$ is supplying that information). The accept state of $M$ corresponds to the accept state  of $V$. %\utkanadd{(with $M$ having $\binom{2^r}{2^{r-1}+1}$ copies of them)}.

We see that $M$ recognizes $L$ by noting that any $w$ is in $L$ if and only if there exists a set of random bit sequences with total probability greater than $\sfrac{1}{2}$ such that all members of this set lead $V$ to acceptance of input $w$ after an interaction with $P$, and $M$ accepts all and only such input strings by design.

For completeness, a formal description of the \tafa{} $M$ is presented in \Cref{sec:tafaconstruction}.
\end{proof}

We will focus on the computational power of constant\-/space verifiers with a fixed private\-/coin budget when they are also allowed to use public coins. Although some of Dwork and Stockmeyer's results (\eg,~\cite[Theorem~3.12]{DS92}) involve such ``mixed\-/coin'' constructions, the asymptotic bounds on private and public coin tosses are equal in their machines. In those setups, all public coins can simply be replaced by private coins by a straightforward modification that increases the runtime of the protocol by a constant factor. The proof of the following lemma is in \Cref{sec:privatesimulatespublicproof}.
%\utkanadd{. It} %\utkanadd{\Cref{lem:privatesimulatespublic} below remarks the relationship between and how private coins are generally superior to the public ones. It
%will also be useful in one of the later proofs.

\begin{restatable}{lemma}{privatesimulatespublic}\label{lem:privatesimulatespublic}
    The following is true for all % \utkanadd{complexity}!!!!!BU LAFA NE GEREK VAR?!??!!!! \utkanadd{UTKAN: Bilmem ki, fonksiyonlar çok çeşitli. Negatif değerli veya parçalı fonksiyonlar var. O tür fonksiyonlar big-O(...) içine yazılınca garip şeyler olur mu diye düşünerek yazmıştım da olmuyor bir şey sanırım.}
    functions %\utkanrem{$s(n), f(n), g(n)$, and $t(n)$} \utkanadd{
    $s$, $f$, $g$, and $t$:
    % !!!!!!!!!!BÜTÜN IP ifadelerinde AYNI SIRADA GÖRÜNSÜNLER, BİRİNDE SPACE ÖNDE ÖBÜRÜNDE COINS ÖNDE OLMASIN!!!!!!!!! Utkan: Yaptim
    \begin{multline*}
        \IP{\OH{s(n)}\spa, \OH{f(n)}\pri\ran, \OH{g(n)}\pub\ran, \OH{t(n)}\tim} \subseteq\\
        \IP{\OH{s(n)}\spa, \OH{f(n)+g(n)}\pri\ran, \OH{t(n)}\tim}.
    \end{multline*}
    This inclusion also holds for the  $\IPhigh{\ldots}$  variants of the classes,  and when worst\-/case time bounds are replaced with those corresponding to expected usage.% \utkanadd{on both ends of the inclusion simultaneously}.  \utkanadd{The worst\-/case private and public coin bounds together may also be replaced with the corresponding expected bounds on both ends simultaneously.}
    %worst\-/case or expectations. \utkanadd{and/or some bounds are given in expected terms, with or without allowing a low probability of unbounded usage}.
\end{restatable}

%\utkanadd{The proof is given in \Cref{sec:privatesimulatespublicproof}.}

\Cref{lem:privatesimulatespublic} implies that allowing the finite\-/state verifiers in the constructions of  \Cref{fact:GS22,fact:SY14} to use a constant number of public coins in addition to their private\-/coin budgets would not enlarge the corresponding classes of verified languages.%:  The added resource cannot make the verifiers less powerful, but also the function of any constant amount of public coins can readily be simulated by more of the existing constant private coin budget according to \Cref{lem:privatesimulatespublic}.
% \begin{theorem}
%     \begin{multline*}
%         \IPhigh[0.9]{\co\spa, \co\pri\ran, \co\pub\ran, \infty\tim} =\\ \IPhigh[0.9]{\co\spa, \co\pri\ran, \infty\tim}.
%     \end{multline*}
% \end{theorem}
% \begin{proof}
%     The following is trivially true, since the left\-/hand side has more resources:
%     \begin{multline*}
%         \IPhigh[0.9]{\co\spa, \co\pri\ran, \co\pub\ran, \infty\tim} \supseteq\\
%         \IPhigh[0.9]{\co\spa, \co\pri\ran, \infty\tim}.
%     \end{multline*}
%     By \Cref{lem:privatesimulatespublic} we also have
%     \begin{multline*}
%         \IPhigh[0.9]{\co\spa, \OH{n}\pri\ran, \OH{n}\pub\ran, \infty\tim} \subseteq\\
%         \IPhigh[0.9]{\co\spa, \paren*{\OH{n}+\OH{n}}\pri\ran, \infty\tim},
%     \end{multline*}
%     which is to say
%     \begin{multline*}
%         \IPhigh[0.9]{\co\spa, \co\pri\ran, \co\pub\ran, \infty\tim} \subseteq\\
%         \IPhigh[0.9]{\co\spa, \co\pri\ran, \infty\tim}.
%     \end{multline*}
%     Thus, the two classes are equal to each other.  By \Cref{fact:SY14}, they are also equal to \NL.
% \end{proof}

Our main result to be presented in the next section involves constant\-/space verifiers that flip a constant number of private coins and superlinear amounts of public coins. We note that expected (rather than worst\-/case) time bounds are appropriate for studying that scenario, since the imposition of any worst\-/case bound on the runtime of a finite\-/state verifier precludes it from benefiting from superlinear amounts of any kind of resource.

%Proceeding towards supplying verifiers with public coin budgets greater than a constant, one notices that a verifier must be careful about keeping its budgets so that it does not get disqualified as a verifier of the respective complexity class by exceeding it.  This has the following implication for the constant\-/space, constant private randomness verifiers with any amount of public coins.
\begin{theorem}\label{thm:strictboundislinear}
    For any time complexity function $t$,
    %$f(n)$ and $g(n)$ such that $\OH{g(n)} \supseteq \OH{n}$,
    \begin{multline*}
        \IP{\co\spa, \infty\pri\ran, \infty\pub\ran, \OH{t(n)}\tim} \subseteq\\
        \IP{\co\spa, \li\pri\ran, \li\pub\ran, \li\tim}.
    \end{multline*}
    The same is true also for the %high error 
    $\IPhigh{\ldots}$ variants of the two classes.
\end{theorem}
\begin{proof}
    % The equality is trivial when $\OH{g(n)} = \OH{n}$.  The former class being a superset of the latter is also trivial.  We will show the other direction of the inclusion for when $\OH{g(n)} \supsetneq \OH{n}$ to complete the proof.
    % 
    %Let $V$ be any verifier of the former class and $w$ be any input to it.
  %There are $n+2$ different input head positions.  The rest have a constant amount of variation.  
    The number of configurations available to a constant\-/space verifier $V$ running on an input of length $n$ is %\utkanadd{$f(n)$ for some $f(n) \in 
    in $\OH{n}$.
    % \utkanrem{less than $k \cdot n$ for some constant $k$.} %BUNUN ESKI HALI KUCUK $n$'ler icin gecerli degildi, ``large enough $n$'' gibi bir sey demek gerekiyordu.

    If $V$'s transition function allows it to enter the same configuration more than once with nonzero probability for some input string, it will do so  arbitrarily many times with positive probability, exceeding any preset time bound. Therefore, all constant\-/space verifiers with a worst\-/case time bound $t(n)$ are actually limited to run within a linear bound. %run for more than %$\utkanadd{f(n) \in} 
    %\utkanadd{some} $\OH{n}$ steps. 
    The linear bounds on coin usage are established by noting that a machine that can run for at most $T$ steps is restricted to flipping at most $T$ coins of any type.
% !!!!!!!!OLMUŞ MU???!!! tesekkurler
    % \utkanadd{UTKAN: \OH{n}'in bir küme olmasından kaynaklı huzursuzluklar yaşıyorum. İlkinde ``in'' sorunu kesin çözüyor, ikincisinde ``some'' çözebiliyor mu emin değilim.}
\end{proof}

\section{A new characterization of \texorpdfstring{\PP}{P}}\label{sec:results}

%Due to \Cref{thm:strictboundislinear}, any strict budget greater than linear cannot be utilized by the finite\-/state, constant private coin verifiers.  We, therefore, turn our attention to statistical bounds.

Let us now examine finite\-/state verifiers employing a constant amount of private coins and an expected polynomial amount of public coins (unless they are tricked to loop forever with some arbitrarily small probability).
This setup turns out to provide a new characterization of the complexity class \PP, corresponding to the collection of languages decidable by deterministic Turing machines in polynomial time and space.
\begin{theorem}\label{thm:P}% !!!!!!!TEK SATIRDA DAHA GÜZEL OLUR?!?!?!!!!!! Utkan: Yaptim
    % \begin{equation*}
%    \utkanadd{The classes $\IP{\co\spa, \co\pri\ran, \po\pex\pub\ran, \po\pex\tim}$, $\IPhigh{\lo\spa, \co\pri\ran, \infty\pub\ran, \infty\tim}$, and $\PP$ are all equal.
 %   \begin{multline*}
  %      \IP{\co\spa, \co\pri\ran, \po\pex\pub\ran, \po\pex\tim} = \\
   %     \IPhigh{\lo\spa, \co\pri\ran, \infty\pub\ran, \infty\tim} = \PP.
    %\end{multline*}
    The following three classes are equal:
    \begin{enumerate}
        \item $\IP{\co\spa, \co\pri\ran, \po\pex\pub\ran, \po\pex\tim}$
        \item $\IPhigh{\lo\spa, \co\pri\ran, \infty\pub\ran, \infty\tim}$
        \item $\PP$
    \end{enumerate}
    % \utkanrem{$\IP[0.9]{\co\spa, \co\pri\ran, \po\pex\pub\ran, \po\pex\tim} = \PP$.}
    % \end{equation*}
\end{theorem}
%\begin{proof}
 %   It is  known~\cite{con89,GKR15} that
    %\begin{equation*}
  %      \IP[0.9]{\lo\spa, \po\pub\ran, \po\tim} = \PP.
   % \end{equation*}
  %  The proof follows from this fact and \Cref{lem:dir1,lem:dir2}.
    % \qed %% LLNCS ONLY
%\end{proof}
%The equality \utkanadd{Equalities??} in
\Cref{thm:P} follows from \Cref{lem:dir1,lem:dir2NEW}.

\begin{lemma}\label{lem:dir1}
    %\begin{multline*}
     %   \IP[0.9]{\lo\spa, \po\pub\ran, \po\tim}
     $\PP \subseteq
        \IP{\co\spa, \co\pri\ran, \po\pex\pub\ran, \po\pex\tim}$.
   % \end{multline*}
%    More specifically, for any $t > 1$,
 %   \begin{multline*}
  %      \IP[0.9]{\lo\spa, \OH{n^t}\pub\ran, \OH{n^t}\tim}\subseteq\\
   %     \IP[0.9]{\co\spa, \co\pri\ran, \OH{n^{t+2}}\pex\pub\ran, \OH{n^{t+2}}\pex\tim}.
    %\end{multline*}
    % \utkanrem{Languages that can be verified using log\-/space, poly\-/time, and infinite public coins \utkanadd{in an IPS (two\-/way)} with a non\-/zero probability of error can \utkanadd{also} be verified using \utkanadd{cons\-/private\-/coins instead of the log\-/space} \utkanrem{constant\-/space, constant private coins, and infinite public coins} with arbitrarily low non\-/zero probability of error. This new verifier is also expected to run for at most a polynomial time, except for some other and independent arbitrarily low non\-/zero probability of error.}
    % Languages that can be verified using logarithmic space (in terms of the input string's length), polynomial time, and infinite public coins with arbitrarily low non\-/zero probability of error can also be verified by using constant\-/space, constant private coins, and infinite public coins with arbitrarily low non\-/zero probability of error.
\end{lemma}

\newcommand{\psimulation}{\ensuremath{p_{\mathrm{head}}}}
\newcommand{\ptimer}{\ensuremath{p_{\mathrm{timer}}}}

\begin{proof}
Let $L$ be any language in $\PP$. As Goldwasser et al.\ have proven in~\cite{GKR15}, there exists a  public\-/coin verifier $V_1$ verifying $L$ with the following properties:
\begin{itemize}
    \item $V_1$ has perfect completeness;
    %, \ie, there exists a prover whose interaction with $V_1$ leads it to accept any string in $L$ with probability $1$;% when it interacts with a truthful \utkanadd{AND EXPERT? YA DA SADECE EXPERT? TRUTHFUL OLMASI YETERLI DEGIL, ISE YARAMAZ DOGRULAR YETMEZ, RESOURCEFUL/KNOWLEDGEABLE DA OLMALI (P3'un "truthful" olmasinin spesifik olarak "istenen bilgileri dogru iletmesi" gibi bir anlami oldugu icin yeterli oluyor ama bu genel olarak her verifier\-/prover etkilesimi icin gecerlidir diyebilir miyiz?)} prover about ;
    \item %For any string $w$ not in $L$, no prover can convince $V_1$ to accept $w$ with probability greater than $\sfrac{1}{2}$ \utkanadd{UTKAN: olasiliklardan bahseden cumleler negatif kurulunca anlamasi cok zorlasiyor. (final itirazina gelen bir ogrencimizi animsadim. sizin okudugunuz ``guaranteed'' kelimesini iceren yargimizin contrapositive'ini almaya kalkinca cumle yanlis bir seye donusmus, belki siz de hatirlarsiniz.) pozitif kursak daha iyi olabilir. MESELA:\\
    For any string $w$ not in $L$ and any prover $P^*$, the probability that $V_1$ is convinced to accept $w$ by $P^*$ is at most $\sfrac12$; and %\utkanadd{OLASILIK ORANININ NEYE BAGLANDIGI SYNTACTIC OLARAK KARISIK. [No prover can convince V to accept any string not in L] with probability greater than 1/2 mi, yoksa No prover can convince V to [accept any string not in L with probability greater than 1/2] mi? (ikincisi tabii ki ama ilki gibi de okunabilmesi okumayi zorlastiriyor)}
    \item There exists an integer
    $t>1$ such that  $V_1$ uses $\OH{\log n}$ space and  $\OH{n^t}$ time for any input of length $n$. % \utkanadd{UTKAN: miktar-cins ve cins-miktar seklinde ters yazmisiz, bu intended mi?} (for some ).
    (These are worst\-/case bounds.) %!!!!BURADA FOR ANY INPUT LAFINI CÜMLENİN EN SONUNA ALMAK LAZIM, DEĞİL Mİ??!!  UTKAN: Başta ve sonda olması arasındaki farkı göremedim.
 \end{itemize}

%Let $V_1$ be a machine that simply runs $V$ $n$ times, and accepts the input if no run ends in rejection.
%The one\-/sided error bound of $V$ can be reduced   easily to $2^{-n}$ by repetition. For any desired positive value $\varepsilon_1$,  one can then obtain a machine 
%$V_1$ operates in time $\OH{n^t}$ (for some $t>1$) and verifies $L$ with one\-/sided error $2^{-n}$. %!!!!!!!!YOKSA BUNU 2ÜSSÜEKSİn BIRAKIP SONDA V3'ÜN STATELERİYLE Mİ OYNASAK!!!!!!!!!!1 Note that one can reduce the error bound of $V_1$ by just increasing the number of its states, without changing its work tape usage. %Furthermore, $V_1$ . Note that one can  obtain a verifier with any desired (positive) value for the one\-/sided error bound by modifying $V_1$  to repeat  the protocol an appropriate number of times. % \utkanadd{that is allowed to be arbitrarily low}. 
   We will assume that exactly
   % \utkanrem{$\ceil*{\log n}$}
   $\floor*{\log(n+2)}$ cells are used in the work tape of $V_1$, and a multi\-/track alphabet (\eg, as in~\cite{H72}) is used to accommodate for the required amount of memory.
    % Let $L$ be a language that can be verified using logarithmic space, polynomial time, and infinite public coins with arbitrarily low non\-/zero probability of error. Then, let $V_1$ be such a verifier with the probability of error $\varepsilon_1 > 0$. We will assume that the working tape of $V_1$ is exactly $\log n$ cells wide but with a stacked alphabet to accommodate for \OH{\log n} memory (where $n$ is the length of the input string), for the sake of simplicity.
    
   % \utkanadd{The language verified in an IPS is determined by the verifier's decisions when it is paired against the most convincing prover, among the infinite domain of provers.  Since the verifier is a fixed and finite entity in this scheme, such a domain includes provers that (assume to) know the verifier's algorithm by heart (albeit possibly being incorrect at that).  Moreover, we can assume that the most convincing prover $P_1$ does indeed know the $V_1$'s algorithm correctly and by heart (since knowing $V_1$'s algorithm correctly cannot make the prover any less convincing).  Since all the coins of $V_1$ are also public, any message that $V_1$ would send to $P_1$ would already be obvious to $P_1$, and redundant.  Therefore, without loss of generality, we will assume that the only information that $V_1$ gives out in the dialogue of interaction is the outcomes of the public coins as they are tossed.}

In the following discussion, let any prover facing $V_1$ be called $P_1$. %Since $V_1$ cannot be fooled into accepting a non\-/member of $L$ with high probability no matter what prover it is facing, it is also immune against any such $P_1$ that ``knows'' $V_1$'s algorithm. Since all coins are public, such a $P_1$ can be assumed to have complete knowledge about $V_1$'s configuration at every step of their interaction.
% Since the definition stipulates that $V_1$ is faced with the ``best'' possible prover (named, say, $P_1$) for this job, one assumes that $P_1$ ``knows'' $V_1$'s algorithm. Since all coins are public, $P_1$ has complete knowledge about $V_1$'s configuration at every step of their interaction.
% As in the proof of \Cref{thm:conspubliconly}, we assume that  $V_1$ sends no information to $P_1$ other than the coin outcomes, without loss of generality.  \utkanadd{UTKAN: Bu cümle doğru, $V_1$'in ``information'' göndermesi gerekmiyor.  Ancak $V_1$ hala ``prover'ın yönlendirmesine ihtiyacım var'' anlamında ``sinyal'' gönderebilmeli.  Eğer onu da göndermiyor dersek bu generality'i bozar.}

%\utkanadd{In line with our definition for IPS'es, possibly among many other, one prover $P_1$ that is the most convincing at making $V_1$ accept its inputs would have complete knowledge of $V_1$'s algorithm.   This renders any further communication from $V_1$'s part obvious and redundant.  Hence, given the existence of such provers, without loss of generality, we will assume that $V_1$ does not write to the communication cell.}

   There exists
    a constant\-/space, public\-/coin, $k$\=/head verifier $V_2$ that can verify $L$ by simply executing $V_1$'s program, simulating $V_1$'s
    logarithmic\-/length work tape by the technique of \Cref{thm:hartmanisextended}. (Note that $k$ depends on the precise worst\-/case memory requirement of $V_1$.) % APPENDIX REINCLUDE in \Cref{sec:hartmanis}.
        % using most of its $k$ heads' positions over the input as a means of mimicking memory using the technique introduced in~\cite{H72},
      %  \item \textbf{public coins} using its own public coins, and
       % \item \textbf{computation} by executing over the input string and accepting whenever the simulated $V_1$ accepts.
    %\end{enumerate}
Since the simulation is direct and does not involve any additional use of randomness, $V_2$ verifies $L$ with the same amount of one\-/sided  error as $V_1$.  The only time overhead is caused by the simulation of the logarithmically bounded memory, so, by \Cref{thm:hartmanisextended}, % APPENDIX REINCLUDE in \Cref{sec:hartmanis},
$V_2$ will complete its execution in \OH{n^{t+1}} time.  $V_2$'s prover, say, $P_2$, is supposed to follow the same protocol as $P_1$.
% Just like $V_1$, $V_2$ sends no information to its prover, say, $P_2$ (which is supposed to follow the same protocol as $P_1$), except the outcomes of its public coins. %Note that $P_2$ need not flip a coin at every step, since the procedure for aligning the heads of $V_2$ to simulate an individual step of $V_1$ (described in Appendix \ref{sec:hartmanisextendedproof}) is deterministic.

    % The runtime of $V_1$ will be longer due to the overhead of simulating memory with the head positions. However, it will still be polynomial in the length of its input. The simulation overhead occurs at every simulated transition of $V_1$, thereby appearing as a factor in the total runtime, and involves the following:
    % \begin{enumerate}
    %     \item Decoding the bit values at the \ith{i} index of simulated memory from a head's position, which requires halving of the heads' positions $i-1$ times (incurring less than $2n$ steps) and checking if the resulting index is even or odd (incurring less than $n$ steps).
    %     \item Possibly modifying the heads' positions to change the value of a bit at a particular index $i$, which requires moving a spare head to the \ith{2^i} index (incurring less than $2n$ steps) and moving the space\-/encoding head by $2^i$ cells (incurring less than $n$ steps). This needs to be done for all the space\-/encoding heads, whose total number is less than $k$. 
    % \end{enumerate}
    % The total overhead factor $3n + 3nk$ is polynomial (even linear) in terms of the input length, allowing $V_2$ to run also in polynomial time, \OH{n^{t+1}}.

    % \threeast{}

We now describe  $V_3$,   a constant\-/space, single\-/head verifier that uses a constant number of private coins, in addition to the public coins that it flips at almost \emph{every} step, %\utkanadd{V1 VE V2 SUREKLI PUBLIC COIN ATMIYOR DA OLABILIR. V3 YINE DE POLY TIMER'I ICIN SUREKLI PUBLIC COIN ATMALI, AMA BU KISMIN (sanki V2'nin simulasyonu icin HER adimda public coin atmak asikar bir gereklilikmis gibi) PARANTEZ ICINDE OLMASI GARIP}) 
to emulate $V_2$'s verification of $L$.\footnote{This is an adaptation of a technique introduced by Say and Yakary\i{}lmaz~\cite{SY14} for simulating a multihead nondeterministic automaton in an interactive proof system whose verifier is a (single\-/head) probabilistic finite automaton.} 

%$V_3$ starts by scanning the input string $w$. If the length of $w$ is at most $n_0$, $V_3$ consults a lookup table to decide whether to accept or reject $w$. Otherwise, 

% \pagebreak[2]
$V_3$  (\Cref{alg:multiheadsimulator}) performs the following $m$\=/round procedure:\footnote{The precise settings of  $V_3$'s parameters (like $m$) will be discussed below.}

Each round begins with $V_3$ flipping $r$ of its private coins.  %\utkanrem{Thanks to} \utkanadd{Through} 
Using this randomness, it picks one of the $k$ heads of $V_2$.  Each head has the same very small probability $p=2^{-r}$ of being selected.  (How $V_3$ operates with the remaining high probability $1 - kp$ will be explained later.) %ESKIDEN "GERIYE KALAN OLASILIK" BU ASAMADA VARLIGI AKLA GELECEK BIR SEY DEGILDI AMA SIMDI $p$'NIN BURACIKTA SABITLENMESIYLE HEMEN AKLA GELIYOR.)} 
$V_3$ then engages in an interaction with its own prover, say, $P_3$, to simulate the execution of $V_2$, including $V_2$'s interaction with $P_2$ about  the input string. 

% $V_3$ does not send any information other than the outcomes of its public coins to $P_3$.
Note that $V_2$ flips coins in some, but not necessarily all of its computation steps, whereas $V_3$ flips a public coin %\utkanadd{its public coins} 
at each step of the %\utkanadd{$V_2$'s} 
simulation. Only some of these coin outcomes are used to stand in for $V_2$'s public coins.
% $V_2$'s messages to $P_2$.
What $V_3$ does with the remaining public random bits will be explained below.

To simulate $V_2$, $V_3$ traces the selected head of $V_2$ with its own single head, and relies on $P_3$ to provide an unbroken stream of information about what the other heads of $V_2$ would be reading at every step of its execution. In its response to any coin flip of $V_3$, $P_3$ is expected to transmit    its claims about the readings of all $k$ heads of $V_2$
    at that step of the simulated interaction (and $P_2$'s response to $V_2$, if the currently simulated transition of $V_2$ emanated from a communication state).
    % Furthermore, when
    When the simulation arrives at an actual coin\-/flipping state of $V_2$, $P_3$ is supposed to interpret the latest coin outcome as a public coin flipped by $V_2$.
    % , and include $P_2$'s response to $V_2$ in its message as well.

%\footnote{\utkanadd{Although $V_3$ tosses public coins at every step to not compromise the secrecy of the mode it is operating in (here, in simulation mode, rather than the timer mode that will be described shortly), in this mode, it will put them into use only when feeding them to the simulated $V_2$ when it needs one, and discard the rest.}}  $P_3$, on the other hand, is expected  to transmit  both
 %   \begin{itemize}
  %      \item what $P_2$ would be transmitting to $V_2$, and
   %     \item %\utkanadd{(unaware of the head picked by $V_3$)}
   %     its claims about the readings of all $k$ heads of $V_2$
   % \end{itemize}
    %at every step of the simulated interaction.
    %\utkanrem{purportedly accepting execution as narrated by the certificate. 
    %This narration consists of the readings of $V_2$'s $k$ heads throughout its execution.
    %Paralleling this narration, $V_1$ also expects the certificate to provide the certificate that would bring $V_2$ (and $V_1$) to its purported acceptance.}
    %This certificate is prepared on\-/the\-/fly by the prover that is also witnessing the outcomes of the public coins as they are tossed.
    
   At any step, $V_3$ checks the part of $P_3$'s claims regarding the head it had chosen in private, and \rejects{} if it sees any discrepancy. If the information sent by $P_3$ has led the simulation of $V_2$ to reach acceptance at the current step,  $V_3$ has not been able to catch a lie up to that point, and if this was not the \ith{m} round, $V_3$ moves its head back to the left end of the input tape without flipping coins, and proceeds to the next round.

\newcommand\mtimer{\ensuremath{M_{\mathrm{timer}}}}
\newcommand\hmode[1]{\ensuremath{H_{#1}}}
\newcommand\tmode{\ensuremath{\mathit{TIMER}}}
\newcommand\mode{\ensuremath{\mathit{mode}}}

  The probability  $\psimulation$ that $V_3$ will attempt to use its head to check the claims of $P_3$ in the manner described above is just  $k p$. 
    %While the $p_i$ are positive for all $i$, their sum, $\psimulation = \sum_{i=1}^k p_i$ is very small by design, \utkanadd{and each are chosen to be 
   % maximally 
 %  close to their mean by being set to equal either
  %  \begin{equation*}
   %     \psimulation \cdot \frac{\floor*{\sfrac{2^r}{k}}}{2^r}
    %    \qquad\text{or}\qquad
     %   \psimulation \cdot \frac{\ceil*{\sfrac{2^r}{k}}}{2^r},
    %\end{equation*}
    %making each $p_i$ strictly greater than $\sfrac{\psimulation}{2k}$.}
\begin{algorithm}[!b]

In the following, $V_2$ is assumed to never utilize its work tape and to always send a dummy symbol (\eg, $\blanksymb$) to its prover when it needs it to update its communication cell, without loss of generality.  $\Set{\hmode1, \hmode2, \dotsc, \hmode{k}, \tmode}$ is the set of different modes in which $V_3$ can operate.

\vspace{\topsep}

\begin{turinglist}%{V_3}{On input $w$:}
    \item Repeat the following for $m$ rounds:
    \begin{turinglist}
        \titem{Use $r$ private coin flips to select $\mode$ to be an element of $\Set{ \hmode1, \dotsc, \hmode{k} }$ with $p = 2^{-r}$ probability for each, %\utkanadd{UTKAN: Bu parantezi kaldırsak mı? Bu şekilde hemen aynı cümlenin devamında parantez dışında "*remaining* probability" dememiz havada kalmış gibi geldi.}
        or \tmode{} with the remaining probability.}
        \titem{Move the input head to the left end\-/marker.} %\utkanadd{\timespec{\OH{n}}}
        \titem{Initialize a simulation of $V_2$, using a field $\gamma_2$ to keep its communication cell content.} %keeping its state $q_2$ and
        \titem{Communicate with $P_3$ for it to reply with its claims about $V_2$'s initial head readings.}
        \titem{If $\mode = \tmode$, initialize a parallel simulation for \mtimer.} %keeping its state.
        \titem{Repeatedly execute the following move, which advances the simulation(s) by one step, until $V_2$ accepts:}
        \begin{turinglist}
            \item{(All actions described for this iteration are executed in a single verifier transition.)}
            \item{If $\mode = \tmode$ and $\mtimer$ has halted, \reject.}
            \item{If $V_2$ has rejected, \reject.}
            \item{Let $\gamma_3$ denote the symbol in the communication cell.}
            \item{Extract $P_3$'s claims about $V_2$'s head readings from $\gamma_3$.}
            \item{If $\mode = \hmode{i}$, and the scanned input symbol does not match $P_3$'s claim, \reject.}
            \item{If %\utkanadd{this is not the first step of $V_2$'s simulation and}
            $V_2$ communicated in the previously simulated step, extract the new value for $\gamma_2$ from $\gamma_3$, retaining its old value otherwise.}
            \item{Flip a public coin.}
            \item{Advance the simulation(s) utilizing the extracted information about $V_2$'s head readings, communication symbol and the coin outcome, moving the input head to imitate the \ith{i} head of $V_2$ if $\mode = \hmode{i}$, and  to imitate $\mtimer$'s head otherwise.}
            % If the simulated $V_2$ step was from a communication state, send $\gamma'$ to $P_3$.
           % Communicate with $P_3$ for it to reply with its claims for $V_2$'s readings, appended with $P_2$'s reply to $V_2$ if its simulation just transitioned from a communication state.
            %\bitem{   If a verification error has occurred, \reject.  Otherwise (meaning $V_2$ has accepted), continue.}
        \end{turinglist}
    \end{turinglist}
    \titem{\Accept.}
\end{turinglist}
\caption{A constant-space, single-head verifier $V_3$.}
\label{alg:multiheadsimulator}
\end{algorithm}
    With the remaining high probability $\ptimer$ (\ie, $1 - \psimulation$), 
    $V_3$ simulates $V_2$ by relying on $P_3$'s claims about the head readings and $P_2$ responses blindly, while using its own head and the public coins it flips to simultaneously %\utkanrem{operate  as a probabilistic timer} 
    simulate a probabilistic finite automaton $\mtimer$ that functions as a timer (\Cref{sec:polyclockproof}) in  this round.  
    % \utkanrem{(\ie, longer than $V_2$'s runtime)}
    This timer has an expected runtime of \OH{n^{t+2}}, and  exceeds $V_2$'s runtime with probability $1-\errprematuretimeout$, for some positive  $\errprematuretimeout$ that can be set to be arbitrarily close to $0$, by the premise of \Cref{lem:polyclock}. % APPENDIX REINCLUDE as demonstrated in \Cref{lem:polyclock} in \Cref{sec:clock}.
    %\utkanrem{with the expected runtime of \OH{n^{t+2}}, running for longer than $V_2$'s polynomial runtime except for some arbitrarily low probability $\errprematuretimeout > 0$ , and timing out eventually with probability 1} 
     If the simulation of $V_2$ fed by $P_3$ reaches acceptance before the timer runs out, then $V_3$ %\utkanadd{again resets its head back to $\lend$ without tossing coins and} 
     completes this  round of verification without rejecting. Otherwise (if the timer runs out before the simulation ends), $V_3$ \rejects.

    $V_3$ \accepts{} if it does not reject for $m$ rounds of verification. The total number of private coins used is $mr$.

    The rest of the proof is an analysis of the error probability and runtime of $V_3$.

    \paragraph{Arbitrarily small verification error.} %\utkanadd{The verification error of $V_3$ is defined as $\verr_3 = \max\paren{{\verracc_3, \verrrej_3}}$.}
    For any input string that is a member of $L$, $P_3$ can simply tell $V_3$ the truth about what $V_2$ would read with its $k$ heads, and emit the messages that $P_2$ would send to $V_2$ alongside those readings. Even when communicating with such a truthful $P_3$, $V_3$ may erroneously reject at any given round, %either due to the  simulated $V_2$ also rejecting,\footnote{Our definitions allow $V_1$, and therefore $V_2$, to reject members of $L$ with some small probability.} or 
    due to a premature timeout of the probabilistic timer.  The probability of that is at most %$\psimulation \cdot \varepsilon_1 +$
    $ \ptimer \cdot \errprematuretimeout$. To accept a member of $L$, $V_3$  must go through $m$ consecutive rounds of verification without %committing \utkanrem{such} %\utkanadd{those} errors. 
this event occurring.    
    The maximum probability with which  $V_3$ can fail to accept a string in $L$ is therefore
    \begin{equation*}
        \verracc_3 \le 1 - \paren*{1 - %\paren*%{%\psimulation \cdot \varepsilon_1 +
        \ptimer \cdot \errprematuretimeout}^m.
    \end{equation*}
    % Since $\psimulation$ and $\errprematuretimeout$ can be lowered arbitrarily to any positive constant, $\varepsilon_3^+$ can also be lowered to any positive constant.
    % Since $\psimulation$ and $\errprematuretimeout$ (and, although unnecessary, $\varepsilon_1$ as well) can be lowered arbitrarily to any positive constant, $\varepsilon_3^+$ can also be lowered to any positive constant.

    \newcommand{\acclabel}{\mathsf{P}}
    \newcommand{\acc}[1][]{\ensuremath{\acclabel^{#1}}}
    % \NewDocumentCommand{\IP}{ e{^_} O{1} m }{\ensuremath{\langclassformat{IP^{\IfValueT{#1}{#1}\IfValueT{#2}{#2}}_{\IfValueT{#2}{\arbitrarilylow}}\paren*{\scalebox{#3}{$#4$}}}}}

    \newcommand{\loo}{%
    \mathchoice
    {% Display style
    \adaptivecircle{$\displaystyle\acclabel$}{0.09em}
    }
    {% Text style
    \adaptivecircle{$\textstyle\acclabel$}{0.08em}
    }
    {% Script style
    \adaptivecircle{$\scriptstyle\acclabel$}{0.062em}
    }
    {% Scriptscript style
    \adaptivecircle{$\scriptscriptstyle\acclabel$}{0.05em}
    }
    }
    
    \newcommand{\adaptivecircle}[2]{%
    \tikz[baseline=(char.base)]{
        \node[shape=circle,draw,inner sep=-1.5*#2,line width=#2] (char) {\phantom{#1}};%
    }%
    }

    \renewcommand{\loo}{\mathsf{L}}

    We will now determine a bound for $\verrrej_3$, the probability with which $V_3$ may fail to reject an input string that is not a member of $L$. 
    Unlike $\verracc_3$,  $\verrrej_3$ is a value that  $P_3$ ``wants'' to maximize, so one has to be careful to include all possible strategies for the prover in this consideration. The program of $V_3$ dictates that there are exactly $m+1$ different ways (corresponding to alternative strategies for $P_3$) in which $V_3$ may fail to reject a string $w$ not in $L$: 

    \begin{itemize}
        \item $V_3$ may pass all $m$ rounds  by carrying out an assisted simulation of $V_2$ that ends in what seems to it  to be acceptance. %, or by ``watching'' the clock and receiving an acceptance claim from the prover. 
        We denote this event by $\acc[m]$.
        \item $V_3$ may be tricked (as will be described below) by $P_3$ into entering an infinite loop during the \ith[st]{j+1} round, after passing the first $j$ rounds,  for $j \in \Set{0, \dotsc, m-1}$. Such an event will be denoted by  $\acc[j]\loo$. %\utkanadd{UTKAN: loop roundunun sembolunu \adaptivecircle{$\textstyle\acclabel$}{0.08em} yerine $\loo$ yaptim. nasil olmus? geri alabilirim.}
    \end{itemize}

Recall that $P_3$ is expected to feed $V_3$ with information about the symbols that would be scanned by $V_2$'s input heads at any point of its execution, and  the transmissions of $P_2$. For $w\notin L$, if $P_3$ always tells the truth about the head readings, %\utkanadd{each round of $V_3$'s simulation will end in at most \OH{n^{t+1}} simulation steps and will be passed} \utkanrem{
$V_3$ can pass any round with the probability (at most $2^{-1}$) that $V_2$ would be deceived about $w$, and finally accept with probability at most $2^{-m}$.
% !!!!DOĞRU; DEĞİL Mİ?!!!! \utkanadd{Utkan: Bence dogru}
Since $P_2$'s messages are optimal for convincing $V_2$, whose program is hard\-/coded in $V_3$, $P_3$'s only option for achieving better odds of deceiving $V_3$ is by lying about the head readings of $V_2$. 

Since $V_3$ is able to track only one of $V_2$'s input heads with its own head, %; \utkanadd{BU NOKTALI VIRGUL DOGRU MU? BAGLACLA BAGLI CUMLECIKLERI COK FAZLA AYIRIYOR GIBI}
it is dependent on $P_3$ to carry out the simulation. If $V_3$ has picked some head $h$ of $V_2$ to track, a lie by $P_3$ about the symbol seen at that step by some other head $h'$ of $V_2$ may be enough to deceive $V_3$ into eventually believing that $V_2$  accepts $w$ in that round. $V_3$ can  catch such a lie only if it has previously %\utkanadd{PREVIOUSLY DERKEN ASLINDA BU TUR, ONCEKI BIR TUR DEGIL. YANLIS ANLASILMIYORSA SORUN DEGIL} 
picked the head that is being lied about, and detects the discrepancy between $P_3$'s claim and its own scan. 

Unlike $P_1$ and $P_2$, which deal with verifiers that have sufficient memory to keep track of their own runtimes, $P_3$ also has the capability to trick $V_3$, whose number of different possible configurations  is, in general, less than the runtime of $V_2$, which it is supposed to be simulating, into running forever. 
 %\utkanadd{DAHA NET BIR SEBEP ``the number of different configurations $V_3$ can be in is, in general, less than the runtime of the interaction it is simulating.'' BELKI YAZILI OLANA EK, OVERARCHING(?) BIR SEBEP OLARAK VERILEBILIR} 
It is common (see, for instance, \Cref{sec:hartmanisextendedproof}) for multihead finite automata to require a head to wait at a tape location while  another head is walking towards a specific symbol, \eg, an end\-/marker. If $V_3$ has  picked to track such a waiting head of $V_2$ with its own  head, it can be tricked into getting stuck in that configuration by an endless stream of claims from $P_3$ that the other head is still walking in the required direction. $V_3$ can  catch such a lie directly, if it has picked the head that is being lied about; or indirectly, if it has been running in ``timer mode''.

Our analysis of the error probabilities associated with the $m+1$ alternative strategies for the prover mentioned above requires setting bounds for the two following values:

\begin{itemize}
    \item  $\verr_{\loo}$, the probability that $V_3$ is tricked to enter an infinite  loop in a particular round.
    \item $\verr_{\acc}$, the probability that $V_3$ is led to complete a round without rejection when $w\notin L$.
\end{itemize}

$V_3$ cannot run forever in any round in which it has selected to act as a timer. %, since it would reject when the clock runs out if $P_3$ has still not announced the end of $V_2$'s run by that point. %\utkanadd{SOLDAKI CUMLE COK UZUN, SADECE VIRGUL ONCESINI TUTUP "CLEARLY" DESEK OLMUYOR MU?} 
The probability that  $V_3$ has selected to simulate $V_2$ with a head % the probability with which it has picked a head 
that $P_3$ would lie about is at least $p$.
%$\sfrac{\psimulation}{2k}$  that .
We conclude that %in any specific round, the probability 
%\utkanrem{$\verr_{\loo}$ %that $V_3$ can be led to loop is at most $\psimulation - p= (k-1) \cdot p$.}
\begin{equation*}
    \verr_{\loo} \le \psimulation - p= (k-1) \cdot p.
\end{equation*}

There are two different ways in which the prover can cause $V_3$ to complete a round without rejection when $w\notin L$:

\begin{enumerate}
    \item $P_3$ can be truthful. With probability at most $\sfrac{1}{2}$, the public coin flips would lead the resulting faithful simulation   to end in  $V_2$'s accept state.  
    \item $P_3$ can lie about a reading of a  head of $V_2$. In this case, $V_3$ may be tricked  into believing that $V_2$ accepts $w$ %\utkanadd{proper,}
    % \utkanadd{[[BUNU "V3 may be tricked into believing that V2 accepts w (as if it was a member of L, \ie, as if V2 were accepting it with probability 1 and not just 1/2)" MANASINDA YAZMAK ISTIYORUZ ASLINDA; "proper," EKLEMEM O ANLAMI SAGLIYOR MU, EMIN DEGILIM]]} !!!!!!!SENİN PROPER'I SİLDİM ÇÜNKÜ BENCE BURADA O MANADA YAZMAK İSTEMİYORUZ, V3'ün V2'nin ÖYLE VEYA BÖYLE ACCEPT ETTİĞİNİ SANMASINI KASTEDİYORUZ!!!!! Utkan: Doğru, artık katılıyorum.
    if $V_3$ has not selected to track that head. The probability of this event is  at most $1-p$. 
\end{enumerate}

In this scenario, $P_3$ can maximize its chances of leading $V_3$ to passing a round by basing its decision about whether to lie or not on the public coin outcomes: Call a public coin sequence \emph{lucky} if it leads  $V_2$ to  accepting $w$ when guided by $P_2$. In such cases, honesty is the best $P_3$ policy, since it involves zero probability of rejection as a result of being caught lying. For unlucky coin sequences, honesty has no chance of success, and $P_3$ would have to lie for $V_3$ to complete the round.

To obtain an easy upper bound for $\verr_{\acc}$, we grant  $P_3$ certain additional ``superpowers'' that allow it to always make the correct decision in the situations described above:  In particular, we assume that $P_3$ can always correctly foresee %intuit \utkanadd{BU KELIMEYI ILK DEFA DUYUYORUM AMA INTUITION ILE PARALEL BIR ANLAMI VARSA EGER, BU YINE BULGUYA DAYALI AMA FORMULIZE ETMESI ZOR BIR SEZGISEL KARAR MEKANIZMASI IFADE EDER; BURADAYSA HICBIR BULGUYA DAYALI OLAMAYACAK BIR TRUE-RANDOMNESS GELECEK-GORUSU TARZI BIR SEY VAR}  
whether the entire public coin sequence of the present round will be a lucky one or not, and that it is able to communicate with $V_3$ (lying about a single $V_2$ head or reporting all head readings truthfully) accordingly. Under these extremely favorable circumstances, $P_3$ can manage to be truthful exactly when the public coin sequence of the present round is a lucky one, and lie in the remaining cases.  The probability of a lucky sequence is, as mentioned before, at most $\sfrac12$.  We therefore have %Furthermore, 
\begin{equation*}
    \verr_{\acc} \le \frac{1}{2} + \paren*{1-\frac{1}{2}} \cdot \paren*{1 - p}=  1 - \frac{p}{2}.
    %\paren*{1 - 2^{-n}} \cdot p.
\end{equation*}

We can now establish bounds for the probabilities of the $m+1$  ``failure to reject'' events listed above. 

    The probability that  $V_3$ is led to accept $w \notin L$ 
    through $\acc[m]$ is %\utkanadd{UTKAN: ayni ifadede ``probability'' ve ``can'' olmasi anlamayi biraz zorlastiriyor. baska ne yapabilirdik? ``The probability that $V_3$ is led to accept...''? hemen asagida ve onceki sayfada $\verr_{\loo}$ ve $\verr_{\acc}$'i tanimladigimiz yerde de bu tip cumlelerden var.}
    \begin{equation*}
        \verr_{\acc[m]} =  \verr_{\acc}^m \le \paren*{1 - \frac{p}{2}}^m.
    \end{equation*}
    
    The  probability that $V_3$ is made to   fail to reject $w$ through $\acc[j]\loo$ ($j \in \Set{0, \dotsc, m-1}$) is
\begin{equation*}
    \verr_{\acc[j]\loo} =  \verr_{\acc}^j\verr_{\loo} \le  \paren*{%1 - \paren*{1 - 2^{-n}} \cdot p
    1 - \frac{p}{2}}^j \cdot \paren*{k - 1}\cdot p.
\end{equation*}

% \utkanadd{NIYE $\verr_{\acc[m]} = \verr_{\acc}^m$ VE $\verr_{\acc[j]\loo} = \verr_{\acc}^j\verr_{\loo}$ DEMIYORUZ?}!!!!!!!HAKLISIN, ÖYLE DİYELİM PLZ!!!!  Utkan: Der olduk

Although the prover has the ability to use the public coins to implement a probabilistic mixture of the $m+1$ mutually exclusive strategies in its repertory, it is easy to see that $P_3$ must focus only on the strategies associated with the highest ``success'' probabilities from its own point of view:
%    Instead of deciding on which one to employ deterministically, $P_3$ can assign probabilities to each one of the $m+1$ strategies available to it and pick one at random.  The failure to reject error probability of $V_3$ would then be
%    \begin{equation*}
 %       \verrrej_3 =
  %      p_{\acc[0]\loo} \cdot \verr_{\acc[0]\loo} +
   %     p_{\acc[1]\loo} \cdot \verr_{\acc[1]\loo} +
    %    \dotsb + 
     %   p_{\acc[m]} \cdot \verr_{\acc[m]},
   % \end{equation*}
%    where $p_{\acc[0]\loo} + p_{\acc[1]\loo} + \dotsb + p_{\acc[m]} = 1$.  This affine combination, however, is again maximized by setting the assigned probability corresponding to the greatest error term to $1$. (BURASI DAHA IYI IFADE EDILEBILIR BELKI, $p_X$ of the greatest $\verr_X$ DEMEK ISTERDIM) Thus, the bound for the failure to reject error is
    \begin{align*}
        \verrrej_3 &\le \max\paren*{\verr_{\acc[0]\loo}, \verr_{\acc[1]\loo}, \dotsc, \verr_{\acc[m]}}.
    \end{align*}
Since $\verr_{\acc[0]\loo}$ is clearly the maximum among all $\verr_{\acc[j]\loo}$'s, we have
    \begin{align*}   
        \verrrej_3 &\le \max\paren*{\verr_{\acc[0]\loo}, \verr_{\acc[m]}}.
    \end{align*}

Let us show that we can tune $V_3$ so that it verifies $L$ with error $\verr_3 = \max\paren*{\verracc_3, \verrrej_3}$, for any  small positive value $\verr_3$. Since
\begin{equation*}
    p=2^{-r} \quad \text{and} \quad \ptimer=1-\psimulation=\frac{2^r-k}{2^r},
\end{equation*}
% $p=2^{-r}$ and $\ptimer=1-\psimulation=\frac{2^r-k}{2^r}$,
we have established that the following three quantities should be less than or equal to the allowed error bound:
% !!!!!!BUNU MAX OLMAKTAN ÇIKARALIM; ALT ALTA O ÜÇ INEQUALITY'Yİ r'li şekilde YAZALIM!!!! Utkan: Yazdim. Substitution'larin dogru oldugundan emin olmak icin orijinal bulgularimizi da kopyaladim. Bence bunlar olmadan takip/teyit etmek (benim icin de, okur icin de) zor. Boyle de cirkin duruyor, silebiliriz; ama silince de okur icin ugrastirici oluyor.
\begin{align*}
    \verracc_3          &\le 1 - \paren*{1 - \ptimer \cdot \errprematuretimeout}^m = 1 - \paren*{1 - \frac{2^r-k}{2^r} \cdot \errprematuretimeout}^m\\
    \verr_{\acc[m]}     &\le \paren*{1 - \frac{p}{2}}^m = \paren*{1 - 2^{-r-1}}^m\\
    \verr_{\acc[0]\loo} &\le \paren*{k - 1}\cdot p = \paren*{k - 1}\cdot 2^{-r}
\end{align*}
    Given some desired $\verr_3$, %the requirement above (!!!!!!!!BUNA BİR NUMARA VEREBİLİRİZ!!!) can be satisfied as follows:
    \begin{enumerate}
        \item Set $r$ to a value that ensures $\verr_{\acc[0]\loo}$ does not exceed $\verr_3$.
        \item Set $m$ to a value that ensures that $\verr_{\acc[m]}$ does not exceed $\verr_3$.
        \item Set $\errprematuretimeout$ so that $\verracc_3$ does not exceed $\verr_3$. This can be achieved by arranging the number of states in the implementation of the probabilistic timer, as described in \Cref{sec:polyclockproof}. %!!!BUNU ORADA SON CÜMLEYLE SÖYLEDİM!!!!! \utkanadd{Utkan: BEN BASKA BIR SON CUMLE ALTERNATIFI EKLEDIM}
    \end{enumerate}
    Note that the parameters $r$, $m$, and $\errprematuretimeout$ are independent, allowing each of them to be set without affecting the values of the other two.

    \paragraph{Polynomial expected runtime with arbitrarily high probability.} With
    % \utkanrem{$\verrloop_3$}
    $\verr_3$ (and thereby also the probability of looping) set to a desired tiny value, $V_3$ will be running for at most $m$  rounds with the remaining high probability.  At each of those rounds, $V_3$ will either complete $V_2$'s simulation in \OH{n^{t+1}} time, or will operate as a probabilistic timer that has expected runtime \OH{n^{t+2}}.  Thus, it is expected to run in \OH{n^{t+2}} time with  probability at least $1-\verr_3$.

    \paragraphEnd{}

    We conclude that $L\in \IP{\co\spa, \co\pri\ran, \po\pex\pub\ran, \po\pex\tim}$.  
\end{proof}

% \begin{theorem}
    % Languages that can be verified using log\-/space, poly\-/time, and inf-public-coins in an IPS with a non-zero probability of error \uline{and perfect completeness} can also be verified using cons-private-coins instead of the log-space with arbitrarily low non-zero probability of error \uline{and perfect completeness}. This new verifier is also expected to run for at most a polynomial time, except for some other and independent arbitrarily low non-zero probability of error.
% \end{theorem}

The following corollary to \Cref{lem:dir1} depicts that the class $\NC$  has finite\-/state verifiers flipping a fixed number of private coins and with a tighter upper bound on their public coin budget compared to those for \PP.

% \Cref{lem:dir1,lem:dir2} together prove the following:
% \begin{theorem}
%     \begin{multline*}
%         \IP[0.7]{\lo\spa, \po\pex\tim, \po\pex\pub\ran} = \\
%         \IP[0.7]{\co\spa, \po\pex\tim, \po\pex\pub\ran, \co\pri\ran}.
%     \end{multline*}
% \end{theorem}
% !!!BUNA COROLLARY DESEK DAHA İYİ!!!! Dedik
\begin{corollary}
    % AYNI SATIRA ALAYIM MI?yes!!!!!
    % \begin{equation*}
        $\NC \subseteq \IP{\co\spa, \co\pri\ran, \OH{n^4}\pex\pub\ran, \OH{n^4}\pex\tim}$.
    % \end{equation*}
\end{corollary}
\begin{proof}
    Fortnow and Lund have proven~\cite{FL93} that 
    \begin{equation}\label{eq:FL93NC}
        \NC \subseteq \IP{\lo\spa, % \ze\pri\ran,
        \OH{n\log^2n}\pub\ran, \OH{n\log^2n}\tim}.
        % !!!!BUNA BİR DENKLEM NUMARASI VERELİM!!! verdim sanirim
    \end{equation}
% !!!!0-private-coins mu? BAŞTA KOYDUĞUMUZ RACONA AYKIRI!!! Utkan: sildim
The verifiers described in~\cite{FL93} for the right\-/hand side of % !!!BENCE BU BİR EQUATION DEĞİL, HİÇ İSİM VERMEDEN (1) DE GEÇ!! o zaman \Cref'i \ref yaptim, oldu hmm parantezi yok...İYİ GECELER- iyi geceler hocam
\eqref{eq:FL93NC} have perfect completeness. One can apply the sequence of constructions described in the proof of  \Cref{lem:dir1}, starting with such a verifier playing the role played by $V_1$ in that proof, to obtain the claimed result. %!!!!!BU DEDİĞİM DOĞRU MU?? O LEMMAYA BAKAN BİR VATANDAŞ TAK TAK GİDEREK BUNU ELDE EDİYOR DEĞİL Mİ?? SADECE t BUNA ÖZGÜ BİR DEĞER ALIYOR, RIGHT?!!! \utkanadd{Utkan: Evet, $t=2$ icin gecerli oluyor. $\OH{n\log^2n} \subseteq \OH{n^2}$ oldugunu not edebiliriz belki kolay anlasilsin diye, ve/veya ``$t=2$ icin'' diyebiliriz.}
% eger bir degisiklik olmadiysa eskiden oyleydi, tekrar bakiyorum
% (Bu dedigim sorun degilmis:) NE YAZIK KI OLMUS, ARTIK O LEMMA DOGRUDAN P ILE KIYAS YAPIYOR, IP'LER ARASI BIR KIYAS YOK
    %    
%        Since $\log^2n\in \OH{n}$ by standard asymptotic analysis, we also have
 %       \NC \subseteq \IP[0.8]{\lo\spa, \ze\pri\ran, \OH{n^2}\pub\ran, \OH{n^2}\tim}.
   % 
  %  The claimed result then follows directly from \Cref{lem:dir1}.
    % \qed %% LLNCS ONLY
\end{proof}
% NASIL???????????????

% \newcommand{\errloop}{\ensuremath{\err_{\text{loop}}}}
% \newcommand{\errpremature}{\ensuremath{p_{\text{long}}}}

\begin{lemma}\label{lem:dir2NEW}
    %\begin{multline*}
        $\IPhigh{\lo\spa, \co\pri\ran, \infty\pub\ran, \infty\tim}\subseteq \PP$.%\\
        %\IP[0.9]{\lo\spa, \po\pub\ran, \po\tim}.
    %\end{multline*}
   % More specifically, for any integer $t > 1$,
    %\begin{multline*}
     %   \IP[0.9]{\co\spa, \co\pri\ran, \OH{n^t}\pex\pub\ran, \OH{n^t}\pex\tim}\subseteq\\
      %  \IP[0.9]{\lo\spa, \OH{n^{t+1}}\pub\ran, \OH{n^{t+1}}\tim}.
    %\end{multline*}
\end{lemma}

\newcommand{\verrdiff}{\ensuremath{\verr_x}}

\begin{proof}
    Let $V_1$ be a log\-/space verifier that uses at most $r$ private coins and an unlimited budget of public coins to verify a language $L$ with some  error bound $\verr_1 < \sfrac{1}{2}$, for some constant $r$. %The two
    % \utkanrem{three}
%    sub\-/types of errors that $V_1$ may commit are that
 %   \begin{itemize}
  %      \item it might fail to accept %\utkanrem{reject} 
   %     a member of $L$ %when communicating with %an honest 
% !!!!GALİBA TANIMIMIZA GÖRE BU DURUMDA HONEST DEMEYE GEREK YOK; HONEST OLMAK ZORUNDA ZATEN!!!! \utkanadd{Utkan: honesty bu noktada ne demek ki? ortada verifier'in niyetli/beklentili framework veren bir tanimi yok. "honest" yerine "as desired" gibi bir sey denebilir diye dusunuyorum. member inputlarla da hata yapabilen bir verifier icin mesela, prover koyulan framework'u abuse ederek (= dishonest (?)) kabul ettirme sansini arttiriyor olabilir, ama bu da hala "as desired", cunku verifier member input'u olabildigince yuksek olasilikla kabul etmek istiyor}
    %   the prover 
%    with some probability $\verracc_1$,
 %       \item it might fail to reject
        % \utkanrem{be tricked  to accept}
  %      a non\-/member of $L$ with some probability $\verrrej_1$.!!!!!BUNLARIN İKİSİNİN ARKASINA DA "<= epsilon1" YAZALIM MI???!!!!
        % \item \utkanrem{it might be tricked to run forever when the input is not a member of $L$ with some probability $\verrloop_1$.}
   % \end{itemize}
    In the following, the prover that $V_1$ interacts with will be named $P_1$. The length of the input string is denoted by $n$.

    % Let $V_1$ be a (single\-/head) constant\-/space verifier that uses at most $r$ private coins and an unlimited budget of public coins to verify a language $L$, for some constant $r$. $V_1$ may run forever with some small probability $\errloop$; excluding those cases, it runs in expected time $f_1(n) \in \OH{n^t}$ where $t > 1$. In the following, the prover that $V_1$ interacts with will be named $P_1$.

We will now demonstrate the existence of a log\-/space verifier $V_2$ that verifies the same language $L$ with an error bound close to that of $V_1$ without using any private coins.  $V_2$ also has an unlimited budget of public coins.  In a nutshell, $V_2$ will interact with its prover, which we name $P_2$, to simulate $V_1$ parallelly for all the $2^r$ different possible private random bit sequences. It will feed all of these simulations with the same public random bit sequence that it will generate on the fly, while checking that $P_2$ is not trying to cheat by supplying messages in a way that would be impossible for $P_1$ (which, unlike $P_2$, does not know the current configuration of the verifier it is communicating with),  
% !!!! YENİ HALİ. OK İSE !!SIZLAŞTIR PLZ!!!!! %\utkanadd{UTKAN: Burada "impossible" cok havada kalmis, bir tur qualification (mesela yazdigim "in a way" veya "[impossible for $P_1$...] to do something" gibi bir sey) faydali olur.}
or by attempting to trick $V_2$ into an unnecessarily long conversation. $V_2$ will be operating a probabilistic clock to ensure that its expected runtime will be within %!!!!BU OMEGA DEĞİL DE O OLMALI; YANILIYOR MUYUM??
$2^{2^{\OHMEGAp{p(n)}}}$, for a suitably chosen polynomial $p$. %!!!!!!BU OMEGA MI BİR DAHA DÜŞÜNELİM EN SONUNDA!!!!!!!%$2^{2^{\OHMEGA{n}}}$. 
When this timer finally runs out, $V_2$ will pick one of the parallel simulations at random to imitate $V_1$'s decision. 
We now present a detailed discussion of each of these points.

%\utkanadd{UTKAN: ``sim'' ne demek, kendi içimizde netleştirelim.  Bence sim = ``simulation of $S_i$.''  Metinde ``sim'' yazan yerlerde gezinip bu şekilde okuyunca garip olan yerlere düzeltme önerilerimi bırakıyorum.}

\paragraph{The interaction between $V_2$ and $P_2$.} Let $\bin_r(i)$ denote the $r$\=/bit binary representation of the natural number $i \in \Set{0, \dotsc, 2^r - 1}$ (padded with $0$'s from left as needed). As mentioned above, $V_2$ will be running  $2^r$ parallel simulations (``sims'') of $V_1$,  where the \ith{i} simulated verifier $S_i$  %employs the \ith[st]{\paren*{i+1}} head of $V_2$, and 
% \utkanadd{UTKAN: Burada $S_i$ isimlerini simülasyonlara vermişiz.  Ama aşağıda $S_i$'ler doğrudan simülasyondaki verifier'ın ismiymişçesine initial state'inden bahsediyoruz, $P_1$'den kendilerine sembol gönderiyoruz.  Algoritmayı yazarken de bu $S_i$ ismini (aşağıdaki gibi) simüle edilen verifier'ların adı olarak kullanmak istiyorum ama burada verifier'ın değil de simülasyonun adı olarak tanımlanmış olmaları tutarsız ve bundan ötürü yanlış hissettirdi.}
is hardwired to use the bits of $\bin_r(i)$ as its private ``random'' bits.  More formally, for each non\-/halting state $q$ of $V_1$, 
each $S_i$ 
has $r+1$ states of the form $(q,\pi)$, one for each postfix $\pi$ of $\bin_r(i)$ to keep track of the remaining unused bits.  % \utkanadd{UTKAN: Bu $p$'leri de $\pi$ yapmaya ne dersiniz?  !!!!İSTİYORSAN YAP; PRIME'LILYA GEREK YOK!!!!!!!!  Eğer sonraki $\pi$'den farklı olsun diyorsanız, sonraki $\pi$'yi $\pi'$ da yapabiliriz.}  
The initial state of $S_i$ is the pair consisting of the initial state of $V_1$ and $\bin_r(i)$ itself. For each transition of $V_1$ that emanates from a state $q$, consumes a private random bit $\bpri\in\Set{0, 1}$, and enters some state $q'$, $S_i$ has corresponding transitions emanating from each state of the form $(q,\bpri\pi)$ (where $\pi$ is a proper postfix of $\bin_r(i)$), employing no private coin, and entering state $(q',\pi)$. % \utkanadd{UTKAN: $V_1$'in private coin atan $q$ state'leri için $(q, \estring)$ tipi state'lerden şu an hiç transition çıkmıyor ama Sec2.1'deki makine tanımımız gereği çıkmalı.  Unreachable oldukları için nereye gittikleri önemli değil ama yine de herhalde reject state daha sağduyulu bir seçim.!!!!DOĞRU AMA BURADAKİ MUHABBETİN DETAYINDA GEREKSİZ GİBİ GELDİ BANA!!!!!!! UTKAN: Bence şu veya bu şekilde tanımlanmaları şart. Paragrafın sonuna bir öneri bıraktım. Her alternatif benim için ok (mesela "tanımsız bırakılan her transition reject state'e gider" de denebilir), yeter ki makine modeline uygun şekilde tanımlı olsun.}
(Each such $S_i$ transition mimics every other aspect, \eg{}, communication symbols, input and work tape head actions, etc.\ 
% \utkanadd{UTKAN: ``etc.'' İngilizce bir kelime sayılır mı yoksa yabancı bir kelime mi?!!!! DİZGİCİLER DÜŞÜNSÜN??!!!!!}
of the corresponding $V_1$ transition faithfully.) Transitions of $V_1$ that do not consume private random bits are simulated by transitions of $S_i$ that do not change the second component of the machine's state.  $V_1$\=/transitions entering halting states are mimicked by $S_i$\=/transitions entering the unique accept and reject states of $S_i$.
% !!! YENİ CÜMLE!!! \utkanadd{Çok iyi, teşekkür ederim}
$S_i$\=/transitions emanating from (unreachable) states of the form $(q, \estring)$, where $q$ is a private\-/coin\-/tossing state of $V_1$, are directed to the reject state. % of $S_i$ that are left to define are those going out of the states \utkanalt{where a private\-/coin\-/tossing state of $V_1$ is paired with the empty string}{$(q, \estring)$ where $q$ is a private\-/coin\-/tossing state of $V_1$}, which are unreachable due to the private coin budget of $V_1$.  For completeness, those transitions are mapped to the reject state of $S_i$.}
% \utkanadd{!!!!!!!YUKARI YAZDIĞIM OK İSE BU MAVİ KISIM UÇSUN!!!!!!
% consists of $r+1$ copies of $V_1$, one for each prefix of $i$'s $r$\=/bit binary representation (padded with $0$'s from left as needed).  The initial state of $S_i$ is the initial state of the copy that corresponds to the longest prefix, \ie, the entire binary representation.  Transitions in each copy mimics the original and stays within the copy, except for the transitions that use a private random bit.  For a copy corresponding to the prefix $b\pi$ where $b$ is in $\Set{0, 1}$, its transitions consuming $b$ as the private random bit and terminating at state $t$ are modified \begin{enumerate*}[label=(\roman*)]
%     \item to terminate at the state that corresponds to $t$ in the copy corresponding to the prefix $\pi$ and
%     \item to not consume any private random bit in doing so,
% \end{enumerate*} whereas the sibling transitions that consume $1-b$ as the private random bit are removed.  In the copy for the prefix $\estring$, transitions emanating from private coin tossing states (which are unreachable due to definition of $V_1$) are rewired to terminate at the reject state of that copy.  To comply with the definition of verifiers, all copies of the accept (resp.\ reject) states are mapped to a new accept (resp.\ reject) state of $S_i$.}

At each step of its interaction with $V_2$, $P_2$ is supposed to supply the responses that $P_1$ would supply to each sim at the corresponding point, according to the convention that will be explained below. Whenever $V_2$ is ready to receive the next $P_2$ symbol, it will use the communication cell to indicate this. As we will see,  $V_2$ will flip public coins in some, but not necessarily all, of its communication steps.

The execution of $V_2$ (\Cref{alg:privatecoinsimulator}) proceeds in \emph{stages} delimited by its (public) coin flips. Each stage consists of one or more \emph{segments} delimited by communication steps.
Originally starting each $S_i$ from its initial configuration, $V_2$  proceeds in each segment by going through all the $S_i$'s sequentially,\footnote{Note that $V_2$'s logarithmic\-/space budget is sufficient to trace the executions of all the $2^r$ sims and store the configurations of those that are paused.} continuing the simulation of each $S_i$  until that sim has accepted, rejected,  entered a communication state, or is determined to have entered an infinite loop. (Note that there exists a polynomial $f$, % \utkanadd{UTKAN: $q$ ismini stateler için de kullandığımız için çakışıyorlar.}!!!! LÜTFEN SAKINCASIZ BİR İSİM BULUP HER YERDE UYGULA!!!!
dependent on $V_1$, such that a sim that runs for more than $f(n)$ steps without entering a communication state must be in an infinite loop. $V_2$ has sufficient memory to count deterministically up to $f(n)$,
\begin{algorithm}[htb]
\begin{turinglist}%{V_2}{On input $w$:}
    \titem{Initialize simulations of $S_0, \dotsc, S_{2^r-1}$ (the sims).} %  Keep their configurations. 
    \titem{Initialize the \emph{communication partition} (that will keep track of groups of sims with identical communication transcripts) with  all sims  in the same block.} %Keep lists of sims that have (so far) communicated identically against $P_1$ as ``communication partitions.''
    \titem{Initialize the Boolean variable to be used by the \TICK{} subroutine that will implement the timer.}
    \titem{Repeatedly execute the following until the timer runs out: (Each iteration of this loop carries out one \emph{stage} of the protocol, as defined in the text.)} %Iterate sims collectively in stages until each one is halted or determined to be looping, for up to $2^{2^{p(n)}}$ expected stages tracked via the \TICK{} subroutine, as follows:
    \begin{turinglist}
        \titem{Initialize the \emph{segment counter} to $0$.}
        \titem{Execute the following until each sim is labeled as halted, looping, or waiting for a coin toss:} %  !!!PARANTEZ İYİ DEĞİL. LABEL'LARI İTALİK FALAN MI YAZSAK DEDİM AMA WAITING OLAN BİRDEN FAZLA KELİME OLDUĞU İÇİN GÜZEL OLMAZ!! \utkanadd{UTKAN: Parantez, olası labelların bu üçünden ibaret olduğunu belirtmek içindi. Bunu ayrı bir yerde de belirtebiliriz, mesela algoritmanın tepesinde (Alg1'deki gibi). Belirtmemek de bir seçenek. Ama belirtirsek daha sonra uzun uzun "not labeled as 1, 2, or 3" demek yerine "not labeled" diyerek kısaltabiliyoruz.}!!!!!!BANA PARANTEZ SEMBOLLERİNİ SİLMEK ANLAMI ASLA DEĞİŞTİRMEZ GİBİ GELİYOR!!!!!!! \utkanadd{Cümlenizi anlayamadım. Bu noktada bir sorun yok aslında, elzem olmayan bir iyileştirme önerisi: "Algoritmada kullanılacak labellar sadece 'halted', 'looping' ve 'waiting for a coin toss' labellarıdır." bilgisini herhangi bir yere yazarsak iyi olabilir diye düşündüm. Bu birinci önerinin üstüne, ama paket olarak değil, ayrı bir öneri olarak: Bu verdiğimiz bilgiyi kullanarak algoritmadaki bazı cümlelerimizi kısaltabiliriz. Önerilerim doğru anlaşıldıktan sonra "ikincisini ..." veya "hiçbirini yapmayalım" derseniz benim için hiç sorun değil.}!!!!ANLAŞTIK!!} % Iterate sims collectively in segments until each one is halted, determined to be looping, or waiting for a public coin toss, for up to $g(n)$ segments tracked via counting, as follows:}
        \begin{turinglist}
           % \titem{Increment the segment counter and \reject{} if its value exceeds $g(n)$.}
         \titem{For each $S_i$ that is not labeled as halted, looping, or waiting for a coin toss, do the following:}
            \begin{turinglist}
                \titem{Proceed with $S_i$'s simulation %\utkanrem{(within a time limit of  $f(n)$ steps)} 
                until it halts, enters  a communication state, or runs for more than  the time limit of $f(n)$ steps without entering a communication state.  If  $S_i$ has halted or exceeded the limit, label it as halted or  looping, respectively. If  $S_i$ has entered a coin\-/tossing state, label it as waiting for a coin toss.}
            \end{turinglist}
            \titem{Advance any unlabeled sims (every such sim is in a non\-/coin\-/tossing communication state) %\utkanadd{UTKAN: Takip edebildiysem sadece ``[any sims that are] not labeled'' da denebilir.} 
            for one more transition, refine the communication partition by arranging sims transmitting different symbols %!!!DEĞİŞMİŞ HALİ!!! \utkanadd{Çok teşekkürler, ellerinize sağlık, çok iyi olmuş} 
            into different blocks, and prompt $P_2$ to provide the corresponding responses.} %!!!OK????!!!
            % \bbitem{!!!!ÖNCESİ OKSE BURAYI SİL!!!Break down the communication partitions by whether sims have just made a communicating transition, and if so, by the symbol sims wish to have communicated.}
            \titem{Use the communication partition to check whether  $P_2$ has attempted to give different responses to sims in the same block. If such a violation is detected, \reject{}. Otherwise, update the configurations of the communicating sims with the new responses.} %!!!OK!!!?
           \titem{Increment the segment counter.}
           \titem{If the value of the segment counter has exceeded $g(n)$, label all as yet unlabeled sims as looping.}
        \end{turinglist}
        \titem{Toss a public coin.  Using its outcome, advance all sims that have been waiting for a coin toss %\utkanadd{= (işi bitmemiş) bütün simler, qualification hafiften redundant.} 
        for one more transition. Unlabel those sims.}
        \titem{Refine the communication partition by arranging sims with different communications % !!!DEĞİŞTİ!!! 
        %   \utkanadd{UTKAN: Burası da Stage 10'daki gibi.  Oraya parantez içinde açıklama eklenirse burası olduğu gibi kalabilir.}
        into different blocks.} % \utkanadd{UTKAN: Bu maddeyi Stage 10'daki gibi üsttekiyle birleştirelim mi? Tabii Stage 12'de "toss a public coin" ve "unlabel those sims" biraz farklı, gözden kaçmamaları önemli, ama kaçmaz herhalde.} %!!!!ÖNCESİ OKSE SONRAYI SİL!!!!Break down the communication partitions by whether sims have just made a communicating transition, and if so, by the symbol sims wish to have communicated.
        \titem{Use the communication partition to check whether  $P_2$ has attempted to give different responses to sims in the same block after the last coin toss. If such a violation is detected, \reject{}. Otherwise, update the configurations of the unlabeled sims % \utkanadd{UTKAN: ``Label'ı kaldırılmış'' mı demek istiyorsak ``de-labeled'' denebilir belki.  ``Unlabeled'' eskiden beri label'ı olmayanları da kapsar (ama Stage 12'de dediğim gibi, Stage 12'deyken bütün sim'lerin label'ı vardı, o sebeple de-labeled ve unlabeled burada aynı şeye denk düşüyor).} 
        with the new responses.} %!!!OK!!!?
        %For sims that just communicated, get their replies from $P_2$ and update their configurations.  Verify that sims of same communication partition receive same replies.}
        \titem{Call the \TICK{} subroutine.}
    \end{turinglist}
    \titem{Toss $r$ public coins and accordingly choose an $S_i$ at random.  If $S_i$ has rejected or is labeled as looping, \reject.  Otherwise, \accept.}
\end{turinglist}
\caption{A log-space verifier $V_2$.}
\label{alg:privatecoinsimulator}
\end{algorithm}
which is simply the number of different possible configurations of $V_1$ on the present input string.) If no sims are in a communication state at the end of this scan of the $S_i$'s, %\utkanadd{UTKAN: aitlik belirten 's gibi durmasin diye ``sims'' desek?}
the overall simulation has finished. Otherwise, one or more sims are ``waiting'' for a response from the prover, and therefore $V_2$ should request the next $P_2$ symbol.

At this point, it may be the case that some sims are waiting in (communication) states that flip a public coin, whereas some other sims are in communication states that do not flip a coin. Since $V_2$ is designed to present the same public coin sequence to every sim, it performs a coin flip (which ends the present stage) only if \emph{all} the waiting sims are in coin\-/flipping states. Otherwise, $V_2$ flips no coin during its request to $P_2$; in this case, the sims waiting at coin\-/flipping states will be suspended in the subsequent segments of this stage, until the occurrence of the flip.

In response to $V_2$'s request, $P_2$ is supposed to provide a $2^r$\=/tuple whose \ith{i} element is the symbol that $P_1$ would send to $S_i$ if $S_i$ is  a waiting sim  that $V_2$ is going to continue simulating in the next segment, and a filler symbol %\utkanadd{UTKAN: ``null symbol'' tanim gerektirmeksizin anlamli bir terim mi?}!!!!!!!!BENCE ÖYLE? BLANK'İ BAŞKA İŞ İÇİN KULLANDIK SANKİ!!!
otherwise. (Since $V_2$ employs no private coins,  $P_2$ has complete information about all of the sims' configurations, as well as all the communication symbols any sim would have transmitted had it been interacting with $V_1$, at this point.) $V_2$ performs some additional controls (to be detailed below under separate headings) upon receipt of this symbol, rejecting if it detects a ``cheating'' attempt by $P_2$ at this juncture. 

After the single coin flip at the end of each stage, $V_2$ calls the %``double exponential timer  tick'' 
\TICK{} subroutine, described in \
\Cref{sec:clock}.\footnote{$V_2$ flips % \utkanrem{up to}
polynomially many additional public coins for each such subroutine call.  $P_2$ waits for the subroutine to finish and the next stage of $V_2$'s execution to resume.} %\utkanadd{UTKAN: Bu tırnak içindeki ismi nerede kullanıyoruz? Neye refer ettiğimiz iyi anlaşılıyor mu? Timer'ın tanıtıldığı yerde isimlendirmemiş gibiyiz, bir isim verip o isimle mi seslensek?}
% !!!!OLMUŞ MU? IF SO, !!'SIZLAŞTIRALIM !!!! \utkanadd{UTKAN: İsmin sabitlenmesi iyi olmuş.  Ancak isimden "timer" kelimesi yok olunca hemen sonra gelen cümledeki "timer" tabiri biraz havada kalmış.  İsmi TIMER TICK olarak değiştirmek bir çözüm olabilir.  Veya "If the timer the TICK emulates does not run out in this call, ..." denebilir.}
If the timer does not run out at  this tick,  simulation proceeds with the next stage. If the timer does run out, $V_2$ considers the overall simulation finished.

When the overall simulation is finished, $V_2$ flips $r$ more coins to select one of the sims.  %!!!EKLEME!!! \utkanadd{Teşekkürler}
(Essentially, where $V_2$ chooses a single computation branch randomly in private during its execution, $V_3$ simulates all those $2^r$ branches publicly and then chooses one of them  at the very end.) If the selected sim has  been determined to have rejected, or to have entered an infinite loop, $V_2$ \rejects. Otherwise, %\utkanadd{UTKAN: ``actually unfinished but considered finished'' simlerin de buna girdigini ekstradan not etmekte fayda olabilir mi?}, !!!!AZ SONRA BUNU DİYORUZ ??!!!!!
it \accepts.

\paragraph{Members of $L$ are accepted by $V_2$ with sufficiently high probability.} Consider an intermediate verifier $V'_1$, which is obtained by modifying $V_1$ so that it calls the same \TICK{} %\utkanadd{UTKAN: iki ust paragrafta ``timer tick'' denmis, burada ``time clock tick'' denmis} 
subroutine that we mentioned above in $V_2$'s description after every actual public coin toss of $V_1$ to determine whether it should time out and accept. Since the only difference between $V_1$ and $V'_1$ is that $V'_1$ may accept some input strings (which can get involved in very long computations that are cut off by the timer) with higher probability than $V_1$, $P_1$ is clearly able to convince $V'_1$ to accept any member of $L$ with probability at least   $1-\verr_1$.\footnote{Strictly speaking, $V'_1$ faces a slightly different  prover that imitates $P_1$ faithfully, pausing this procedure only to wait for the coin tosses of the \TICK{} subroutines to be completed.} We claim that $P_2$ will be able to convince $V_2$ to accept any member of $L$ with the same probability, by simply obeying the protocol described above and transmitting precisely the symbols that $P_1$ would have transmitted in response to communications from $V_1$ at every step: In this case, $P_2$'s interaction with $V_2$ would be a perfectly faithful imitation of $P_1$'s interaction with $V'_1$, with identical probabilities of acceptance, determined by the public coins flipped during the simulations, and the $r$ final public random bits that stand for the $r$ private coin flips of $V'_1$.

\paragraph{$P_2$ cannot trick $V_2$ into looping forever within a stage.} We have already noted that each sim can have at most $f(n)$ different configurations, for some polynomial $f$.
Consider the $2^r$\=/tuple consisting of the configurations of all the sims, which we will refer to as the ``collective configuration''. Clearly, the number of different possible collective configurations  is $g(n) \in \OH{f(n)^{2^r}}$, which is another polynomial in $n$. Considering the collective configurations at the end of each segment, a stage which has continued for more than $g(n)$ segments  must have therefore  repeated an end\-/of\-/segment collective configuration.  Note that, within any particular stage of the execution of $V_2$, $P_2$ has complete, deterministic control of the evolution of the collective configuration between subsequent communication steps, through the responses it sends. This would give a malevolent prover a chance to trick the verifier into looping forever within a stage. Since any ``honest'' prover\-/verifier interaction that eventually terminates with the verifier reaching a halting state can be shortened to avoid all such collective configuration repetitions without changing the outcome, $V_2$ expects $P_2$ to respect this convention, and concludes any stage that is seen to contain more than $g(n)$ segments by labeling any active sims that have not reached a coin-flipping state by that time as looping. % !!!BU CÜMLE DEĞİŞTİ!!!! \utkanadd{Sadece sonunu değiştirerek çözebileceğimizi sanmıyordum ama olmuş (meğer önceki cümleler hep ince qualify edilmiş), teşekkürler}

\paragraph{$P_2$ will be caught if it does not mimic $P_1$.} As noted above, $P_2$ has complete information about all the sims that it is communicating with. This is quite different than the situation of $P_1$, which does not see the private bits of the verifier. It is sometimes possible for $P_1$ to deduce information about the private random bits used by $V_1$ by noting the communication symbols that it transmits during the interaction, but $P_1$ has no way of distinguishing two probabilistic branches of $V_1$ corresponding to distinct private coin sequences that have the same communication transcript, \ie, that have both communicated exactly the same string of symbols, punctuated with public coin flips in identical locations, up to the present point. This presents a devious $P_2$ an opportunity to attempt to achieve a lower rejection probability than $P_1$ by transmitting different messages to sims that would be indistinguishable to (and would therefore have to receive the same message from) $P_1$. As a countermeasure, $V_2$ maintains an up\-/to\-/date list of groups of sims that should be presently indistinguishable to $P_1$,  and \rejects{} whenever it sees that the $P_2$ symbol is trying to convey different messages to different sims in the same group. As a result, any interaction where $P_2$ attempts to send messages that are not ``$P_1$\=/compliant'' in this sense leads to rejection with probability $1$.

% can operate similarly to keep the runtime of each sim throughout the entire simulation without resetting, limiting it to \OH{n^{t_2}}.

% the simulation of each $S_i$, $V_2$ will use 
% , to constrain the runtime by some polynomial $c \cdot n^t$, which is ought to be enough (as will be proven below) for the prover demonstrate membership

\paragraph{Double exponential runtime is enough for  $V_2$ to reject non\-/members of $L$  with sufficiently high probability.} It remains to analyze  the probability that $V_2$ fails to reject a non\-/member of $L$. Since $P_2$ is unable to make $V_2$ loop forever, and any attempt to deviate from the protocol described above would be caught, all we need to focus on is the additional error possibly introduced by $V_2$ cutting off sims with very long computations, and regarding them as having accepted. %Fortunately, Condon and Lipton have shown~\cite{CL89,C93b} that a verifier whose space consumption is logarithmically bounded \utkanadd{UTKAN: CL05/th6.1 bunu 2pfa private-coin verifierlar icin soyluyor, C93b/th6 bunu log-space public-coin verifierlar icin en basta parantez icinde gecistirilen bir argumanla soyluyor (ustelik double-exponential da degil yeterince buyuk bir exponential'in yeterli olacagini soyluyor); log-space private-coin verifierlar icin bunu soyleyen yer neresi?} halts \utkanadd{UTKAN: halt edecegini degil de ``benzer bir hata ile bu kadar surede halt eden bir versiyonun yapilabilecegini'' soyluyorlar, degil mi?} with high probability in double exponential time.
Fortunately, Condon and Lipton have shown~\cite{CL89,C93b} that the execution of a verifier whose space consumption is logarithmically bounded can be cut off after a double\-/exponentially large number of steps without a significant change in the acceptance and rejection probabilities.\footnote{The verifier model in~\cite{CL89} uses only private coins. The validity of this result for our verifiers is a simple consequence of the argument in the proof of \Cref{lem:privatesimulatespublic}.}  %Theorems~6.1, 6.2, and 6.3 
By the reasoning in the proof of Theorem~6.1 of~\cite{CL89}, for any desired small positive value of $\verrdiff$, there exists a polynomial %\utkanrem{$p(n)$} \utkanadd{
$p$ such that the log\-/space verifier $V_1$ will have rejected any non\-/member input string of length $n$ with probability at least $1-\verr_1-\verrdiff$ within at most $2^{2^{p(n)}}$ computational steps. By tuning the parameters (\Cref{sec:clock}, \Cref{sec:dir2appendices}) of the double exponential timer, we can therefore set the error bound of $V_2$ to any desired value in the interval $\paren*{\verr_1, \sfrac{1}{2}}$. %\utkanadd{UTKAN: $0.5$ yazsak ifadenin acik aralik oldugu daha kolay anlasilir mi?}.!!!INTERVAL DEDİM!!!

% !!!BURADA BİR SATIR BOŞLUK OLMASI HAYIRLI OLUR!???!!!!!!!!!!!!!! \utkanadd{UTKAN: Bir boşluk koydum ama diğer \texttt{\textbackslash paragraph\{...\}} boşluklarıyla aynı miktarda değil. Ne yazık ki "diğerleriyle aynı miktarda boşluk koy" diye bir şey yazmak da mümkün gözükmüyor. Sadece bizim bu dosya için o boşluğun ne kadar olduğunu tespit edip aynı miktarda koyabilirim (ama başka bir formata geçince oradaki boşluk farklılaşırsa bu hala aynı kalır), hatta hemen aşağıda, comment-out edilmiş bir şekilde o alternatif duruyor.}

% \bigskip
% \vspace{3.25ex plus 1ex minus .2ex}
\paragraphEnd{}

We conclude the proof by noting that the language $L$ verified by $V_2$ is in $\PP$, due to the fact that 
\begin{equation*}
    \IPhigh{\lo\spa, \infty\pub\ran, \infty\tim} = \PP,
\end{equation*}
% !!!!!!!!!!LÜTFEN BUNU YILDIZSIZ IP YAPALIM!!!!!!!!!!!! %\utkanadd{Utkan: yaptim}
% $\IP[0.9]{\lo\spa, \po\pub\ran, \po\tim}\subseteq \PP$,!!!!!!BUNU ORTALAYIP SONRAKİ SATIRI SOLA YASLA PLZ!!!!!! Utkan: yaptim
which was proven by Condon in~\cite{con89}.%!!!!!!!!!!!REFERANSI NİYE DEĞİŞTİRDİN???????!!!!!!!!!! \utkanadd{Utkan: statement'i da degistirdim (eski statement yeterli degil sanki? makinemiz poly-time'da calismiyor). statement'i degistirince ayni referans calismiyor olabilir diye dusundum. bende con89'un kendisi yok, icinde ne var bilmiyorum. lemma'yi bu sekilde guncelleyelim dediginiz sirada c93b/theorem6'ya bastan siz isaret etmistiniz sanki.}
%\\
        %.
% 
% \utkanadd{Using some more tracks in its log\-/space, $V_3$ can time its run and ensure that it does not run for more than the necessary polynomial amount. AMA BUNUN DA BIR OVERHEAD'I VAR. ACABA $V_2$'NIN TIMER KAFALARINI DA TAKIP EDIVERSEK MI? HEM ZATEN LOG-SPACE'TE DOGRUDAN POLY-TIMER YAPARKEN TUTTUGUMUZ SAYILARLA $V_2$'NIN TIMER KAFALARININ INDEKSLERI BIREBIR AYNI}
% \qed %% LLNCS ONLY
\end{proof}

%Would increasing the public coin budget to superpolynomial levels enlarge the class of verifiable languages beyond \Cref{thm:P}? We only have a partial answer to this question. For the special case of verifiers with perfect completeness, no amount of public coins would help a finite\-/state verifier with a constant number of private coins verify a language outside $\PP$. 
% with any fixed error bound $\epsilon$. 
%To see this, note that an interactive proof system (IPS) with a verifier $V$ with no limitation on its use of public coins but obeying the constraints described above corresponds to a debate system (DS), where the prover  of the DS corresponds precisely to the prover of the IPS, the verifier ($V_D)$ of the DS is a constant\-/space machine that only flips a constant number of private coins and runs a program that is identical to that of $V$, with the refuter of the DS supplying the ``public coin outcomes''. (Our argument does not require the refuter  to make sure that those are ``fair'' coin outcomes.) Thanks to the perfect completeness of $V$, the language $L$ verified by $V$ is precisely the language  which has  debates checkable by $V_D$. We conclude by Fact \ref{fact:debinp} that $L \in \PP$.

Since it is obvious that
\begin{multline*}
    \IP{\co\spa, \co\pri\ran, \po\pex\pub\ran, \po\pex\tim}\subseteq\\\IPhigh{\lo\spa, \co\pri\ran, \infty\pub\ran, \infty\tim},
\end{multline*}
% \IP{\co\spa, \co\pri\ran, \po\pex\pub\ran, \po\pex\tim}\subseteq\IPhigh{\lo\spa, \co\pri\ran, \infty\pub\ran, \infty\tim}$,
\Cref{lem:dir2NEW} concludes the proof of \Cref{thm:P}.

\section{Concluding remarks}\label{sec:conc}

%Although the construction in the proof of Lemma \ref{lem:dir1} starts from a verifier with perfect completeness, this is not a crucial requirement, and a straightforward modification of the analysis is sufficient to prove the equality $\IP{\co\spa, \co\pri\ran, \po\pex\pub\ran, \po\pex\tim}=$, \ie, that the additional freedom of failing to accept some members of the verified language does not increase the power of such verifiers.

We end with some remaining open questions on finite\-/state verifiers.

%Teoremden sildiğimiz eşitliği (bizim class ile standart logspace verifierlarınki arasında olanı) burada note edebiliriz!!!!!!!!!!ETMEYEBİLİRİZ::

\newcommand{\commaglue}{,\hskip 0pt plus 0.4ex\relax}

Focusing on verifiers which halt with probability $1$, Dwork and Stockmeyer~\cite{DS92} proved that the class  $\IPhigh{\co\spa\commaglue \po\rex\pri\ran\commaglue \po\rex\tim}$ contains all context\-/free languages and some \NP\Complete{} languages, but whether $\PP \subseteq \IPhigh{\co\spa, \po\rex\pri\ran, \po\rex\tim}$ or not is still an open question.\footnote{Dwork and Stockmeyer do construct finite\-/state verifiers that halt with probability $1$ for all languages in $\DTIME{2^{\OH{n}}}$, but those verifiers have double exponential runtime.~\cite{DS92}} On the other hand, our \Cref{thm:P} establishes that  polynomial\-/time finite\-/state machines (even those with our additional restrictions on private\-/coin usage) which are allowed to be tricked into looping with an arbitrarily small probability have the capability to verify all languages in $\PP$. Does imposing the requirement that a constant\-/space machine should halt with probability $1$ preclude it from verifying some languages?

It has been proven~\cite{GMS87} that every language verifiable with arbitrarily low error by a polynomial\-/time verifier (with no bound imposed on the space consumption) is also verifiable with such a verifier that has perfect completeness. We do not know whether a counterpart of this result holds for finite\-/state verifiers. (Note that the construction of \Cref{lem:dir1} creates verifiers without perfect completeness.) %Although we used \Cref{fact:debinp}  to show that finite\-/state verifiers with perfect completeness and a constant private\-/coin budget are restricted to languages in $\PP$ in \Cref{sec:results}, that fact cannot help one deduce that \emph{every} language in $\PP$ is verifiable by such verifiers, since the existence of a debate system for a language $L$ does not necessarily imply that the same language has an interactive proof system with a constant bound $\verr < 1$ on its probability of accepting input strings that are not members of $L$. 

It would be interesting to characterize the classes that are associated by reducing the public coin budgets in \Cref{sec:results}, like %\IP{\co\spa, \co\pri\ran, \infty\pub\ran, \infty\tim} and 
\IP{\co\spa, \co\pri\ran, \log\pex\pub\ran, \infty\tim}. 
Does Condon and Ladner's result showing  that logarithmic\-/space verifiers that flip only logarithmically many public coins cannot verify any language outside the class \LOGCFL{}~\cite{conlad95} have a counterpart for the constant\-/space case?

We also note that the following question, posed more than 30 years ago by Dwork and Stockmeyer~\cite{DS92}, is still open: 

Is there a nonregular language in \IPhigh{\co\spa, \po\rex\pub\ran, \po\rex\tim}?

%Class of languages that can be verified by the constant\-/space, constant\-/randomness, one\-/way\-/input verifiers, which \cref{thm:SY} proves to be equal to the class of languages verified by multi\-/head finite automaton, encompasses the rest.  
%The classes \vsri{\consX}{\consX}{\rtX} and \ODFAK{3} have members \nonpal{} and $\lfrac$ as exclusives from one another, respectively, as shown in \cref{theorem:nonpal,theorem:bomb2}.  Moreover, \cref{theorem:1dfa2subset} shows that the class \ODFAK{2} resides at their intersection.  Yet another language $\lmatch$ that can be verified with a two\-/headed one\-/way finite automaton is shown to be beyond real\-/time verification using a constant\-/space and a constant number of random bits in \cref{theorem:bombastik}.

%Unlike $\lfrac$, it may be impossible to recognize $\lmatch$ with a multi\-/head one\-/way finite automaton with any number of heads, which is an open question.  Likewise, $\lfrac$, while seems to be outside of \ONFAK{2}, is not yet proven to be so.  There are in total 9 regions in \cref{fig:conclusionvenn} without a definitive member written in it, not counting the inside of \ODFAK{2} and the outside of $\bigcup_k \ONFAK{k}$ which are trivially non\-/empty; it would be interesting to see that any one of them is actually empty.

%We think that the conjectures $\lmatch \notin \bigcup_k \ODFAK{k}$, $\Set{ww | w \in \Set{0, 1}^*} \notin \ODFAK{2}$, and even $\Set{ww | w \in \Set{0, 1}^*} \notin \vsri{\consX}{\consX}{\rtX}$ can be proved via a technique similar to that of \cref{theorem:bombastik} (???).

\acknowledgments{}

% \begin{acknowledgments}
% \acknowledgements
% This research was partially supported by Boğaziçi University Research Fund Grant Number 19441.
% Utkan Gezer's participation in this work is supported by the Turkish Directorate of Strategy and Budget under the TAM Project number 2007K12-873.
We  thank the user with the pseudonym \emph{obscurans} at \emph{Mathematics Stack Exchange} for their help with the analysis in the proof of \Cref{lem:polyclock}.
We are grateful to the anonymous referees for their helpful comments.
% We thank the reviewers of ICTCS '23 for their comments on~\cite{GS23}, the paper that precedes this substantially extended version.

% An earlier version of this paper~\cite{GS23} was presented in the 24th Italian Conference on Theoretical Computer Science, Palermo, Italy, September 13--15, 2023.  We thank the reviewers of that conference for their comments.  This is a substantially extended version.
% \end{acknowledgments}

% \nocite{*}
% \bibliographystyle{abbrv}
\bibliographystyle{abbrvnat}
% % use the following instead if you encounter problems 
% \bibliographystyle{alpha}
% \bibliography{sample-dmtcs}
% \label{sec:biblio}

% \begingroup
% \raggedright
% % \bibliographystyle{elsarticle-num}
% \bibliographystyle{plainnat}
\bibliography{references} 
% \endgroup

% \end{document} % CUTS OFF APPENDIX

% \begin{filecontents*}{appendix.tex}

% \renewcommand{\appendixpagename}{Appendix}

\clearpage

\appendix
% Unique anchors for appendix sections:
\renewcommand\theHsection{appendix.\Alph{section}}%
% If you have subsections in the appendix, keep their anchors unique too:
\renewcommand\theHsubsection{appendix.\Alph{section}.\arabic{subsection}}%
\crefalias{section}{appendix}

%% ELSARTICLE
% \appendixpage

%% LLNCS
\section*{\LARGE Appendix}
% \section*{\huge Appendix}

% \renewcommand{\thesection}{A.\arabic{section}}

\section{Proof of \texorpdfstring{\Cref{thm:hartmanisextended}}{Theorem \ref{thm:hartmanisextended}}}\label{sec:hartmanisextendedproof}

\hartmanisextended*

\begin{proof}
    Let $M$ be any logarithmic\-/space Turing machine as described in the theorem statement, and $L$ be its language.  Without loss of generality, let the work tape alphabet of $M$ be the set of $c$\=/digit binary numbers from $0$ to $2^c - 1$, where $0$ corresponds to the blank symbol of the alphabet.
    %by the language homomorphism \utkanadd{(DO WE NEED A REFERENCE?)}.  
    
    We will construct a $\paren*{c+5}$\=/head finite automaton $R$, whose heads will be denoted $\head_1, \dotsc, \head_c$, $\head_{\alpha}$, $\head_{\beta}$, $\head_{\gamma}$, $\head_{\delta}$, and $\head_{\epsilon}$. $R$ is to recognize $L$ by simulating $M$.
    The locations (indices) of heads $\head_1$ through $\head_c$ will encode the contents of $M$'s work tape, with the  \ith{i} head's index matching the number whose
    % \utkanrem{$\ceil{\log n}$\=/bit} 
    $\floor{\log(n+2)}$\=/bit binary representation is written %\utkanadd{UTKAN: Bence burada yine work tape'e doğru aktiflik ifade eden fiiller kullanmamanın yollarını arayalım. Aynı ifade ``read'' fiiliyle yazılabilir belki.} 
    in reverse on the \ith{i} track, \ie, the sequence obtained by concatenating the \ith{i} bit of each symbol of the work tape string. (The   leftmost input tape cell containing \lend{} has index $0$.) 
    $\head_\alpha$ will mimic $M$'s input tape head exactly, and $\head_\beta$ %\utkanadd{UTKAN: Input kafası subscript'ini $\sigma$ ($\delta$ tanımlarında input sembolüne verdiğimiz harf), work tape kafası subscript'ini $\phi$ ($\delta$ tanımlarında work tape sembolüne verdiğimiz harf) yapmayı düşünebiliriz. Fakat bunlar bu şekilde ``anlamlı'' olurlarsa diğer dandik olanlardan da insanın anlam çıkarası gelir. Onlara belki iyice dandik duracak şekilde $\head_{\mathrm{(i)}}$, $\head_{\mathrm{(ii)}}$ ve $\head_{\mathrm{(iii)}}$ diyebiliriz (ama $\head_i$ ile karışmayacağından emin olalım).} 
    will mimic the distance of $M$'s work tape head from the left end in its position, albeit on the input tape of $R$.  The remaining three heads will be used as auxiliaries to facilitate the manipulations of the indices of $\head_1$ through $\head_c$ required for the simulation of $M$'s memory.

    We now recall some multihead automaton programming techniques that will be used in the machine  described below.
% \item \textbf{Storing or multiplying an index.} 
Our algorithms sometimes require a head to walk from its original position to the left end of the tape in order to control the number of iterations of a  loop, or to perform calculations based on information encoded in its index. This original index position may also be needed for future operations.  
    For instance, since the content of the \ith{i} track is represented by the index of $h_i$ in $R$,  one has to make sure that this information is stored elsewhere when $h_i$ needs to take such a walk.   This is achieved by making some auxiliary head $h_a$ (that is originally at position $0$)  mirror $h_i$, \ie, $h_a$ moves right whenever $h_i$ moves to the left during this procedure. For ease of reference, one can then simply swap the identities of $h_a$ and $h_i$ to restore $h_i$'s index to its original value before the walk. %\utkanadd{UTKAN: Çarpma olayı ayrı madde de olabilir.} 
    This mirroring method can be modified easily to double or halve %\utkanadd{UTKAN: Neden double tırnak içinde değil ama halve tırnak içinde? Bölümün sonucu buçuklu çıkabilir diye mi? (Eğer öyleyse, bizde olmasa gerek ama ikiyle çarparken de inputun uzunluğu aşılabilir.)} 
    the index of a given head: Starting with $h_i$ at index $x$ and $h_a$ at index $0$, move $h_a$ twice as fast (resp.\ slow) as $h_i$, so that $h_a$ arrives at index $2x$ (resp.\ $\floor*{\sfrac{x}{2}}$) when $h_i$ arrives at the left end. In the algorithms to be presented below, we will use a separate font to denote these \DOUBLE{} and \HALVE{} procedures  as a reminder that they involve loops, and therefore take more than unit time to execute, although they are specified in a single line of the pseudocode.  %\utkanadd{UTKAN: Burada da identity swap yapılmalı.} %Note that all these variants of the basic procedure take linear time, since they terminate when  a head finishes  a single one\-/directional sweep of the input tape. \utkanadd{UTKAN: Bu sanki alttaki madde için de geçerli, sadece burada yazması bi garip.}
 %   &\item \textbf{Increasing or decreasing an index by another index.} In order to add (resp.\ subtract) the present index  of  $h_i$ to (resp.\ from) the index of  $h_j$, walk $h_i$ all the way to the left  end\-/marker, while matching every move of $h_i$ with a move of  $h_j$ to the right (resp.\ left). 
%\end{itemize}

%\paragraph{Exponentiation.} Starting with head $h_i$ at some index $x$, we can use two additional heads, say, $h_a$ and $h_b$ to locate index $2^x$ as follows: After moving $h_a$ to index 1, perform a loop which doubles the index of $h_a$ (using  $h_b$ as auxiliary) for every leftward move of $h_i$, until $h_i$ reaches the left end-marker.
    %!!!BURAYA MİNİK BAŞLIKLAR ALTINDA "COPYING AN INDEX", "INCREMENTING/DECREMENTING AN INDEX BY ANOTHER INDEX" "HALVING/DOUBLING AN INDEX", "LOCATING THE $2^b$'TH INDEX, GIVEN A HEAD AT THE B'TH INDEX" gibi NUMARALARI ANLATIP ZAMANLARINI DA VERSEK SONRASINI BASİTLEŞTİREBİLİRİZ????

    $R$ will simulate each transition of  $M$ using a sequence of transitions of its own, and report the same decision as $M$. 
      Each transition of $M$
    % \utkanrem{(and any Turing machine in general)} CUNKU COIN ATILABILIRDI, COMMUNICATION CELL ILE IS YAPILABILIRDI, VS.
    consists of two parts:
    \begin{enumerate}
        \item ``Sensing'' by the current configuration; which involves taking account of the internal state of the machine and reading the symbols under both the input and work tape heads.
        \item ``Acting'' on the current configuration; which can involve changing the internal state, the symbol under the work tape head, and the positions of both the input and work tape heads.
    \end{enumerate}

% !!!AŞAĞIDAKİ KODLARI BU ŞEKİLDE ÇOK DAHA HIGH-LEVEL HALE GETİRELİM BENCE. ŞİMDİKİ HALLERİNDE 1.1.2'nin hesaba 0 adım eklemesi gibi normal psödokod analizine alışmış insanları mahveden ayrıntılar var!!!!!!!!!!!!!!

 %   \utkanadd{UTKAN: Fakat o halinin complexity overhead'i ve kullandığı kafa sayısı farklı olur. Biz burada bu ikisini de olabildiğince küçük tutan (aralarında tradeoff var, mesela sanırım kafa sayısını $2c+5$ yaparak overhead küçültülebiliyordu (sanırım yine \OH{n}'di), biz $c+5$ kalmasını daha makul bulmuştuk) bir algoritma çıkarıp onu analiz ediyorduk. Kesin olmamakla beraber bu bahsettiğiniz şekilde yapınca da overhead muhtemelen \OH{n} mertebesinde kalır ama tight analizi kaybederiz.
    
  %  Reviewer'lar bence bu konuda bir eleştiri yapmamışlar. (R2 sadece "mevcut numaralandırma şekli aldatıcı olmuş" demiş. Analiz konusunda hiçbir eleştirisi olmamış; okumuş, anlamış ve onaylamış gibi duruyor.) "Eleştirilmeyen şeylere dokunmayalım" shortcut'ını almak isterseniz diye hatırlatmak istedim.}
    
    The rest of the proof describes the simulation of these parts. %\utkanrem{the ``sense'' and ``act'' parts of each simulated transition.}  \utkanadd{UTKAN: Ama her birinin devamında süre analizi de var. Üstüne bir de toplam süre analizi ve nihai ``multiplicative runtime overhead'' claim'imiz var.}

    \paragraph{Sensing.} $R$ traces $M$'s current state in its  own finite memory and reads its input tape with $\head_\alpha$ to obtain the symbol $M$ is scanning with its input head.

%    \utkanadd{QUESTION: CAN M ATTEMPT TO PUSH ITS WORK TAPE HEAD BEYOND LOG N (and fail)?}

    Obtaining the symbol under $M$'s work tape head requires sophistication.
    % At the beginning of this stage, $\head_\beta$'s location on the input tape of $R$ equals the location of  $M$'s work tape head in the corresponding point of the simulation.
    Let $b$ denote the index of $\head_\beta$, which can range from $0$ (when $\head_\beta$ is on $\lend$) up to $\floor*{\log(n+2)}-1$ (since $M$ uses at most $\floor*{\log(n+2)}$ work tape cells).
    % \utkanrem{Let $b$ denote the index of $\head_\beta$ (which is $0$ when $\head_\beta$ is on $\lend$), which is at the same index as $M$'s work tape head in $M$'s work tape.}
    To decode the symbol under  $M$'s work tape head from the positions of  heads $\head_1$ through $\head_c$, essentially, the least significant $b$ bits of their indices will be truncated and the last bits of what remains will be retrieved. %(which are the \ith{b} !!BURADA "LEAST SIGNIFICANT" MI DEMEK İSTİYORSUN?!!!! \utkanadd{Evet} BU PARANTEZ KOMPLE SİLİNSE NE KAYBEDERİZ???!!!!!! \utkanadd{Uygundur, sunduğu bilgi çok bariz, evet}     bits of the original indices). 
    %This will be performed sequentially for each head \utkanadd{UTKAN: Bu cümle parçasında aslında "sequentially"de bir vurgu olmalı (önceki cümlenin üzerine bu cümle parçasının vermeyi amaçladığı tek bilgi o) ama sanki yazı o vurguyu veremiyor ("performed sequentially for each head"in tamamı bilgi olarak veriliyormuş gibi duruyor). Acaba "sequentially"i italik yapsak çözülür mü?!!! ONUN YERİNE BU CÜMLEYİ DİPNOTSUZ ŞEKİLDE 
    This is  performed sequentially for each head, although a machine with more auxiliary heads could perform this in parallel to expedite the sensing routine. 
    %DESEK?????!! Bence çok iyi}, in order to back\-/up and restore the original positions of the heads.\footnote{It is possible to use more heads to perform this in parallel and expedite the sensing routine.} 
    In \Cref{alg:sensing}, the loop controlled by \Cref{stg:understandingindexdroploop} walks $\head_i$ to the left so that its original index $j_i$  is reduced to  $\floor*{\sfrac{j_i}{2^b}}$ at the end of this walk. Each iteration of this loop  runs for half as many steps as the previous one, while performing the \HALVE{} procedure described above. Since the runtime of the following  parity check stage (\ref{stg:understandingparity}) is just $\floor*{\sfrac{j_i}{2^b}}$, we conclude that  
    ``reading'' the symbol under the simulated work tape head with this procedure incurs at most
    \begin{equation*}
        \sum_{i=1}^c \paren[\Big]{\underbrace{\textstyle j_i + \floor*{\frac{j_i}{2}} + \floor*{\frac{j_i}{4}} + \dotsb + \floor*{\frac{j_i}{2^{b-1}}}}_{\text{\Cref{stg:understandingindexdroploop}}} +
        \underbrace{\textstyle \floor*{\frac{j_i}{2^b}}}_{\mathclap{\text{\Cref{stg:understandingparity}}}}}
        \le
        \sum_{i=1}^c 2j_i
        \le 2cn + 2c
    \end{equation*}
    % \utkanremlong[]{
    % \begin{equation*}
    %     \sum_{i=1}^c \paren{\underbrace{j_i + \floor*{\sfrac{j_i}{2}} + \floor*{\sfrac{j_i}{4}} + \dotsb + \floor*{\sfrac{j_i}{2^{b-1}}}}_{\text{\Cref{stg:understandingindexdroploop}}} +
    %     \underbrace{\floor*{\sfrac{j_i}{2^b}}}_{\mathclap{\text{\Cref{stg:understandingparity}}}}}
    %     \le
    %     c \cdot \paren{n + \sfrac{n}{2} + \sfrac{n}{4} + \dotsb + 1} \le c + 2cn
    % \end{equation*}
    % }
    % \begin{equation*}
    %     c \cdot \paren{{\underbrace{n + \sfrac{n}{2} + \sfrac{n}{4} + \dotsb + 1}_{\text{\Cref{stg:understandingindexdroploop,stg:understandingparity}}}}} \le c + 2cn
    % \end{equation*}
    steps. %Sensing the  current simulated state and the input symbol can be handled parallelly, and would incur no additional time cost.
    
    The reader should note that the execution of the procedure of \Cref{alg:sensing} leaves $\head_1, \dotsc, \head_c$, and $\head_{\beta}$  (after the reidentifications) at their original locations. We will justify  the statement at \Cref{stg:understandinginitialization} of \Cref{alg:sensing} by the end of this proof.
    
   % The procedure above completes the ``sense'' part of the simulation of an $M$\=/transition.    

% \newcommand\annotatesteps[2]{%
% \begin{minipage}{\linewidth-5em}%
% #1%
% \end{minipage}%
% \hfill\vline\hfill%
% \begin{minipage}{4em}%
% #2%
% \end{minipage}%
% }

% \hfill\vline\hfill%
% \begin{minipage}{4em}%
% #2%
% \end{minipage}%
% }

% \newcommand\annotatesteps[2]{%
% \schema[closed]{\begin{minipage}{\linewidth-5em}%
% #1%
% \end{minipage}%
% }{%
% \begin{minipage}{4em}%
% #2%
% \end{minipage}%
% }%
% }

\begin{algorithm}[htb]
    \begin{minipage}{\linewidth-\annotationwidth}
    \begin{turingmultienum}[start=0]
        \item\label{stg:understandinginitialization} 
        (At this point, $\head_{\gamma}$, $\head_{\delta}$, and $\head_{\epsilon}$ are guaranteed to be on \lend.)
        \item\label{stg:understandingreadmemory} Do the following for $i = 1, \dotsc, c$:
        \begin{turingmultienum}
\annotatesteps{
            \item\label{stg:understandingindexdroploop} 
             Walk $\head_{\beta}$ to \lend, using $\head_{\gamma}$ to store $\head_{\beta}$'s original index, and repeating the following parallelly with every leftward move of $\head_{\beta}$: %\utkanadd{UTKAN: Bu da bence pek conventional bir loop başlığı olmamış.  Bence conventional olsun istiyorsak ya burada sadece bir for-loop statement (Stage 1'deki gibi) ya da sadece bir while-loop statement (eski Stage 1.1'in ikinci cümlesindeki gibi) yazmalı.  Aksiyonların tamamı loop içerisinde yazmalı.  Ayrıca... ``Stage 1.1.2.'nin tutması gereken +1 adım nerede?'' sorunu burada da var, sadece ustaca (adım ve loop statement'larını iç içe giriştirerek) bu satıra gizlenmiş ama bence bu hakiki bir çözüm değil; resmen gizlemişiz, görememelerini umarak çözüyoruz, gördükleri anda yine aynı soru ile baş başa kalıyoruz. Esas çözüm, gerçekten aklımızda olan ve analizi yaparken düşündüğümüz şey: Bu ``move $\head_{\beta}$ one step to the left'' hareketinin paralel olarak işletiliyor olduğu gerçeği. Eğer ortada böyle bir sorunumuz varsa çözümü açık açık bunu yazmak olmalı.}
            \begin{turingmultienum}
                \item\label{stg:understandingindexhalving}
                \HALVE{} the index of $\head_i$ (from  $x$ to $\floor*{\sfrac{x}{2}}$), using $\head_{\delta}$ as auxiliary. In the first execution of this stage (\ie, the first iteration of loop~\ref{stg:understandingindexdroploop}), use  $\head_{\epsilon}$ as auxiliary to parallelly store the original index of $\head_i$.
               % Move $\head_i$ to the left such that its index is halved, \ie, is reduced to $\floor*{\sfrac{j}{2}}$ from its original value $j$.\footnote{In detail, move $\head_i$ towards \lend{} while moving $\head_{\delta}$ to the right in every second movement of $\head_i$.  Then, reidentify $\head_{\delta}$ as $\head_i$, and vice versa.} \utkanadd{\timespec{\OH{n}} (Burada da aşağıda acting algoritmasında bahsettiğim durum var.)}
            \end{turingmultienum}
}{\OH{n} steps}
\annotatesteps{
            \item\label{stg:understandingparity}
            Determine the parity of $\head_i$'s index by walking $\head_i$ to \lend%\utkanrem{: The parity is odd if $\head_i$ reaches \lend{} in an odd number of steps, and even otherwise}
            . 
            If the parity is 
            even, ``read'' the \ith{i} bit of the symbol under $M$'s simulated work tape head as $0$.  Otherwise, read it as $1$. %\utkanadd{\timespec{\OH{n}} (Bu analizde ``ekstra $n$ adım'' olarak gözükmüyor, çünkü in the worst case, Stage 2'deki loop $h_i$'nin indeksini ikiye böle böle en sola kadar getiriyor.)}
}{\OH{n} steps}
            \item\label{stg:reid}
            Reidentify the auxiliary heads $\head_\epsilon$ and $\head_\gamma$ to set $\head_i$ and $\head_\beta$ back to their locations at the beginning of the procedure.%\footnote{To do so, $\head_\gamma$ is made to mirror !!!!MIRROR LAFI TANIMLI MI? TERS YÖNDE GİTMEK Mİ DEMEK? AÇIKÇA YAZILMALI!!!! the movements of $\head_\beta$ throughout, and $\head_\epsilon$ is made to mirror the movements of $\head_i$ as it is moving towards $\lend$ during the first execution  of  \Cref{stg:understandingindexhalving}.  Finally, the mirror\-/pairs are reciprocally reidentified in \Cref{stg:reid}.}
        \end{turingmultienum}
    \end{turingmultienum}
    \end{minipage}
    \caption{Using $R$'s heads to read a symbol on the work tape of $M$.}
    \label{alg:sensing}
\end{algorithm}

    \paragraph{Acting.} Updating the internal state and head positions of the simulated machine is again straightforward:  $R$ updates the simulated machine's state in its own finite  memory.  It also moves $\head_{\alpha}$ and $\head_{\beta}$, which  mimic $M$'s input and work tape heads, in the required directions, after simulating the changes on $M$'s work tape contents at that step.

    Changes on the work tape are accomplished as follows.  Once again, let $b$  denote the index of $\head_\beta$, ranging from $0$ to $\floor*{\log(n+2)}-1$.  If the %!!!BURALARDA DA LEAST SIGNIFICANT MI DEMEK LAZIM???!!!!! \utkanadd{Bence hayır, buradaki "bit" çok basamaklı bir sayı içindeki basamak olan "bit" değil.  Buna "bit" yerine "symbol" bile diyebiliriz. Precisely work tape'teki $i$th track'in $b$'inci sembolünü (= $b$'inci sembolün $i$'inci component'ını) kastediyoruz. O sembol aynı zamanda tanımı gereği happens to be the $b$th least significant bit of the number represented in the $i$th track, ama burada tanıtıcı unsuru o değil, tanıtıcı unsur teybin soldan $b$'inci sembolü olması.} 
    bit on the \ith{i} track of the simulated work tape symbol at that index  needs to be changed from $1$ to $0$, $h_i$ is moved $2^b$ cells to the left. If that bit is to change from $0$ to $1$, $h_i$ is moved $2^b$ cells to the right. In \Cref{alg:acting}, \Cref{stg:manipulatingoffsetpreparationinit,stg:manipulatingoffsetpreparationloop} have the combined effect of moving $\head_{\epsilon}$ to the \ith{2^b} index on the input tape. The exponentiation is achieved by repeated doubling. The loop controlled by \Cref{stg:manipulatingsimulatedmemory} then updates all relevant head indices by  $2^b$ units in the required directions parallelly. 
    
    The loop controlled by \Cref{stg:manipulatingoffsetpreparationloop} is iterated up to
    % \utkanrem{$\log n$}
    $\floor{\log(n+2)}-1$ times, with each iteration taking twice as much time as its predecessor. The runtime of \Cref{stg:manipulatingsimulatedmemory} equals $2^b$, which is not greater than
    % \utkanrem{the input length $n$}
    $\sfrac{n}{2}+1$. The %, where $b$ is the index of $\head_{\beta}$ on the input tape.     
    procedure therefore incurs up to
    \begin{equation*}
        \underbrace{1}_{\mathclap{\text{\cref{stg:manipulatingoffsetpreparationinit}}}} + \underbrace{2 + 4 + 8 + \dotsb + 2^b}_{\text{\Cref{stg:manipulatingoffsetpreparationloop}}} + \underbrace{2^b}_{\mathclap{\text{\cref{stg:manipulatingsimulatedmemory}}}}
        \le
        3 \cdot 2^b - 1 \le \sfrac{3n}{2} + 2
    \end{equation*}
    % \utkanremlong[]{
    % \begin{equation*}
    %     \underbrace{1}_{\mathclap{\text{\cref{stg:manipulatingoffsetpreparationinit}}}} + \underbrace{2 + 4 + 8 + \dotsb + 2^b}_{\text{\Cref{stg:manipulatingoffsetpreparationloop}}} + \underbrace{2^b}_{\mathclap{\text{\cref{stg:manipulatingsimulatedmemory}}}}
    %     \le
    %     1 + 2 + 4 + \dotsb + n + n \le 1 + 3n
    % \end{equation*}
    % }
    % \begin{equation*}
    %     {\underbrace{1 + 2 + 4 + 8 + \dotsb + n}_{\text{\Cref{stg:manipulatingoffsetpreparationinit,stg:manipulatingoffsetpreparationloop}}}} + {\underbrace{n}_{\text{\cref{stg:manipulatingsimulatedmemory}}}} \le 1 + 3n
    % \end{equation*}
        steps.% for $b>0$.
        \footnote{Note that, when $b=0$, \ie, when $h_{\beta}$ is scanning $\lend$, $R$ can proceed to \Cref{stg:manipulatingsimulatedmemory} directly without performing any transition associated with \Cref{stg:manipulatingoffsetpreparationloop}. (``\Cref{stg:revertworkhead}'' always incurs $0$ steps.)} The task of updating the simulated machine state and the simulated input and work heads' positions would incur a single step by themselves, and are handled in parallel with the  procedure of \Cref{alg:acting}, requiring no additional time.  

\begin{algorithm}[htb]
    \begin{minipage}{\linewidth-\annotationwidth}
    \begin{turingmultienum}[start=0]
        \item\label{stg:manipulatinginitialization} 
        (At this point, $\head_{\gamma}$, $\head_{\delta}$, and $\head_{\epsilon}$ are  on \lend.)
\annotatesteps{
        \item\label{stg:manipulatingoffsetpreparationinit} 
        Move $\head_{\epsilon}$ one step to the right.
        \item\label{stg:manipulatingoffsetpreparationloop} 
             Walk $\head_{\beta}$ to \lend, using $\head_{\gamma}$ to store $\head_{\beta}$'s original index, and repeating the following parallelly with every leftward move of $\head_{\beta}$: %\utkanadd{UTKAN: [sensing loop'undaki yorumumun tekrarı]}
            \begin{turingmultienum}
                \item
                \DOUBLE{} the index of $\head_\epsilon$, using $\head_{\delta}$ as auxiliary. 
               % Move $\head_i$ to the left such that its index is halved, \ie, is reduced to $\floor*{\sfrac{j}{2}}$ from its original value $j$.\footnote{In detail, move $\head_i$ towards \lend{} while moving $\head_{\delta}$ to the right in every second movement of $\head_i$.  Then, reidentify $\head_{\delta}$ as $\head_i$, and vice versa.} \utkanadd{\timespec{\OH{n}} (Burada da aşağıda acting algoritmasında bahsettiğim durum var.)}
            \end{turingmultienum}
}{\OH{n} steps}
        \item\label{stg:revertworkhead}
         Reidentify the auxiliary head $\head_\gamma$ to set  $\head_\beta$ back to its location at the beginning of the procedure.
\annotatesteps{
         \item\label{stg:manipulatingsimulatedmemory}
         Walk $\head_{\epsilon}$ to \lend,  repeating the following parallelly with every leftward move of $\head_{\epsilon}$: %\utkanadd{UTKAN: Burada da yine aynı ``gizlenmiş adım'' ve ``adım ve loop statement'ının birleştirilmesi'' sorunu var.}
        % \utkanrem{$i$ between $1$ and $c$ where the transition being simulated changes the \ith{i} bit of the symbol under the simulated work tape head, do the following:}
        \begin{turingmultienum}
            \item\label{stg:manipulatingsimulatedmemoryiter} 
                If the  bit on the \ith{i} track should change from $0$ to $1$, move $\head_i$ one step to the right. If the  bit on the \ith{i} track should change from $1$ to $0$, move $\head_i$ one step to the left. All such heads move simultaneously in the same transition. %\utkanadd{UTKAN: Bence 4 ve 4.1 adımlarını birleştirelim. Burada sadece birer adım atılıyor. Üstelik yukarıda anlattığımız ``increase/decrease by'' rutini tam burada kullanmalık, hatta sanırım başka yerde de kullanmıyoruz, muhtemelen tam olarak burada kullanmak için yazmışsınız.}
        \end{turingmultienum}
        % \item \utkanrem{Move $\head_\epsilon$ to $\lend$.}\footnote{\utkanrem{This, in fact, is done by simply not moving $\head_\delta$ and keeping it on $\lend$ as $\head_\epsilon$ is moving to $\lend$ during the last time} \Cref{stg:manipulatingsimulatedmemoryiter} \utkanrem{is being performed.}}
}{\OH{n} steps}
    \end{turingmultienum}
    \end{minipage}
    \caption{Rearranging $R$'s heads to simulate a change on the worktape of $M$.}
    \label{alg:acting}
\end{algorithm}

    The two routines for reading and changing the simulated memory contents are carried out back to back, and they both leave the heads $\head_{\gamma}$, $\head_{\delta}$, and $\head_{\epsilon}$ on the left end\-/marker as they finish. Moreover, these heads are on the left end\-/marker also when $R$ begins its execution, by definition.  The assumptions in \Cref{stg:manipulatinginitialization} of both routines are thus justified. %Therefore, the assumptions at the beginning of each procedure holds, without those procedures having to spend any time on moving those heads.

    The two parts together take at most
    % \utkanrem{$1 + c + 3n + 2cn$}
    $\paren*{2c + \sfrac{3}{2}}n + 2c + 2$ steps, which is an upper bound to the multiplicative runtime overhead of $R$ simulating $M$. Thus, $R$ completes its simulation within
    % \utkanrem{$t(n) \cdot \paren*{1 + c + 3n + 2cn}$}
    $t(n) \cdot \paren*{\paren*{2c + \sfrac{3}{2}}n + 2c + 2}$ steps.
\end{proof}

\section{Proof of \texorpdfstring{\Cref{lem:polyclock}}{Lemma \ref{lem:polyclock}}}\label{sec:polyclockproof}

\polyclock*

\begin{proof}
    By the implications of $f(n) \in \OH{n^t}$, let $c > 0$ and $n_0 > \errprematuretimeout^{-1}$ be  integers large enough to satisfy $f(n) \le c \cdot n^t$ for all $n \ge n_0$.  The extra constraint upon $n_0$ to be larger than $\errprematuretimeout^{-1}$ will be evident by the end of the proof.
    We will analyze the runtime requirements of each stage of  the probabilistic finite automaton $M_{t,c,n_0}$ presented in \Cref{alg:polyclock}.
%    let $M_{t,c,n_0}$ be a probabilistic finite automaton embodying the following algorithm:

\begin{algorithm}[htb]
    \begin{turingmultienum}%{M_{t,c,n_0}}{On input $w$:}
        \item\label{stg:polyclockshortfilter} If $n < n_0$, %\footnote{\utkanrem{Note that this check brings the input head to the right end\-/marker.}}
        pause for % \utkanrem{$\max\paren*{f(n), \ceil*{c \cdot n^{t+1}}}$}
        $f(n)$ time\-/steps and \emph{halt}. % \utkanadd{\timespec{\OH{n}}}
        \item\label{stg:polyclocksublinearfilter} If $t = 1$, % $t \le 1$,
        move the input head back to \lend{}. Then, move the input head right until it reaches \rend{},  while pausing it for $c - 1$ time\-/steps after each movement, then \emph{halt}. % \utkanadd{\timespec{\OH{n}}}
        \item\label{stg:polyclockbegin} Repeat the following $t$ % $\ceil{t}$
        times:
        \begin{turingmultienum}
            \item\label{stg:polyclockheadreset}
            Move the input head to the beginning of  the input string. % \utkanadd{\timespec{\OH{n}}} %\utkanadd{Bu adim alttakinin altindaydi, buraya aldim.}
            \item\label{stg:polyclockrandwalk}
            Perform a random walk with the input head (\ie, repeatedly move the input head towards left or right by the flips of a coin) until the input head reaches \lend{} or \rend{}. Pause for $c - 1$ time\-/steps after each step moving the input head. % \utkanadd{\timespec{\OH{n}}}
        \end{turingmultienum}
        \item If any of the last $t$ % $\ceil{t}$
        random walks have ended with the input head reaching \lend{}, go back to \Cref{stg:polyclockbegin}.
        \item \emph{Halt}.
    \end{turingmultienum}
    \caption{The probabilistic finite automaton $M_{t,c,n_0}$.}
    \label{alg:polyclock}
\end{algorithm}

    \Cref{stg:polyclockshortfilter} starts with $\min\paren{n + 2, n_0 + 1}$ steps to check whether $n < n_0$. If so, $M_{t,c,n_0}$ runs for a total of $n + 2 + f(n)$ time\-/steps and then halts, which yields a runtime within desired bounds with certainty.
    % The extra $n$ steps here will not be an issue with the longer inputs whereby $M_{t,c,n_0}$ will halt on a subsequent stage, since $n \in \OH{n^{t+1}}$ for all $t > 0$.

    When $n \ge n_0$ and $t=1$, % $t \le 1$,
    $M_{t,c,n_0}$ halts after running for $2n_0 + cn + 2$ steps in \Cref{stg:polyclockshortfilter,stg:polyclocksublinearfilter} combined, which is again within the expected and minimum time limits, and with certainty.
    
    The rest of the proof will examine the claims for $n \ge n_0$ and $t > 1$.

    \paragraph{Expected runtime.} Each random walk at \Cref{stg:polyclockrandwalk} is commonly known as the \emph{gambler's ruin}, and the probability of each ending at the right\-/hand side of the input tape is $\paren*{n+1}^{-1}$ by their Markov chain analysis~\cite[Section~7.1]{KS60}. Consequently, the probability of having a \emph{terminating batch} (\ie, a batch of $t$ % $\ceil{t}$
    random walks where each of them ends up at the right\-/hand side of the input tape) is $\paren*{n+1}^{-t %\ceil{t}
    }$.

    The expected number of independent trials to observe a binomial event is the reciprocal of its occurrence probability~\cite[Section~C.4]{CLRS09}. Therefore, the expected number of batches until termination is $\paren*{n+1}^{t %\ceil{t}
    }$, and the same for random walks is $t %\ceil{t}
    \cdot \paren*{n+1}^{t %\ceil{t}
    }$.

    The expected runtime of each random walk is $cn$ steps~\cite[Section~7.1]{KS60}. The reset in \Cref{stg:polyclockheadreset} takes exactly $n_0 + 2$ steps before the first random walk, and it is expected to take
    % \utkanrem{another}
    \begin{equation*}
        {\underbrace{\paren{n+1}^{-1} \cdot (n + 2)}_{\text{resetting from \rend}}} + {\underbrace{\paren{1-\paren{n+1}^{-1}} \cdot 1}_{\text{resetting from \lend}}} = 2
    \end{equation*}
    steps
    % \utkanrem{with each random walk}
    before every subsequent one. Taking \Cref{stg:polyclockshortfilter} into account as well, the expected runtime of $M_{t,c,n_0}$ is
    \begin{equation*}
        t %\ceil{t}
        \cdot \paren*{n+1}^{t %\ceil{t}
        } \cdot \paren*{cn + 2} + 2n_0 + 1 \in \OH{n^{t %\ceil{t}
        +1}}.
    \end{equation*}
% !!!!BUNU TAM ANLAMIYORUM. İLK RANDOM WALK'TAN ÖNCE n0+1 sağa, n0 + 2 de sola adım atıyor değil mi? Bu ifade exact mi? Kaç random walk varsa o kadar da 2 adımlık dönüş süresi sayarsan 2 adım fazla saymış olmuyor musun? Bu mikroskopiklik yerine yukarıda "at most" falan desek iş kolaylaşır mı?!!!!! \utkanadd{}DUR DUR ANLADIM GALİBA, 2 SOLDAN SAĞA GİTMİŞ; HESAP DOĞRU evet o fazlalık iki sağdan çıkmış, ben de dün bakarken "kesin burada bir hata vardır" dedim ve yokmuş, kendime hayret ettim... yine de (for our purposes) gereksiz precise tabi bunların hepsi

    \paragraph{Minimum runtime.} For simplicity, we note that $\paren*{n+1}^{-t %\ceil{t}
    }$, the probability of a terminating batch occurring, is less than $n^{-t}$.%\footnote{The reader should note that we are loosening the bounds and inequalities here and throughout the analysis for the sake of simplicity.}
    %  (given $n \ge n_0 > \errprematuretimeout^{-1} > 0$)
    
    It is evident that the execution of a terminating batch takes more than $%\ceil{t} \cdot nc \ge
    tnc$ time\-/steps, and thus, in a time frame of $f(n) \le c \cdot n^t$ steps, there are at most $\frac{n^{t-1}}{t}$ opportunities for terminating batches to occur, which is less than $n^{t-1}$ (given $t > 1$).
    
    The probability of an event with an occurrence probability less than $n^{-t}$ happening at least once given less than $n^{t-1}$ opportunities, is less than
    \begin{align*}
        g_t(n) &= 1 - \paren*{1 - n^{-t}}^{n^{t-1}}.\\
        \intertext{Consider the following binomial expansion: %series:
        }
        g_t(n) &= 1 - \paren*{\sum_{k=0}^{n^{t-1} %\infty
        } \paren*{-1}^k \cdot \binom{n^{t-1}}{k} \cdot n^{-tk}}\\
        &= \frac{n^{t-1}}{n^t} - \frac{n^{t-1}}{n^t} \cdot \frac{n^{t-1} - 1}{2n^t} + \frac{n^{t-1}}{n^t} \cdot \frac{n^{t-1} - 1}{2n^t} \cdot \frac{n^{t-1} - 2}{3n^t} - \dotsb\\
        \intertext{Since the terms are % series is
        alternating and the magnitudes are decreasing, we have}
        g_t(n) &< \frac{n^{t-1}}{n^t} = n^{-1}.
    \end{align*}

    $g_t(n)$ constitutes an upper bound to the ``error'' of $M_{t,c,n_0}$ halting sooner than the desired lower bound of $f(n)$ time\-/steps. Since $g_t(n)$ is monotone decreasing for positive $n$, $g_t(\bar{n}) < \errprematuretimeout$ follows for all $\bar{n} \ge n_0$ (recalling that $n_0 > \errprematuretimeout^{-1}$).  With \Cref{stg:polyclockshortfilter} ensuring that $M_{t,c,n_0}$ waits long enough for the strings shorter than $n_0$ with probability 1, the probability of $M_{t,c,n_0}$ halting prematurely will be less than $\errprematuretimeout$ for strings of any length. Note that $\errprematuretimeout$ can be set to any desired small positive value by simply employing a suitably large number of states (to make $n_0$ appropriately large) in the implementation of \Cref{stg:polyclockshortfilter}.  
%    \utkanadd{Note that decreasing $\errprematuretimeout$ only affects $n_0$ by potentially increasing it.  To accommodate for that increase, $M_{t,c,n_0}$ only needs an equal amount of more states for the counting in \Cref{stg:polyclockshortfilter}.}
%
%    \paragraph{Finite runtime} \utkanrem{Since the probability of a terminating batch occurring is positive, such a batch will eventually happen, making the probability of $M_{t,c,n_0}$ running forever zero.}
    % \qed %% LLNCS ONLY
\end{proof}

\section{The double exponential timer}\label{sec:dir2appendices}

%\paragraph{Double\-/exponential counter.}

\Cref{alg:tick} restates the subroutine \TICK{} described in \Cref{sec:clock}. Recall that we also maintain a global Boolean variable $T$ that tracks whether  all the red coins flipped since the last reset of this variable came out heads. The integers $c$ and $t$ are parameters that we can use for setting the expected number of ticks. \Cref{stg:polycoins} uses logarithmic space in the work tape %\footnote{\utkanadd{Same feat can be achieved using $t$ heads if one seeks to use the \TICK{} subroutine in a multihead finite automaton.}} !!!!BÖYLE BİR ŞEY YAPAN BİR ÇOKKAFALI MAKİNEMİZ YOKSA BUNU DEMEYELİM!!!!
to count the coin tosses.

\begin{algorithm}[htb]
\begin{turingmultienum}
    \item\label{stg:accumulate} Toss a red coin and update the Boolean variable $T$ according to the outcome.
    \item\label{stg:polycoins} Toss $p(n) = c \cdot n^t$ blue coins.% by doing the following: Move all $t$ heads to the beginning of the input. Sweep $t$ heads across the input in a hierarchy akin to a $t$\=/handed clock, until they all are at the end of input. For some constant $c$, toss $c$ coins at each tick of the fastest head.
    \item\label{stg:pass} If any blue coin outcome at \Cref{stg:polycoins} was tails, \emph{return} %\utkanadd{UTKAN: bunu Appendix A.2'de \emph{halt}'i yazdigimiz gibi veya Lemma 10'da \emph{accept} ve \emph{reject}'leri yazdigimiz gibi, \emph{return} olarak yazmaya ne dersiniz?}!!!!!!!!!HARİKA FİKİR; ÖYLE YAPALIM LÜTFEN!!!!!!
    with the timer still running. %skip the remaining stages of this round's check (retaining the accumulation in $T$ for the next check).
    \item\label{stg:judgement} (Arriving at this stage implies that all the blue coins of \Cref{stg:polycoins} came up heads.) If $T$ indicates that one or more  red coins flipped since the last reset came out tails, reset $T$ and \emph{return} with the timer still running.%skip the remaining stages of this round's check.
    \item\label{stg:timeup} (Arriving at this stage implies that all red coins flipped  at \Cref{stg:accumulate} since %!!!İYİ DE BU STAGE DAHA BİR SANİYE ÖNCE REACH EDİLDİ?!?!?!?!the last time \Cref{stg:judgement} was reached (\ie, 
    the last reset of $T$ were %all
    heads.) \emph{Return} and announce that the timer has run out.
\end{turingmultienum}
\caption{The \TICK{} subroutine.}
\label{alg:tick}
\end{algorithm}

To analyze the expected amount of subroutine calls (ticks) before the timer runs out, we define the following random variables to represent the described counts:
\begin{itemize}
    \item[$X$:] Times the subroutine is called.
    \item[$Y$:] Times the tick  advances control to \Cref{stg:judgement}, \ie, all $p(n)$ blue coins come up heads.
    \item[$Z$:] Number of red coins flipped since the last reset  as the tick advances to \Cref{stg:judgement} (at any given time).
    \item[$Z_i$:] Number of red coins flipped since the last reset  as the tick advances to \Cref{stg:judgement} for the \ith{i} time.
\end{itemize}
% \begin{center}
%     \begin{tabular}{r|p{32em}}
%         \makecell[r]{\textbf{Random}\\\textbf{Variable}} & \textbf{Description} \\\midrule
%         $X$ & Times the subroutine is called.\\
%         $Y$ & Times the tick  advances control to \Cref{stg:judgement}, \ie, all $p(n)$ blue coins come up heads.\\
%         $Z$ & Number of red coins flipped since the last reset  as the tick advances to \Cref{stg:judgement} (at any given time).\\
%         $Z_i$ & Number of red coins flipped since the last reset  as the tick advances to \Cref{stg:judgement} for the \ith{i} time.
%     \end{tabular}
% \end{center}

Let \exv{O} denote the expected value of a random variable $O$.  The expected value that we are after is \exv{X}.

The probability that any tick advances to \Cref{stg:judgement} (\ie, $p(n)$ blue coins come up heads) is $2^{-p(n)}$. % \utkanrem{By~\cite[Section~C.4]{CLRS09}}
Hence, we expect there to be $2^{p(n)}$ attempts, each one flipping another red coin, since the last reset of variable $T$ %and} 
before  control  advances to \Cref{stg:judgement} in some subroutine call.  Thus, we have
\begin{equation*}
    \exv{Z} = 2^{p(n)}.
\end{equation*}

We can say the following for $X$, trivially by the fact that each tick flips exactly one red coin at its \Cref{stg:accumulate}:
\begin{equation*}
    X = Z_1 + Z_2 + \dotsb + Z_Y.
\end{equation*}
Since the expected value operator is a linear operator, we have
\begin{equation*}
    \exv{X} = \exv{Z_1} + \exv{Z_2} + \dotsb + \exv{Z_Y}.
\end{equation*}
Since each \exv{Z_i} trivially equals \exv{Z}, for a particular $Y = y$, we have
\begin{equation*}
    \exv{X \given Y=y} = 2^{p(n)} \cdot y,
\end{equation*}
and by the Law of Total Expectation, we have
\begin{equation*}
    \exv{X} = \exv{\exv{X \given Y}} = \exv{2^{p(n)} \cdot Y} = 2^{p(n)} \cdot \exv{Y}.
\end{equation*}

To determine \exv{Y}, 
consider a subroutine call in which control has just advanced to \Cref{stg:judgement}.  Let $z$ denote the number of red coins flipped since the last  reset if the program had reached  \Cref{stg:judgement} before, and since the beginning of the program's run otherwise.  Such a subroutine call will advance further to \Cref{stg:timeup}, indicating time\-/out, if all $z$ coin outcomes were heads.  The probability of this is $2^{-z}$.  If we had $Z = z$ at any given time \Cref{stg:judgement} was reached, then %by~\cite[Section~C.4]{CLRS09} 
we would expect  \Cref{stg:timeup} to be reached after \Cref{stg:judgement} having been reached  the following number of times:
\begin{equation*}
    \exv{Y \given Z = z} = 2^z.
\end{equation*}
Then, by the Law of Total Expectation, we have
\begin{equation*}
    \exv{Y} = \exv{\exv{Y \given Z}} = \exv{2^Z}.
\end{equation*}
By Jensen's Inequality, we have
\begin{equation*}
    \exv{2^Z} \ge 2^{\exv{Z}}.
\end{equation*}
If we multiply both sides with $2^{p(n)}$ and then make the substitutions by the equalities we established, we obtain our lower bound:
\begin{align*}
    2^{p(n)} \cdot \exv{2^Z} &\ge 2^{p(n)} \cdot 2^{\exv{Z}},\\
    2^{p(n)} \cdot \exv{Y} &\ge 2^{p(n)} \cdot 2^{2^{p(n)}},\\
    \exv{X} &\ge 2^{p(n)} \cdot 2^{2^{p(n)}}.
\end{align*}
This proves that the expected number of subroutine calls is in $2^{2^{\OHMEGA{n^{t}}}}$. %\utkanadd{OMEGANIN ICINE UZERI-t EKLEDIM. ``is at least $2^{2^{p(n)}}$'' DE DIYEBILIRIZ?}

\section{\texorpdfstring%
{Construction of \tafa{} $M$ in the proof of \Cref{thm:conspubliconly}}%
{Construction of \tafa{} M in the proof of Theorem \ref{thm:conspubliconly}}}
\label{sec:tafaconstruction}

Given a constant\-/space verifier $V = \paren*{Q_V, \Sigma, \Phi, \Gamma, \delta_V, q_{V,0}}$ which flips at most $r$ public coins, we will construct a \tafa{} $M$ as described in the proof of \Cref{thm:conspubliconly}.  Let $\Qpub$, $\Qcom$, and $\Set{\qacc[V], \qrej[V]}$ be the subsets of $Q_V$ that correspond to public coin flips, communications, and halting, respectively, as described in \Cref{subs:vers}.  Since $V$ can simulate a constant\-/space work tape in its finite state, we will assume that $V$ repeatedly writes $\blanksymb$ with its work tape head without ever moving it, without loss of generality.

For any given set of coin\-/flip sequences $S$, where each such sequence is a string in $\Set{0,1}^*$ %!!!ŞİMDİ DOĞRU MU???!!! \utkanadd{UTKAN: Bence dogru.} %\utkanadd{UTKAN: "each such sequence is a string \textbf{over} X" dediğimizde X'in alfabe olması gerekir bence. $\Set{0,1}^+$ ise bir string kümesi, dil.  "Over" yerine "in" yazmamız halinde de:  Kümenin $\Set{0,1}^+$ değil de $\Set{0,1}^*$ olmasının, yani bir sequence'ın empty string olabilmesinin sakıncası var mı? Ben düşündüğümde yok gibi gelmişti. (Verifier'ın para atmaksızın kabul ettiği inputlara denk geliyor.)  Hatta sanki "bazı (kısa) inputlar için deterministik karar veren" türde ön filtreleri olan verifier'lar için sequence'ın empty string olarak verilebilmesi ($S = \Set{ \estring }$ tahmininin yapılabilmesi) gerekiyor.} 
representing the outcomes of a fair coin, % made up of symbols from some set of symbols $B$, 
let $\prefix{S}$ (resp.\ $\pprefix{S}$) be the set of all prefixes (resp.\ proper prefixes) of sequences in $S$.  %!!MALUM KISMI ATTIM!!! \utkanadd{UTKAN: Tamamdir}  %, \ie,
%\begin{align*}
 %   \prefix{S} &= \Set{s | s \cdot B^* \in S},\\
  %  \pprefix{S} &= \Set{s | s \cdot B^+ \in S}.\\
   % \shortintertext{VEYA}
    %\pprefix{S} &= \prefix{S} \setminus S.
%\end{align*}
%(UTKAN: BU \ie{} ILE BASLAYAN KISIM GEREKLI DEGILSE ATABILIRIZ. ORAYI ATINCA "where each sequence is made up of symbols from some set of symbols $B$" DEMEYE DE GEREK KALMIYOR.)
% !!!!!!!!YUKARIDA YAPTIĞIM DEĞİŞİKLİK OK İSE BU KISMI ATABİLİRSİN!!!!!!  \utkanadd{UTKAN: Yukarıda (eğer ``over'' yerine ``in'' yazarsak) eskiye kıyasla efektif olarak $B$'yi $\Set{0, 1}$ olarak sabitlemiş ve sequence'ları empty string olamaz hale getirmiş oluyoruz.  Bu değişiklikler bu kısmı nasıl etkiliyor ki?  Bundan bağımsız olarak atılabilirdi bence de ama yapılan değişikliğin bu kısmı nasıl daha fazla atılabilir kıldığını göremedim.}

Let $\binleq{r}$ denote the set of all coin\-/flip %!!!BU CÜMLEYİ DEĞİŞTİRDİM!!! \utkanadd{UTKAN: Bence ok.} 
sequences up to  length $r$ and let $\maj{r}$ be the set of all prefix\-/free subsets of $\binleq{r}$ which have total probability %\utkanadd{UTKAN: Bu total'ın her subset için, subsetin elemanlarının total'ı olduğu açık mı?} 
greater than $\sfrac12$, \ie,
\begin{gather*}
    \binleq{r} = \Set{x | x \in \Set{0, 1}^*, \abs{x} \le r},\\
    \maj{r} = \Set{S | S \subseteq \binleq{r},S \cap \pprefix{S} = \emptyset, \textstyle\sum_{x \in S} 2^{-\abs{x}} > \sfrac12}.
\end{gather*} % !!!!!!!!!SUBSETLİ ŞARTI DA DİREĞİN SAĞ TARAFINA YAZMAK DAHA İYİ OLUR!!!!!!!
%UTKAN: PROPER-PREFIX tanimini \binleq{r} ve \maj{r} taniminin oncesine aldim, cunku \maj{r} tanimini yukaridaki gibi PROPER-PREFIX kullanarak yapinca bence daha dogal/anlasilir oluyor.

\newcommand\qmain{\ensuremath{Q_{\mathrm{main}}}}
\newcommand\qflip{\ensuremath{Q_{\mathrm{flip}}}}

Let $M = \paren*{Q, \Sigma, \delta, q_0}$.  We define $Q$ as the union of four mutually exclusive sets.
\begin{gather*}
    Q = \Set{q_0} \cup \qmain \cup \qflip \cup \Set{\qacc, \qrej},
\shortintertext{where}
    % \qmain = \maj{r} \times Q_V \times \Gamma \times \binleq{r}, \text{ and}\\
    % \qflip = \maj{r} \times \Qpub \times \Gamma \times \binleq{r} \times \Set{0, 1}.\\
    \qmain = \Set{ \paren*{S, q_V, \gamma, p} | S \in \maj{r}, q_V \in Q_V, \gamma \in \Gamma, p \in \prefix{S} }, \\
    % \utkanrem{\qflip = \Set{ \paren*{S, q_V, \gamma, p, \bpub} | S \in \maj{r}, q_V \in \Qpub, \gamma \in \Gamma, p \in \pprefix{S}, \bpub \in \Set{0, 1} }.}\\
    % \\
    % \qflip = \Set{ \paren*{S, q_V, \gamma, p, \bpub} | \begin{array}{l}
    %     S \in \maj{r}, q_V \in \Qpub, \gamma \in \Gamma, \\
    %     p \in \pprefix{S}, \bpub \in \Set{0, 1}
    % \end{array} }.
    % \\
    % \begin{multlined}
    %     \qflip = \\
    %     \Set{ \paren*{S, q_V, \gamma, p, \bpub} | S \in \maj{r}, q_V \in \Qpub, \gamma \in \Gamma, p \in \pprefix{S}, \bpub \in \Set{0, 1} }.
    % \end{multlined}\\
    % \utkanadd{
    \qflip = \set{ \paren*{S, q_V, \gamma, p, \bpub} | S \in \maj{r}, q_V \in \Qpub, \gamma \in \Gamma, p \cdot \bpub \in \prefix{S}, \bpub \in \set{0, 1} },
    % }
\end{gather*}
% \utkanadd{
and $\qacc$ and $\qrej$ are the accepting and rejecting states, respectively.
% }

The set of universal states, \Qun{}, is the following  subset of $\qmain$: 
\begin{equation*}
    \Qun = \Set{\paren*{S, q_V, \gamma, p} | S \in \maj{r}, q_V \in \Qpub, \gamma \in \Gamma, p \in \pprefix{S}}.
\end{equation*}
All other non\-/halting states of $M$ are existential.

At every step, the first four components of the states from $\qmain$ and $\qflip$ will keep
 the following information:
\begin{enumerate}
    \item A set of  coin sequences that are supposed to lead $V$'s simulation to acceptance, which has been guessed by the initial transition and will remain constant for the rest of the execution.
    \item The simulated state of $V$.
    \item The content of the simulated communication cell.
    \item The history of  coin outcomes used in the simulation up to this point. (This component will initially contain the empty string, \estring.)
\end{enumerate}

The states in \Qun{} and $\qflip$ will  be used in a two\-/step procedure to execute a universal branching (to simulate both possible outcomes of a coin toss of $V$ parallelly when necessary), immediately followed by an existential branching (to simulate the prover response to $V$), with the last component of states in $\qflip$ holding that coin outcome during the procedure, as will be detailed below.

We now present the transition function of $M$ in parts.  For all $q\in Q \setminus \Set{\qacc, \qrej}$ and $\sigma\in\Sigma_{\bowtie}$,  $\delta(q,\sigma)$ has the value $\Set{ \qrej }$, unless specified otherwise below.

\paragraph{The initial transition.}
The first step of $M$, which guesses a set $S$ from \maj{r}, is realized by
\begin{equation*}
    \delta(q_0, \lend) = \Set{\paren*{\paren*{S, q_{V,0}, \blanksymb, \estring}, 0} | S \in \maj{r} }.
\end{equation*}%!!!!$\in \qmain \times \Delta$ FASLINI SİLDİM; OK??!!!!!!!! \utkanadd[80]{UTKAN: Bence bu yanlış olmuş, kümede sadece tek bir eleman varmış gibi durmuş ($\Set{x}$ kümesi $x$ belirsiz bırakılmışsa bile "($x$ ne bilmiyorum ama) tek elemanı $x$ olan bir küme" şeklinde okunur bence).  Dik çizgi ve sağına $S \in \maj{r}$ belirtecini ekledim, bence böylesi ok.}

\paragraph{Transitions simulating the non\-/communicating transitions of $V$.} We have %\utkanadd[80]{UTKAN: Bu zaten tek elemanlı bir kümeymiş, böylesi daha sade:}
\begin{equation*}
    % \delta\paren*{\paren*{S, q_V, \gamma, p}, \sigma} = \Set{\paren*{\paren*{S, q'_V, \gamma, p}, d_i} \in \qmain \times \Delta | \delta_V\paren*{q_V, \sigma, \blanksymb, \gamma} = \paren*{q'_V, \blanksymb, d_i, 0}}
    \delta\paren*{\paren*{S, q_V, \gamma, p}, \sigma} = \Set{\paren*{\paren*{S, q'_V, \gamma, p}, d_i}} \text{ where } \delta_V\paren*{q_V, \sigma, \blanksymb, \gamma} = \paren*{q'_V, \blanksymb, d_i, 0},
\end{equation*}
for all $\paren*{\paren*{S, q_V, \gamma, p}, \sigma} \in \qmain \times \Sigma_{\bowtie}$, where $q_V$ is from $\qneg{\Qcom} \setminus \Set{\qacc[V], \qrej[V]}$.

\paragraph{Transitions simulating the communicating transitions of $V$ that do not flip a coin.} We have
%\utkanadd[80]{UTKAN: Burada $\forall\gamma'' \in \Gamma$ yazmak gereksizmiş, filtrelenmediği sürece zaten soldan gelen her alternatif (soldaki alternatiflerin nereden geldiği bilgisini ($\qmain \times \Delta$) yeni ekledim) kümeye dahil olacak:}!!!!!!!BÜTÜN $\in \qmain \times \Delta$'ları SİLSEN BENCE DAHA KOLAY ANLAŞILIR OLACAK!!!!!!!!!!!!
\begin{equation*}
    \delta\paren*{\paren*{S, q_V, \gamma, p}, \sigma} = \Set{\paren*{\paren*{S, q'_V, \gamma'', p}, d_i} %\utkanadd[80]{\in \qmain \times \Delta}
    | \gamma'' \in \Gamma, \delta_V\paren*{q_V, \sigma, \blanksymb, \gamma} = \paren*{q'_V, \blanksymb, \gamma', d_i, 0} %\utkanrem{\text{ and } \forall\gamma'' \in \Gamma}
    }
\end{equation*}
for all $\paren*{\paren*{S, q_V, \gamma, p}, \sigma} \in \qmain \times \Sigma_{\bowtie}$, where $q_V$ is from $\qneg{\Qpub} \cap \Qcom$.  Here, we note that the content of the simulated communication cell switches into $\gamma''$ (the  response from the simulated prover) %\utkanadd{UTKAN: Burada "the guessed response from the \textbf{imagined} prover" yazarken "guess" eylemini $M$'nin yaptigini ($M$'nin hayal etmekte oldugu prover icin "ne derdi" diyerekten tahminde bulunmasi gibi bir senaryo) soylemeye calisiyordum (bilmiyorum tam oyle anlasiliyor muydu, belki de "[...] response \textbf{of} the prover" yazmaliydim).  Simdi "the guessed response from the \textbf{simulated} prover" seklinde degistirilince "guess" eylemini prover yapiyormus gibi okuyorum/anliyorum (neden bilmiyorum, belki "imagined" derken daha $M$'nin hayalinin icinde gibiydik, simdi "simulated" deyince $M$'nin "kafasindan" ciktik, o yuzden ozne Prover olmaya dogru kaydi).  Fakat prover'a "guess" eylemini yakistiramadim, bence "simulated prover" yazacaksak "guess"i de cikaralim.} 
directly, without ever becoming $\gamma'$ (the communication from $V$).  This transition is existential, meaning that $M$ will reach acceptance if and only if there is a symbol $\gamma'' \in \Gamma$ that, when communicated to $V$, can lead $V$ to acceptance.

\paragraph{Transitions simulating a coin toss and the ensuing communication.} As previously mentioned, $V$'s transitions involving a  coin toss are simulated by $M$ in two consecutive transitions.  The first one is to ensure that all the coin flip sequences in $S$ are covered universally:
\begin{equation*}
    \delta\paren*{\paren*{S, q_V, \gamma, p}, \sigma} = %\begin{cases}
        \Set{\paren*{\paren*{S, q_V, \gamma, p, \bpub}, 0}  | \bpub \in \Set{0, 1}, p \cdot \bpub \in \prefix{S} %\utkanrem{p \cdot \bpub \cdot \Set{0, 1}^* \in S}
        }% & \text{if } p \notin S,\\
       % \Set{\qrej} & \text{if } p \in S,\\
   % \end{cases}
\end{equation*}
%for all $\paren*{\paren*{S, q_V, \gamma, p}, \sigma} \in \qmain \times \Sigma_{\bowtie}$ where $q_V$ is from $\Qpub$, \hl{hence we have $\paren*{S, q_V, \gamma, p} \in \Qun$ if $p \in \pprefix{S}$.} 
% !!!BUNU BÖYLE DEMESEK DE ŞÖYLE DESEK?!!  "
for all $\paren*{\paren*{S, q_V, \gamma, p}, \sigma} \in \Qun \times \Sigma_{\bowtie}$. %!!SONRA DA YUKARIDAKİ İKİ SATIRLI ŞEYİN İLK SATIRINI SİLSEK VE AŞAĞIDA BİR YERE "YUKARIDA ÇIKIŞI TANIMLANMAMIŞ HER STATE-SEMBOL ÇİFTİ REJECT STATE'E GİDER" DEYİP KESİP ATSAK????!!!!!
Note that this transition merely introduces the simulated  coin outcome into the state, turning it into a member of $\qflip$.

% \hl{Furthermore, note that we have}
% \begin{equation*}
%     \delta\paren*{\paren*{S, q_V, \gamma, p}, \sigma} = \emptyset
% \end{equation*}
% \hl{when $p$ is a member of the} prefix\-/free \hl{$S$ (since no continuation of $p$ can then be a member of $S$).  When that is the case, $\paren*{S, q_V, \gamma, p}$ is no longer a member of \Qun{} by definition, and therefore such ``branches'' of $M$} \reject{} \hl{(whereas the lack of alternatives would lead a universal transition to accept).}!!!!!!!???BU SON PARANTEZİ VE ÖNCESİNDEKİ CÜMLENİN therefore'a KADARKİ KISMINI KOMPLE SİLSEK BİR ŞEY KAYBEDİYOR MUYUZ? LABELING FONKSIYONU ÇOCUĞU OLMAYAN BİR UNİVERSAL KONFİGÜRASYONU LABEL ETMİYOR DİYE DÜŞÜNÜYORUM AMA BU PARANTEZİ SİLEREK BUNLARDAN HİÇ BAHSETMEDEN ANLAŞABİLİYORSAK EN İYİSİ O!!!!!!!!!!!!
%BUNU NEREDE YAZDIK? SONRASINDA BİR ACCEPTING KONFİGURASYONA GİTMİYORSA NEDEN ACCEPT EDİYOR??!?!!! \utkanadd[80]{UTKAN: Labeling fonksoynunun tanımı o şekilde (ve bence o şekilde de olmalı).  Büyük $\bigwedge$/$\bigvee$/$\sum$/$\prod$ operatörleri empty set üzerinde ilgili işlemin etkisiz elemanına, yani \True{}/\False{}/0/1 değerlerine gider.  Debate analojisinde de refuter'ın hamlesiz kalması ("prover'ın yaptığı stratejik hamlelerle refuter'ı hamlesiz bırakması" da denebilir) durumuna, o şekilde refuter'ın debate'i kaybetmesine ve prover'ın kazanmasına tekabul ediyor.}

Following that first transition, the second transition finishes simulating the interaction of $V$ with its prover:
\begin{multline*}
    \delta\paren*{\paren*{S, q_V, \gamma, p, \bpub}, \sigma} = \\
        \Set{\paren*{\paren*{S, q'_V, \gamma'', p \cdot \bpub}, d_i} %\utkanadd[80]{\in \qmain \times \Delta}
        | \gamma'' \in \Gamma, \delta_V\paren*{q_V, \sigma, \blanksymb, \gamma, \bpub} = \paren*{q'_V, \blanksymb, \gamma', d_i, 0} %\utkanrem{\text{ and } \forall\gamma'' \in \Gamma}
        }
\end{multline*}
for all $\paren*{\paren*{S, q_V, \gamma, p, \bpub}, \sigma} \in \qflip \times \Sigma_{\bowtie}$.  Since  this transition is existential, every prover response %any one 
$\gamma'' \in \Gamma$ is
considered, as described for other transitions from communicating states of $V$ above. %lead $M$ to acceptance. 
The coin value $\bpub$ used in the present branch is added to the history of coin flips carried in the fourth component of the \tafa{} state.
% !!!BU İKİ CÜMLEYİ DEĞİŞTİRDİM??????!!!!!!!  \utkanadd{UTKAN: Tamamdir}

Note that states of the form $\paren*{S, q_V, \gamma, p}$ such that $q_V\in \Qpub$ and $p\in S$ are directed to \reject{} immediately, since they correspond to a scenario where the verifier tosses an additional coin that is not covered by the set $S$. %  it attains a state whose  current history of public coin outcomes is a member of S, since the prefix-free S then clearly does not allow for any more public coin tosses to be made, rendering the current guess of S invalid."!!!!!!!!!!

\paragraph{Halting.}
If the simulation of $V$ reaches a halting state, $M$ also halts with the same decision in its next transition, \ie,
\begin{equation*}
    \delta\paren*{\paren*{S, q_V, \gamma, p}, \sigma} = \begin{cases}
        \Set{\qacc} & \text{if } q_V = \qacc[V],\\
        \Set{\qrej} & \text{if } q_V = \qrej[V],
    \end{cases}
\end{equation*}
for all $\paren*{\paren*{S, q_V, \gamma, p}, \sigma} \in \qmain \times \Sigma_{\bowtie}$ where $q_V$ is either $\qacc[V]$ or $\qrej[V]$.
% !!!!!!!!!BURADA KARARI SANA BIRAKIYORUM; SEVDİĞİN ALTERNATİFİ KESİNLEŞTİR LÜTFEN!!!! \utkanadd{UTKAN: Hiç tercihim yok, olduğu gibi kalsın diyorum.}
% \utkanadd{Alternative:
% \begin{align*}
%     \delta\paren*{\paren*{S, \qacc[V], \gamma, p}, \sigma} &= \Set{\qacc},\\
%     \delta\paren*{\paren*{S, \qrej[V], \gamma, p}, \sigma} &= \Set{\qrej},
% \end{align*}
% for all $S \in \maj{r}$, $\gamma \in \Gamma$, $p \in \prefix{S}$, and $\sigma \in \Sigma_{\bowtie}$.

% UTKAN: Hatta "tanımsız her şeyi $\Set{ \qrej }$'e gönderme" kuralından ötürü $\qrej[V]$'li satır atılabilir bile.  Öyle yapılırsa belki "halting"leri de "accepting" yapmak gerekir.  (Bence $\qrej[V]$'li satırın kalması daha iyi.)
% }

\section{Proof of \texorpdfstring{\Cref{lem:privatesimulatespublic}}{Lemma \ref{lem:privatesimulatespublic}}}\label{sec:privatesimulatespublicproof}

\privatesimulatespublic*

\begin{proof}
    Let $V_1 = \paren*{Q_1,\Sigma,\Phi,\Gamma_1,\delta_1,q_0}$
    be a verifier that uses \OH{f(n)} private and \OH{g(n)} public random bits for some functions $f$ and $g$.  We construct a verifier $V_2 = \paren*{Q_2,\Sigma,\Phi,\Gamma_2,\delta_2,q_0}$ that uses \OH{f(n)+g(n)} private random bits and no public coins.
    The protocol that $V_2$ engages in with the prover is almost identical to that of $V_1$, with the exception that $V_1$'s public coin outcomes are emulated by private random bits of $V_2$ that are sent through the communication cell. 
    
    $Q_2$ is a superset of $Q_1$, where all the states that flip public coins in $V_1$ have been rebranded as  states that flip only private coins in $V_2$. 
    %to simulate $V_1$ directly (\ie, transition by transition), without flipping any public coins, as follows:
$V_2$'s program inherits all transitions  of $V_1$ emanating from states that do not flip public coins. $V_2$ simulates each transition of $V_1$ from a state $q$ that involves such a flip as described below.

If $q$ does not flip a private coin, the $V_1$ transition (cf.\ \Cref{tab:transitionfunction})
    \begin{align*}
        \delta_1\paren{q,\sigma,\phi,\gamma,\bpub} &= \paren{q',\phi',\gamma',d_i,d_w}
    \intertext{is replaced in $V_2$ by the transition}
        \delta_2\paren{q,\sigma,\phi,\gamma,\bpri} &= \paren{q',\phi',\paren{\gamma',\bpri},d_i,d_w},
    \end{align*}
making use of an additional symbol in $\Gamma_2$ that conveys both the original communication symbol $\gamma'$ and the simulated public coin outcome in a single package.  Note that $q$ remains a communication state in $V_2$ in this case.

If, on the other hand, $q$ flips a private coin as well, the  corresponding $V_1$ transition 
  \begin{align*}
    \delta_1\paren{q,\sigma,\phi,\gamma,\bpri,\bpub} &=
        % \paren{q',\phi',\paren{\gamma',\bpub},d_i,d_w}\\%\midrule
        \paren{q',\phi',\gamma',d_i,d_w}
%        \delta_2\paren{q,\sigma,\phi,\gamma,0,\bpub} &= \paren{q',\phi',\paren{\gamma',\bpub},d_i,d_w},
    \intertext{is simulated by two $V_2$ transitions}
        \delta_2\paren{q,\sigma,\phi,\gamma,\bpri_1} &= \paren{(q,\bpri_1),\phi,0,0},\\
        \delta_2\paren{(q,\bpri_1),\sigma,\phi,\gamma,\bpri_2} &= \paren{q',\phi',\paren{\gamma',\bpri_2},d_i,d_w},        
    \end{align*}
that flip two private coins, utilize new states in $Q_2$ to store the first coin outcome temporarily, and then send the required information through the communication cell.
In this case, $q$ becomes a non\-/communication state in $V_2$.  The two new states $(q,0)$ and $(q,1)$ (introduced for each such $q$) are branded as  communication states that also flip (only) private coins.
% !!!BURADAKİ MATERYALİ DOKÜMANIN DIŞINA KOPYALADIM 
   
   For each public coin flip removed from $V_2$'s transition function (and thereby its execution), exactly one private
    % public????!!!!PRIVATE???!!! Utkan: evet, duzelttim, affedersiniz
    coin flip is introduced.  $V_2$ emulates $V_1$, operating with the same verification error.  Its runtime is at most double that of $V_1$.  $V_2$ uses the same amount of work tape as $V_1$.
\end{proof}

% Before proceeding with the remaining of the proof, we make the remark that, by~\cite[Theorem~6.1]{CL89}, a double\-/exponential number of steps, $f_1(n)$, is enough for $V_1$ to reject nonmember input strings with a probability close to $1 - \verr_1$.
% Before proceeding with the remaining of the proof, we make the remark that, by~\cite[Theorem~6.1]{CL89}, for some linear function $l$ and any large constant $k$, $V_1$ will have rejected any nonmember input string of length $n$ in $f_1(n) = k^2 \cdot 2^{2^{l(n)}}$ steps with high probability that is at least $1 - \paren*{\verr_1 + 2^{-k} + k^{-1}}$.

% \end{filecontents*}

% \include{appendix.tex}

\end{document}